 \documentclass[preprint,prd,tightenlines, superscriptaddress]{revtex4-1}

\usepackage{graphicx} 
\usepackage{dcolumn}  
\usepackage{colordvi}
\usepackage{color}
\usepackage{epstopdf}
\usepackage{pstricks}
\usepackage{amssymb}
\usepackage{url}
\graphicspath{{ps}}
\usepackage{hyperref}
\usepackage{tabularx}
\usepackage{multirow}
\usepackage{units}
\usepackage{upgreek}
\usepackage{siunitx}
\usepackage{hyphenat}
\usepackage{subfigure}
\usepackage[italic]{hepnames}

\renewcommand{\PBzero}{\ensuremath{\HepParticle{\PB}{}{}^0}\xspace}
\renewcommand{\APBzero}{\ensuremath{\HepParticle{\APB}{}{}^0}\xspace}
\renewcommand{\PKzero}{\ensuremath{\HepParticle{\PK}{}{}^0}\xspace}
\renewcommand{\APDzero}{\ensuremath{\HepParticle{\APD}{}{}^0}\xspace}
\renewcommand{\Pgpz}{\ensuremath{\HepParticle{\Pgp}{}{}^0}\xspace}
\renewcommand{\PDzero}{\ensuremath{\HepParticle{\PD}{}{}^0}\xspace}
\renewcommand{\PKzS}{\ensuremath{\HepParticle{\PK}{}{}^0_{\rm S}}\xspace}

\begin{document}

\def\belletwo {\it {Belle~II}}

\clubpenalty = 10000  
\widowpenalty = 10000 

\vspace*{-3\baselineskip}
\resizebox{!}{3cm}{\includegraphics{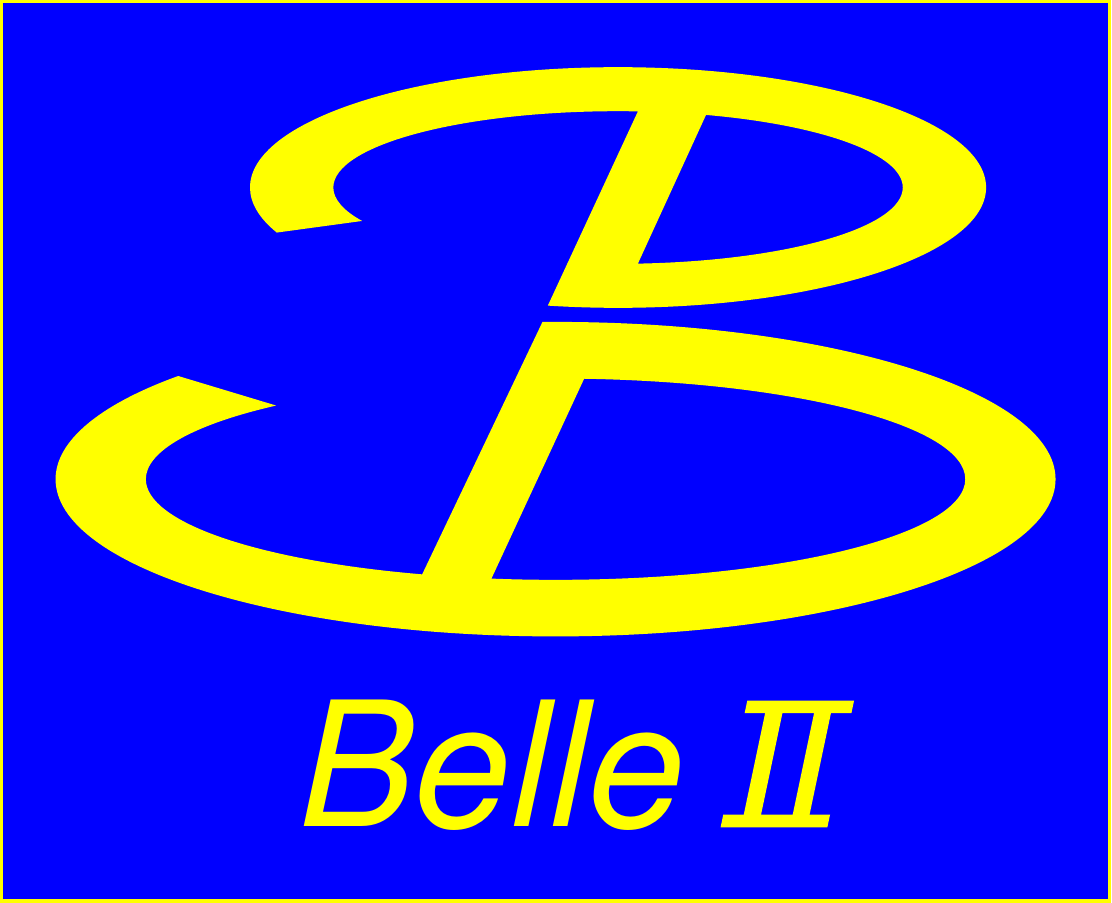}}

\vspace*{-5\baselineskip}
\begin{flushright}
BELLE2-CONF-PH-2020-012

\today
\end{flushright}

\quad\\[0.5cm]

\title {Measurements of branching fractions and CP-violating charge asymmetries in charmless $B$ decays reconstructed in 2019--2020 Belle~II data}

\newcommand{\instSinica}{Academia Sinica, Taipei 11529, Taiwan}
\newcommand{\instCPPM}{Aix Marseille Universit\'{e}, CNRS/IN2P3, CPPM, 13288 Marseille, France}
\newcommand{\instBeihang}{Beihang University, Beijing 100191, China}
\newcommand{\instBUAP}{Benemerita Universidad Autonoma de Puebla, Puebla 72570, Mexico}
\newcommand{\instBNL}{Brookhaven National Laboratory, Upton, New York 11973, U.S.A.}
\newcommand{\instBINP}{Budker Institute of Nuclear Physics SB RAS, Novosibirsk 630090, Russian Federation}
\newcommand{\instCMU}{Carnegie Mellon University, Pittsburgh, Pennsylvania 15213, U.S.A.}
\newcommand{\instCinvestavIPN}{Centro de Investigacion y de Estudios Avanzados del Instituto Politecnico Nacional, Mexico City 07360, Mexico}
\newcommand{\instPrague}{Faculty of Mathematics and Physics, Charles University, 121 16 Prague, Czech Republic}
\newcommand{\instChiangMai}{Chiang Mai University, Chiang Mai 50202, Thailand}
\newcommand{\instChiba}{Chiba University, Chiba 263-8522, Japan}
\newcommand{\instChonnam}{Chonnam National University, Gwangju 61186, South Korea}
\newcommand{\instConacyt}{Consejo Nacional de Ciencia y Tecnolog\'{\i}a, Mexico City 03940, Mexico}
\newcommand{\instDESY}{Deutsches Elektronen--Synchrotron, 22607 Hamburg, Germany}
\newcommand{\instDuke}{Duke University, Durham, North Carolina 27708, U.S.A.}
\newcommand{\instITAR}{Institute of Theoretical and Applied Research (ITAR), Duy Tan University, Hanoi 100000, Vietnam}
\newcommand{\instENEA}{ENEA Casaccia, I-00123 Roma, Italy}
\newcommand{\instEri}{Earthquake Research Institute, University of Tokyo, Tokyo 113-0032, Japan}
\newcommand{\instJuelich}{Forschungszentrum J\"{u}lich, 52425 J\"{u}lich, Germany}
\newcommand{\instFuJen}{Department of Physics, Fu Jen Catholic University, Taipei 24205, Taiwan}
\newcommand{\instFudan}{Key Laboratory of Nuclear Physics and Ion-beam Application (MOE) and Institute of Modern Physics, Fudan University, Shanghai 200443, China}
\newcommand{\instGoettingen}{II. Physikalisches Institut, Georg-August-Universit\"{a}t G\"{o}ttingen, 37073 G\"{o}ttingen, Germany}
\newcommand{\instGifu}{Gifu University, Gifu 501-1193, Japan}
\newcommand{\instSOKENDAI}{The Graduate University for Advanced Studies (SOKENDAI), Hayama 240-0193, Japan}
\newcommand{\instGyeongsang}{Gyeongsang National University, Jinju 52828, South Korea}
\newcommand{\instHanyang}{Department of Physics and Institute of Natural Sciences, Hanyang University, Seoul 04763, South Korea}
\newcommand{\instKEK}{High Energy Accelerator Research Organization (KEK), Tsukuba 305-0801, Japan}
\newcommand{\instJPARC}{J-PARC Branch, KEK Theory Center, High Energy Accelerator Research Organization (KEK), Tsukuba 305-0801, Japan}
\newcommand{\instHSE}{Higher School of Economics (HSE), Moscow 101000, Russian Federation}
\newcommand{\instIISER}{Indian Institute of Science Education and Research Mohali, SAS Nagar, 140306, India}
\newcommand{\instIITBhubaneswar}{Indian Institute of Technology Bhubaneswar, Satya Nagar 751007, India}
\newcommand{\instIITGuwahati}{Indian Institute of Technology Guwahati, Assam 781039, India}
\newcommand{\instIITHyderabad}{Indian Institute of Technology Hyderabad, Telangana 502285, India}
\newcommand{\instIITMadras}{Indian Institute of Technology Madras, Chennai 600036, India}
\newcommand{\instIndiana}{Indiana University, Bloomington, Indiana 47408, U.S.A.}
\newcommand{\instIHEPRussia}{Institute for High Energy Physics, Protvino 142281, Russian Federation}
\newcommand{\instHEPHYVienna}{Institute of High Energy Physics, Vienna 1050, Austria}
\newcommand{\instIHEPChina}{Institute of High Energy Physics, Chinese Academy of Sciences, Beijing 100049, China}
\newcommand{\instChennai}{Institute of Mathematical Sciences, Chennai 600113, India}
\newcommand{\instIPP}{Institute of Particle Physics (Canada), Victoria, British Columbia V8W 2Y2, Canada}
\newcommand{\instIOP}{Institute of Physics, Vietnam Academy of Science and Technology (VAST), Hanoi, Vietnam}
\newcommand{\instIFIC}{Instituto de Fisica Corpuscular, Paterna 46980, Spain}
\newcommand{\instFrascati}{INFN Laboratori Nazionali di Frascati, I-00044 Frascati, Italy}
\newcommand{\instNapoliINFN}{INFN Sezione di Napoli, I-80126 Napoli, Italy}
\newcommand{\instPadovaINFN}{INFN Sezione di Padova, I-35131 Padova, Italy}
\newcommand{\instPerugiaINFN}{INFN Sezione di Perugia, I-06123 Perugia, Italy}
\newcommand{\instPisaINFN}{INFN Sezione di Pisa, I-56127 Pisa, Italy}
\newcommand{\instRomaINFN}{INFN Sezione di Roma, I-00185 Roma, Italy}
\newcommand{\instRomaTreINFN}{INFN Sezione di Roma Tre, I-00146 Roma, Italy}
\newcommand{\instTorinoINFN}{INFN Sezione di Torino, I-10125 Torino, Italy}
\newcommand{\instTriesteINFN}{INFN Sezione di Trieste, I-34127 Trieste, Italy}
\newcommand{\instJAEA}{Advanced Science Research Center, Japan Atomic Energy Agency, Naka 319-1195, Japan}
\newcommand{\instMainz}{Johannes Gutenberg-Universit\"{a}t Mainz, Institut f\"{u}r Kernphysik, D-55099 Mainz, Germany}
\newcommand{\instGiessen}{Justus-Liebig-Universit\"{a}t Gie\ss{}en, 35392 Gie\ss{}en, Germany}
\newcommand{\instKarlsruhe}{Institut f\"{u}r Experimentelle Teilchenphysik, Karlsruher Institut f\"{u}r Technologie, 76131 Karlsruhe, Germany}
\newcommand{\instKennesaw}{Kennesaw State University, Kennesaw, Georgia 30144, U.S.A.}
\newcommand{\instKitasato}{Kitasato University, Sagamihara 252-0373, Japan}
\newcommand{\instKISTI}{Korea Institute of Science and Technology Information, Daejeon 34141, South Korea}
\newcommand{\instKorea}{Korea University, Seoul 02841, South Korea}
\newcommand{\instKSU}{Kyoto Sangyo University, Kyoto 603-8555, Japan}
\newcommand{\instKyotoU}{Kyoto University, Kyoto 606-8501, Japan}
\newcommand{\instKyungpook}{Kyungpook National University, Daegu 41566, South Korea}
\newcommand{\instLPI}{P.N. Lebedev Physical Institute of the Russian Academy of Sciences, Moscow 119991, Russian Federation}
\newcommand{\instLNNU}{Liaoning Normal University, Dalian 116029, China}
\newcommand{\instLMU}{Ludwig Maximilians University, 80539 Munich, Germany}
\newcommand{\instISUni}{Iowa State University,  Ames, Iowa 50011, U.S.A.}
\newcommand{\instLuther}{Luther College, Decorah, Iowa 52101, U.S.A.}
\newcommand{\instMNITJaipur}{Malaviya National Institute of Technology Jaipur, Jaipur 302017, India}
\newcommand{\instMPP}{Max-Planck-Institut f\"{u}r Physik, 80805 M\"{u}nchen, Germany}
\newcommand{\instMPGHLL}{Semiconductor Laboratory of the Max Planck Society, 81739 M\"{u}nchen, Germany}
\newcommand{\instMcGill}{McGill University, Montr\'{e}al, Qu\'{e}bec, H3A 2T8, Canada}
\newcommand{\instMETU}{Middle East Technical University, 06531 Ankara, Turkey}
\newcommand{\instMEPhI}{Moscow Physical Engineering Institute, Moscow 115409, Russian Federation}
\newcommand{\instNagoya}{Graduate School of Science, Nagoya University, Nagoya 464-8602, Japan}
\newcommand{\instNagoyaKMI}{Kobayashi-Maskawa Institute, Nagoya University, Nagoya 464-8602, Japan}
\newcommand{\instNagoyaIAR}{Institute for Advanced Research, Nagoya University, Nagoya 464-8602, Japan}
\newcommand{\instNaraWu}{Nara Women's University, Nara 630-8506, Japan}
\newcommand{\instUNAM}{National Autonomous University of Mexico, Mexico City, Mexico}
\newcommand{\instNTUTaiwan}{Department of Physics, National Taiwan University, Taipei 10617, Taiwan}
\newcommand{\instNUUTaiwan}{National United University, Miao Li 36003, Taiwan}
\newcommand{\instKrakow}{H. Niewodniczanski Institute of Nuclear Physics, Krakow 31-342, Poland}
\newcommand{\instNiigata}{Niigata University, Niigata 950-2181, Japan}
\newcommand{\instNSU}{Novosibirsk State University, Novosibirsk 630090, Russian Federation}
\newcommand{\instOkinawa}{Okinawa Institute of Science and Technology, Okinawa 904-0495, Japan}
\newcommand{\instOsakaCity}{Osaka City University, Osaka 558-8585, Japan}
\newcommand{\instRCNP}{Research Center for Nuclear Physics, Osaka University, Osaka 567-0047, Japan}
\newcommand{\instPNNL}{Pacific Northwest National Laboratory, Richland, Washington 99352, U.S.A.}
\newcommand{\instPanjab}{Panjab University, Chandigarh 160014, India}
\newcommand{\instPeking}{Peking University, Beijing 100871, China}
\newcommand{\instPanjabPAU}{Punjab Agricultural University, Ludhiana 141004, India}
\newcommand{\instRIKENMSL}{Meson Science Laboratory, Cluster for Pioneering Research, RIKEN, Saitama 351-0198, Japan}
\newcommand{\instRIKEN}{Theoretical Research Division, Nishina Center, RIKEN, Saitama 351-0198, Japan}
\newcommand{\instXavier}{St. Francis Xavier University, Antigonish, Nova Scotia, B2G 2W5, Canada}
\newcommand{\instSeoul}{Seoul National University, Seoul 08826, South Korea}
\newcommand{\instShandong}{Shandong University, Jinan 250100, China}
\newcommand{\instSPU}{Showa Pharmaceutical University, Tokyo 194-8543, Japan}
\newcommand{\instSoochow}{Soochow University, Suzhou 215006, China}
\newcommand{\instSoongsil}{Soongsil University, Seoul 06978, South Korea}
\newcommand{\instLjubljanaJSI}{J. Stefan Institute, 1000 Ljubljana, Slovenia}
\newcommand{\instKyiv}{Taras Shevchenko National Univ. of Kiev, Kiev, Ukraine}
\newcommand{\instTata}{Tata Institute of Fundamental Research, Mumbai 400005, India}
\newcommand{\instTUM}{Department of Physics, Technische Universit\"{a}t M\"{u}nchen, 85748 Garching, Germany}
\newcommand{\instECUTUM}{Excellence Cluster Universe, Technische Universit\"{a}t M\"{u}nchen, 85748 Garching, Germany}
\newcommand{\instTelAviv}{Tel Aviv University, School of Physics and Astronomy, Tel Aviv, 69978, Israel}
\newcommand{\instToho}{Toho University, Funabashi 274-8510, Japan}
\newcommand{\instTohoku}{Department of Physics, Tohoku University, Sendai 980-8578, Japan}
\newcommand{\instTitech}{Tokyo Institute of Technology, Tokyo 152-8550, Japan}
\newcommand{\instTokyoMetropolitan}{Tokyo Metropolitan University, Tokyo 192-0397, Japan}
\newcommand{\instUAS}{Universidad Autonoma de Sinaloa, Sinaloa 80000, Mexico}
\newcommand{\instNapoliUNIV}{Dipartimento di Scienze Fisiche, Universit\`{a} di Napoli Federico II, I-80126 Napoli, Italy}
\newcommand{\instNapoliUNIVA}{Dipartimento di Agraria, Universit\`{a} di Napoli Federico II, I-80055 Portici (NA), Italy}
\newcommand{\instPadovaUNIV}{Dipartimento di Fisica e Astronomia, Universit\`{a} di Padova, I-35131 Padova, Italy}
\newcommand{\instPerugiaUNIV}{Dipartimento di Fisica, Universit\`{a} di Perugia, I-06123 Perugia, Italy}
\newcommand{\instPisaUNIV}{Dipartimento di Fisica, Universit\`{a} di Pisa, I-56127 Pisa, Italy}
\newcommand{\instRomaUNIV}{Universit\`{a} di Roma ``La Sapienza,'' I-00185 Roma, Italy}
\newcommand{\instRomaTreUNIV}{Dipartimento di Matematica e Fisica, Universit\`{a} di Roma Tre, I-00146 Roma, Italy}
\newcommand{\instTorinoUNIV}{Dipartimento di Fisica, Universit\`{a} di Torino, I-10125 Torino, Italy}
\newcommand{\instTriesteUNIV}{Dipartimento di Fisica, Universit\`{a} di Trieste, I-34127 Trieste, Italy}
\newcommand{\instMontreal}{Universit\'{e} de Montr\'{e}al, Physique des Particules, Montr\'{e}al, Qu\'{e}bec, H3C 3J7, Canada}
\newcommand{\instIJCLab}{Universit\'{e} Paris-Saclay, CNRS/IN2P3, IJCLab, 91405 Orsay, France}
\newcommand{\instIPHC}{Universit\'{e} de Strasbourg, CNRS, IPHC, UMR 7178, 67037 Strasbourg, France}
\newcommand{\instAdelaide}{Department of Physics, University of Adelaide, Adelaide, South Australia 5005, Australia}
\newcommand{\instBonn}{University of Bonn, 53115 Bonn, Germany}
\newcommand{\instUBC}{University of British Columbia, Vancouver, British Columbia, V6T 1Z1, Canada}
\newcommand{\instCincinnati}{University of Cincinnati, Cincinnati, Ohio 45221, U.S.A.}
\newcommand{\instFlorida}{University of Florida, Gainesville, Florida 32611, U.S.A.}
\newcommand{\instHamburg}{University of Hamburg, 20148 Hamburg, Germany}
\newcommand{\instHawaii}{University of Hawaii, Honolulu, Hawaii 96822, U.S.A.}
\newcommand{\instHeidelberg}{University of Heidelberg, 68131 Mannheim, Germany}
\newcommand{\instLjubljanaUniLJ}{Faculty of Mathematics and Physics, University of Ljubljana, 1000 Ljubljana, Slovenia}
\newcommand{\instLouisville}{University of Louisville, Louisville, Kentucky 40292, U.S.A.}
\newcommand{\instMalaya}{National Centre for Particle Physics, University Malaya, 50603 Kuala Lumpur, Malaysia}
\newcommand{\instLjubljanaUM}{University of Maribor, 2000 Maribor, Slovenia}
\newcommand{\instMelbourne}{School of Physics, University of Melbourne, Victoria 3010, Australia}
\newcommand{\instMississippi}{University of Mississippi, University, Mississippi 38677, U.S.A.}
\newcommand{\instUOM}{University of Miyazaki, Miyazaki 889-2192, Japan}
\newcommand{\instNovaGorica}{University of Nova Gorica, 5000 Nova Gorica, Slovenia}
\newcommand{\instPittsburgh}{University of Pittsburgh, Pittsburgh, Pennsylvania 15260, U.S.A.}
\newcommand{\instUSTC}{University of Science and Technology of China, Hefei 230026, China}
\newcommand{\instSAlabama}{University of South Alabama, Mobile, Alabama 36688, U.S.A.}
\newcommand{\instSCarolina}{University of South Carolina, Columbia, South Carolina 29208, U.S.A.}
\newcommand{\instSydney}{School of Physics, University of Sydney, New South Wales 2006, Australia}
\newcommand{\instTabuk}{Department of Physics, Faculty of Science, University of Tabuk, Tabuk 71451, Saudi Arabia}
\newcommand{\instUTokyo}{Department of Physics, University of Tokyo, Tokyo 113-0033, Japan}
\newcommand{\instIPMU}{Kavli Institute for the Physics and Mathematics of the Universe (WPI), University of Tokyo, Kashiwa 277-8583, Japan}
\newcommand{\instVictoria}{University of Victoria, Victoria, British Columbia, V8W 3P6, Canada}
\newcommand{\instVPI}{Virginia Polytechnic Institute and State University, Blacksburg, Virginia 24061, U.S.A.}
\newcommand{\instWayneState}{Wayne State University, Detroit, Michigan 48202, U.S.A.}
\newcommand{\instYamagata}{Yamagata University, Yamagata 990-8560, Japan}
\newcommand{\instYerevan}{Alikhanyan National Science Laboratory, Yerevan 0036, Armenia}
\newcommand{\instYonsei}{Yonsei University, Seoul 03722, South Korea}
\affiliation{\instCPPM}
\affiliation{\instBeihang}
\affiliation{\instBNL}
\affiliation{\instBINP}
\affiliation{\instCMU}
\affiliation{\instCinvestavIPN}
\affiliation{\instPrague}
\affiliation{\instChiangMai}
\affiliation{\instChiba}
\affiliation{\instChonnam}
\affiliation{\instConacyt}
\affiliation{\instDESY}
\affiliation{\instDuke}
\affiliation{\instITAR}
\affiliation{\instEri}
\affiliation{\instJuelich}
\affiliation{\instFuJen}
\affiliation{\instFudan}
\affiliation{\instGoettingen}
\affiliation{\instGifu}
\affiliation{\instSOKENDAI}
\affiliation{\instGyeongsang}
\affiliation{\instHanyang}
\affiliation{\instKEK}
\affiliation{\instJPARC}
\affiliation{\instHSE}
\affiliation{\instIISER}
\affiliation{\instIITBhubaneswar}
\affiliation{\instIITGuwahati}
\affiliation{\instIITHyderabad}
\affiliation{\instIITMadras}
\affiliation{\instIndiana}
\affiliation{\instIHEPRussia}
\affiliation{\instHEPHYVienna}
\affiliation{\instIHEPChina}
\affiliation{\instIPP}
\affiliation{\instIOP}
\affiliation{\instIFIC}
\affiliation{\instFrascati}
\affiliation{\instNapoliINFN}
\affiliation{\instPadovaINFN}
\affiliation{\instPerugiaINFN}
\affiliation{\instPisaINFN}
\affiliation{\instRomaINFN}
\affiliation{\instRomaTreINFN}
\affiliation{\instTorinoINFN}
\affiliation{\instTriesteINFN}
\affiliation{\instISUni}
\affiliation{\instJAEA}
\affiliation{\instMainz}
\affiliation{\instGiessen}
\affiliation{\instKarlsruhe}
\affiliation{\instKitasato}
\affiliation{\instKISTI}
\affiliation{\instKorea}
\affiliation{\instKSU}
\affiliation{\instKyungpook}
\affiliation{\instLPI}
\affiliation{\instLNNU}
\affiliation{\instLMU}
\affiliation{\instLuther}
\affiliation{\instMNITJaipur}
\affiliation{\instMPP}
\affiliation{\instMPGHLL}
\affiliation{\instMcGill}
\affiliation{\instMEPhI}
\affiliation{\instNagoya}
\affiliation{\instNagoyaKMI}
\affiliation{\instNagoyaIAR}
\affiliation{\instNaraWu}
\affiliation{\instNTUTaiwan}
\affiliation{\instNUUTaiwan}
\affiliation{\instKrakow}
\affiliation{\instNiigata}
\affiliation{\instNSU}
\affiliation{\instOkinawa}
\affiliation{\instOsakaCity}
\affiliation{\instRCNP}
\affiliation{\instPNNL}
\affiliation{\instPanjab}
\affiliation{\instPeking}
\affiliation{\instPanjabPAU}
\affiliation{\instRIKENMSL}
\affiliation{\instSeoul}
\affiliation{\instSPU}
\affiliation{\instSoochow}
\affiliation{\instSoongsil}
\affiliation{\instLjubljanaJSI}
\affiliation{\instKyiv}
\affiliation{\instTata}
\affiliation{\instTUM}
\affiliation{\instTelAviv}
\affiliation{\instToho}
\affiliation{\instTohoku}
\affiliation{\instTitech}
\affiliation{\instTokyoMetropolitan}
\affiliation{\instUAS}
\affiliation{\instNapoliUNIV}
\affiliation{\instPadovaUNIV}
\affiliation{\instPerugiaUNIV}
\affiliation{\instPisaUNIV}
\affiliation{\instRomaUNIV}
\affiliation{\instRomaTreUNIV}
\affiliation{\instTorinoUNIV}
\affiliation{\instTriesteUNIV}
\affiliation{\instMontreal}
\affiliation{\instIJCLab}
\affiliation{\instIPHC}
\affiliation{\instAdelaide}
\affiliation{\instBonn}
\affiliation{\instUBC}
\affiliation{\instCincinnati}
\affiliation{\instFlorida}
\affiliation{\instHawaii}
\affiliation{\instHeidelberg}
\affiliation{\instLjubljanaUniLJ}
\affiliation{\instLouisville}
\affiliation{\instMalaya}
\affiliation{\instLjubljanaUM}
\affiliation{\instMelbourne}
\affiliation{\instMississippi}
\affiliation{\instUOM}
\affiliation{\instPittsburgh}
\affiliation{\instUSTC}
\affiliation{\instSAlabama}
\affiliation{\instSCarolina}
\affiliation{\instSydney}
\affiliation{\instUTokyo}
\affiliation{\instIPMU}
\affiliation{\instVictoria}
\affiliation{\instVPI}
\affiliation{\instWayneState}
\affiliation{\instYamagata}
\affiliation{\instYerevan}
\affiliation{\instYonsei}
  \author{F.~Abudin{\'e}n}\affiliation{\instTriesteINFN} 
  \author{I.~Adachi}\affiliation{\instKEK}\affiliation{\instSOKENDAI} 
  \author{R.~Adak}\affiliation{\instFudan} 
  \author{K.~Adamczyk}\affiliation{\instKrakow} 
  \author{P.~Ahlburg}\affiliation{\instBonn} 
  \author{J.~K.~Ahn}\affiliation{\instKorea} 
  \author{H.~Aihara}\affiliation{\instUTokyo} 
  \author{N.~Akopov}\affiliation{\instYerevan} 
  \author{A.~Aloisio}\affiliation{\instNapoliUNIV}\affiliation{\instNapoliINFN} 
  \author{F.~Ameli}\affiliation{\instRomaINFN} 
% ...
\collaboration{Belle II Collaboration}


\begin{abstract}
We report on first measurements of  branching fractions~($\mathcal{B}$) and CP-violating charge asymmetries~($\mathcal{A}$) in charmless $B$ decays at Belle~II.  We use a sample of electron-positron collisions collected in 2019 and 2020 at the $\Upsilon(4S)$ resonance and  corresponding to $34.6$\,fb$^{-1}$ of integrated luminosity. We use simulation to determine optimized event selections. The $\Delta E$ distributions of the resulting samples, restricted in $M_{\rm bc}$, are fit to determine signal yields ranging from 35 to 450 decays for the channels 
\mbox{$B^0 \to K^+\pi^-$},
 \mbox{$B^+ \to K^+\pi^0$},
\mbox{$B^+ \to \PKzS\pi^+$},
\mbox{$B^0 \to \PKzS\pi^0$},
\mbox{$B^0 \to \pi^+\pi^-$}, 
\mbox{$B^+ \to \pi^+\pi^0$}, 
\mbox{$B^+ \to K^+K^-K^+$},  and \mbox{$B^+ \to K^+\pi^-\pi^+$}.  Signal yields are corrected for efficiencies determined from simulation and control data samples to obtain the following results:
\begin{center}
$\mathcal{B}(B^0 \to K^+\pi^-) = [18.9 \pm 1.4(\rm stat) \pm 1.0(\rm syst)]\times 10^{-6}$,
\end{center}
\begin{center}
$\mathcal{B}(B^+ \to K^+\pi^0) = [12.7 ^{+2.2}_{-2.1} (\rm stat)\pm 1.1(\rm syst)]\times 10^{-6}$,
\end{center}
\begin{center}
$\mathcal{B}(B^+ \to K^0\pi^+) = [21.8 ^{+3.3}_{-3.0}(\rm stat) \pm 2.9(\rm syst)]\times 10^{-6}$,
\end{center}
\begin{center}
$\mathcal{B}(B^0 \to K^0\pi^0) = [10.9^{+2.9}_{-2.6} (\rm stat)\pm 1.6(\rm syst)]\times 10^{-6}$,
\end{center}
\begin{center}
$\mathcal{B}(B^0 \to \pi^+\pi^-) = [5.6 ^{+1.0}_{-0.9}(\rm stat) \pm 0.3(\rm syst)]\times 10^{-6}$,
\end{center}
\begin{center}
$\mathcal{B}(B^+ \to \pi^+\pi^0) = [5.7 \pm 2.3 (\rm stat)\pm 0.5(\rm syst)]\times 10^{-6}$,
\end{center}
\begin{center}
$\mathcal{B}(B^+ \to K^+K^-K^+) = [32.0 \pm 2.2(\rm stat.) \pm 1.4 (\rm syst)]\times 10^{-6}$, 
\end{center}
\begin{center}
$\mathcal{B}(B^+ \to K^+\pi^-\pi^+) = [48.0 \pm 3.8 (\rm stat)\pm 3.3 (\rm syst)]\times 10^{-6}$,
\end{center}
\begin{center}
$\mathcal{A}_{\rm CP}(B^0 \to K^+\pi^-) = 0.030 \pm 0.064 (\rm stat) \pm 0.008(\rm syst)$,
\end{center}
\begin{center}
$\mathcal{A}_{\rm CP}(B^+ \to K^+\pi^0) = 0.052 ^{+0.121}_{-0.119} (\rm stat)\pm 0.022(\rm syst)$,
\end{center}
\begin{center}
$\mathcal{A}_{\rm CP}(B^+ \to K^0\pi^+) = -0.072 ^{+0.109}_{-0.114}(\rm stat) \pm 0.024(\rm syst)$,
\end{center}
\begin{center}
$\mathcal{A}_{\rm CP}(B^+ \to \pi^+\pi^0) = -0.268 ^{+0.249}_{-0.322} (\rm stat)\pm 0.123(\rm syst)$,
\end{center}
\begin{center}
$\mathcal{A}_{\rm CP}(B^+ \to K^+K^-K^+) = -0.049 \pm 0.063(\rm stat) \pm 0.022 (\rm syst)$, and 
\end{center}
\begin{center}
$\mathcal{A}_{\rm CP}(B^+ \to K^+\pi^-\pi^+) = -0.063 \pm 0.081 (\rm stat)\pm 0.023(\rm syst)$.
\end{center}
 These are the first measurements in charmless decays reported by Belle~II. Results are compatible with known determinations and show detector performance comparable with the best Belle results offering a reliable basis to assess projections for future reach.
 
\keywords{Belle~II, charmless, phase 3}
\end{abstract}

\pacs{}

\maketitle

{\renewcommand{\thefootnote}{\fnsymbol{footnote}}}
\setcounter{footnote}{0}



\section{Introduction and motivation}

The study of charmless $B$ decays is a keystone of the worldwide flavor program. Processes mediated by $b\to u\bar{u}d$ transitions offer direct access to the unitarity angle $\upphi_2/\upalpha$ and probe contributions of non-standard-model dynamics in loops.  However, reliable extraction of weak phases and unambiguous interpretation of measurements involving loop amplitudes is spoiled by large hadronic uncertainties,  which are rarely tractable in perturbative  calculations. Appropriately chosen combinations of measurements from decay modes related by flavor symmetries are used to reduce the impact of such  unknowns. An especially  fruitful approach consists in combining measurements of decays related by isospin symmetries. For instance, the combined analysis of branching fractions and CP-violating asymmetries of the whole set of   $B \to \pi\pi$ isospin partners (with $B$ and $\pi$ charged or neutral) enables a determination of~$\upphi_2/\upalpha$~\cite{Gronau:1990ka}. Similarly, isospin constraints between $B\to K\pi$ decays result in simple additive relationships  between branching fractions and CP-violating asymmetries, which may offer a stringent null test of the standard model sensible to the presence of non-SM dynamics~\cite{Gronau:2005kz}.

The Belle  II  physics  program,  featuring the  {\it unique}  capability  of  studying  jointly,  and within a consistent experimental environment, all relevant two-, three-, and multi-body final states is therefore particularly promising. This ability can enable significant advances, including an improved determination of the quark-mixing-matrix angle $\upphi_2/\upalpha$,  a conclusive understanding of long-standing anomalies such as the so-called $K\pi$ CP-puzzle, and a thorough investigation of charge-parity-violating asymmetries localized in the phase space of three-body $B$ decays.

The Belle~II detector, complete with its vertex detector, started its collision operations on March 11 2019 and continued until July 1, 2020.  The sample of electron-positron collisions used in this work corresponds to an integrated luminosity of $34.6\,\si{fb^{-1}}$~\cite{Abudinen:2019osb}  and was collected at the $\Upsilon(4{\rm S})$~resonance as of May 14, 2020.
This \mbox{document} reports on the first measurement of branching fractions and CP-violating charge asymmetries in  charmless decays at Belle~II, which follows the first reconstruction of charmless $B$ decays in Belle II data~\cite{Benedikt:2019,CharmlessMoriond:2020}.

We focus on two- and three-body charmless decays with branching fractions of $10^{-6}$, or larger, into final states sufficiently simple to obtain visible signals in the current data set with a relatively straightforward reconstruction.  The target decay modes are  \mbox{$B^0 \to K^+ \pi^-$},    \mbox{$B^+ \to K^+\pi^0(\to \gamma\gamma)$}, \mbox{$B^+ \to \PKzS(\to \pi^+\pi^-) \pi^+$}, \mbox{$B^0 \to \PKzS(\to \pi^+\pi^-) \pi^0(\to \gamma\gamma)$},  \mbox{$B^0 \to \pi^+\pi^-$},  \mbox{$B^+ \to \pi^+\pi^0(\to \gamma\gamma)$}, \mbox{$B^+ \to K^+K^-K^+$},  and \mbox{$B^+ \to K^+\pi^-\pi^+$}.
Charge-conjugate processes are implied in what follows except when otherwise stated.\par
The reconstruction strategy and procedures are developed and finalized in simulated data. They are then applied and refined on a data subset corresponding to 1/4 of the sample prior to applying it to the full sample.
Most of the analysis uses the following variables, which are known to be strongly discriminating between $B$ signal and background from $e^+e^- \to q\bar{q}$ continuum events, where $q$ indicates any quark of the first or second family (i.e., $u$, $d$, $s$, and $c$), and (in the case of $\Delta E$) background from non-signal $B$ decays:
\begin{itemize}
    \item the energy difference $\Delta E \equiv E^{*}_{B} - \sqrt{s}/2$ between the total energy of the reconstructed $B$ candidate and half of the collision energy, both in the $\Upsilon(4S)$ frame;
    \item the beam-energy-constrained mass $M_{\rm bc} \equiv \sqrt{s/(4c^4) - (p^{*}_B/c)^2}$, which is the invariant mass of the $B$ candidate where the $B$ energy is replaced by the (more precisely known) half of the center-of-mass collision energy. 
\end{itemize}

\section{The Belle~II detector}
Belle~II is a $4\pi$ particle-physics spectrometer~\cite{Kou:2018nap, Abe:2010sj}, designed to reconstruct the products of electron-positron collisions produced by the SuperKEKB asymmetric-energy collider~\cite{Akai:2018mbz}, located at the KEK laboratory in Tsukuba, Japan. Belle~II comprises several subdetectors arranged around the interaction space-point in a cylindrical geometry. The innermost subdetector is the vertex detector, which uses position-sensitive silicon layers to sample the trajectories of charged particles (tracks) in the vicinity of the interaction region to extrapolate the decay positions of their long-lived parent particles. The vertex detector includes two inner layers of silicon pixel sensors and four outer layers of silicon microstrip sensors. The second pixel layer is currently incomplete and covers only a small portion of azimuthal angle. Charged-particle momenta and charges are measured by a large-radius, helium-ethane, small-cell central drift chamber, which also offers charged-particle-identification information through a measurement of particles' energy-loss by specific ionization. A Cherenkov-light angle and time-of-propagation detector surrounding the chamber provides charged-particle identification in the central detector volume, supplemented by proximity-focusing, aerogel, ring-imaging Cherenkov detectors in the forward regions. A CsI(Tl)-crystal electromagnetic calorimeter allows for energy measurements of electrons and photons.  A solenoid surrounding the calorimeter generates a uniform axial 1.5\,T magnetic field filling its inner volume. Layers of plastic scintillator and resistive-plate chambers, interspersed between the
magnetic flux-return iron plates, allow for identification of $K^0_{\rm L}$ and muons.
The subdetectors most relevant for this work are the silicon vertex detector, the tracking drift chamber, the particle-identification detectors, and the electromagnetic calorimeter.

\section{Selection and reconstruction}
\label{sec:selection}

We reconstruct the two-body decays

\begin{itemize}
    \item $B^0 \to K^+\, \pi^-$,
    \item $B^+ \to K^+\, \pi^0 (\to \gamma\gamma)$,    
     \item $B^+ \to K_{\rm S}^0(\to \pi^+\pi^-)\, \pi^+$,
    \item $B^0 \to \PKzS(\to \pi^+\pi^-)\pi^0(\to \gamma\gamma)$     
    \item $B^0 \to \pi^+\, \pi^-$,
    \item $B^+ \to \pi^+\, \pi^0(\to \gamma\gamma)$,
\end{itemize}
and three-body decays
\begin{itemize}
    \item $B^+ \to K^+\, K^-\, K^+$,
    \item $B^+ \to K^+\, \pi^+\, \pi^-$.
\end{itemize}
In addition, we use the control channels
\begin{itemize}
\item \mbox{$B^+ \to \overline{D}^0 (\to K^+ \pi^- \pi^0)\, \pi^+$}, 
\item \mbox{$B^+ \to \overline{D}^0 (\to K^+ \pi^-)\, \pi^+$}, 
\item \mbox{$B^0 \to D^{*-}(\to \overline{D}^0 (\to K^+ \pi^- \pi^0)\, \pi^-)\, \pi^+$}, 
\item \mbox{$B^0 \to D^{*-}(\to \overline{D}^0 (\to K^+ \pi^-)\, \pi^-)\, \pi^+$}, 
\item \mbox{$D^+ \to \PKzS\pi^+$},
\item \mbox{$\PDzero \to K^-\pi^+$}, 
\end{itemize}
for validation of continuum-suppression discriminating variables; optimization of the $\pi^0$ selection; determination of $\pi^0$ selection efficiency; assessment of data-simulation discrepancies in the distributions of drift-chamber hits, particle-identification likelihoods, and continuum-background suppression variables; and determination of instrumental asymmetries. 


\subsection{Simulated and experimental data}
We use generic simulated data to optimize the event selection and compare the distributions observed in experimental data with expectations. We use signal-only simulated data to model relevant signal features for fits and determine selection efficiencies. 
Generic simulation consists of Monte Carlo samples that include $B^0\overline{B}^0$, $B^+B^-$, $u\bar{u}$, $d\bar{d}$, $c\bar{c}$, and $s\bar{s}$ processes in realistic proportions and corresponding in size to 2--20  times the $\Upsilon$(4S) data. In addition, $2\times 10^6$ signal-only events are generated for each channel~\cite{Ryd:2005zz}.
Three-body decays are generated assuming a simplified Dalitz plot structure where major resonances are present but no interferences are simulated. \par 
As for experimental data, we use all 2019--2020  $\Upsilon$(4S) good-quality runs collected until May~14, 2020 and corresponding to an integrated luminosity of $34.6\,\si{fb^{-1}}$. All events are required to satisfy loose data-skim selection criteria, based on total energy and charged-particle multiplicity in the event, targeted at reducing sample sizes to a manageable level with negligible impact on signal efficiency. All data are processed using the Belle~II analysis software framework~\cite{Kuhr:2018lps}.

\subsection{Reconstruction and baseline selection}
We form final-state particle candidates by applying loose baseline selection criteria and then combine candidates in kinematic fits consistent with the topologies of the desired decays to reconstruct intermediate states and $B$ candidates. \par We reconstruct charged pion and kaon candidates by starting from the most inclusive charged-particle classes  and by requiring fiducial criteria that restrict them to the full polar-angle acceptance in the central drift chamber ($\SI{17}{\degree}<\theta<\SI{150}{\degree}$) and to loose ranges of displacement from the nominal interaction space-point (radial displacement $|dr|<\SI{0.5}{cm}$ and longitudinal displacement $|dz|<\SI{3}{cm}$) to reduce beam-background-induced tracks, which do not originate from the interaction region preferably.
We reconstruct neutral-pion candidates by combining photons with energies greater than about $20$\,MeV in pairs restricted in diphoton mass and excluding extreme helicity-angle values  to suppress combinatorial background from collinear soft photons. The mass of the $\pi^0$ candidates is constrained to its known value in subsequent kinematic fits. 
For $K_{\rm S}^0$ reconstruction, we use pairs of oppositely charged particles that originate from a common space-point and have dipion mass consistent with a $K_{\rm S}^0$. To reduce combinatorial background, we apply additional requirements, dependent on $K_{\rm S}^0$~momentum, on the distance between trajectories of the two charged-pion candidates, the $K^0_{\rm S}$~flight distance, and the angle between the pion-pair momentum and the direction of the $K^0_{\rm S}$~flight.

The resulting $K^+$, $\pi^+$, $\pi^0$, and $\PKzS$ candidates are combined through kinematic simultaneous fits of the whole decay chain into each of our target signal channels, consistent with the desired topology. A constraint on the position of the interaction region is used in fits of candidates with a final-state $\pi^0$. In addition, we reconstruct the vertex of the accompanying tag-side $B$ mesons using  all tracks in the tag-side and identify the flavor, which is used as  input to the continuum-background discriminator, using a category-based flavor tagger~\cite{Abudinen:2018}.
The reconstruction of the control channels is conceptually similar.
\par
Simulation is used to identify and suppress contamination from peaking backgrounds, that is, misreconstructed events clustering in the signal region $M_{\rm bc} > 5.27$\,GeV/$c^2$ and \mbox{$-0.15 < \Delta E < 0.15$ GeV}. 

 Sizable peaking backgrounds affect the $B^0 \to K^+K^-K^+$ and $B^+ \to K^+\pi^-\pi^+$ samples.
Dominant \mbox{$B^0 \to \overline{D}^0(\to K^+K^-)K^+$}, \mbox{$B^0 \to \eta_c(\to K^+K^-)K^+$}, and \mbox{$B^0 \to \chi_{c1}(\to K^+K^-)K^+$} contributions to the  
$B^0 \to K^+K^-K^+$ sample are suppressed by excluding the two-body mass ranges \mbox{$1.84 < m(K^+K^-) < 1.88$ GeV/c$^2$}, \mbox{$2.94 < m(K^+K^-) < 3.05$ GeV/c$^2$}, and \mbox{$3.50 < m(K^+K^-) < 3.54$ GeV/$c^2$}, respectively. \\
The $B^+ \to K^+\pi^-\pi^+$ channel is contaminated by $B$ decays proceeding through charmed intermediate states, such as  \mbox{$B^+ \to \overline{D}^0(\to K^+\pi^-)\pi^+$}, \mbox{$B^+ \to \eta_c(\to \pi^+\pi^-)K^+$}, $B^+ \to \chi_{c1}(\to \pi^+\pi^-)K^+$, and 
\mbox{$B^+ \to \eta_c(2S)(\to \pi^+\pi^-)K^+$}, and intermediate resonances decaying to muons misidentified as pions such as \mbox{$B^+ \to J/\psi(\to \mu^+\mu^-)K^+$} and 
\mbox{$B^+ \to \psi(2S)(\to \mu^+\mu^-)K^+$}.
These are suppressed by excluding the two-body mass ranges  \mbox{$1.8 < m(K^+\pi^-) < 1.92$ GeV/c$^2$},\linebreak
\mbox{$0.93 < m(\pi^+\pi^-) < 3.15$ GeV/c$^2$}, \mbox{$3.45 < m(\pi^+\pi^-) < 3.525$ GeV/c$^2$}, $62 < m(\pi^+\pi^-) < 3.665$ GeV/c$^2$, \mbox{$3.67 < m(\pi^+\pi^-) < 3.72$ GeV/c$^2$}. In addition, we veto the genuine charmless \mbox{$B^+ \to K^*(892)^0\pi^+$} subcomponent by excluding candidates with $0.82 < m(K^+\pi^-) < 0.98$ GeV/c$^2$ to be able to compare our results consistently with the branching fraction reported in Ref.~\cite{PDG} where this component is not included.


\subsection{Continuum suppression}
The main challenge in reconstructing significant charmless signals is the large contamination from continuum background. To discriminate against such background, we use a binary boosted decision-tree classifier that combines nonlinearly 39 variables known to provide statistical discrimination between $B$-meson signals and continuum and to be loosely correlated, or uncorrelated,  with $\Delta E$ and $M_{\rm bc}$. The variables include quantities associated to event topology (global and signal-only angular configurations), flavor-tagger information, vertex separation and uncertainty information, and kinematic-fit quality information. We train the classifier to identify statistically significant signal and background features using unbiased simulated samples. 

\begin{figure}[h!]
 \centering
    \centering
    \subfigure{\includegraphics[width=0.49\textwidth]{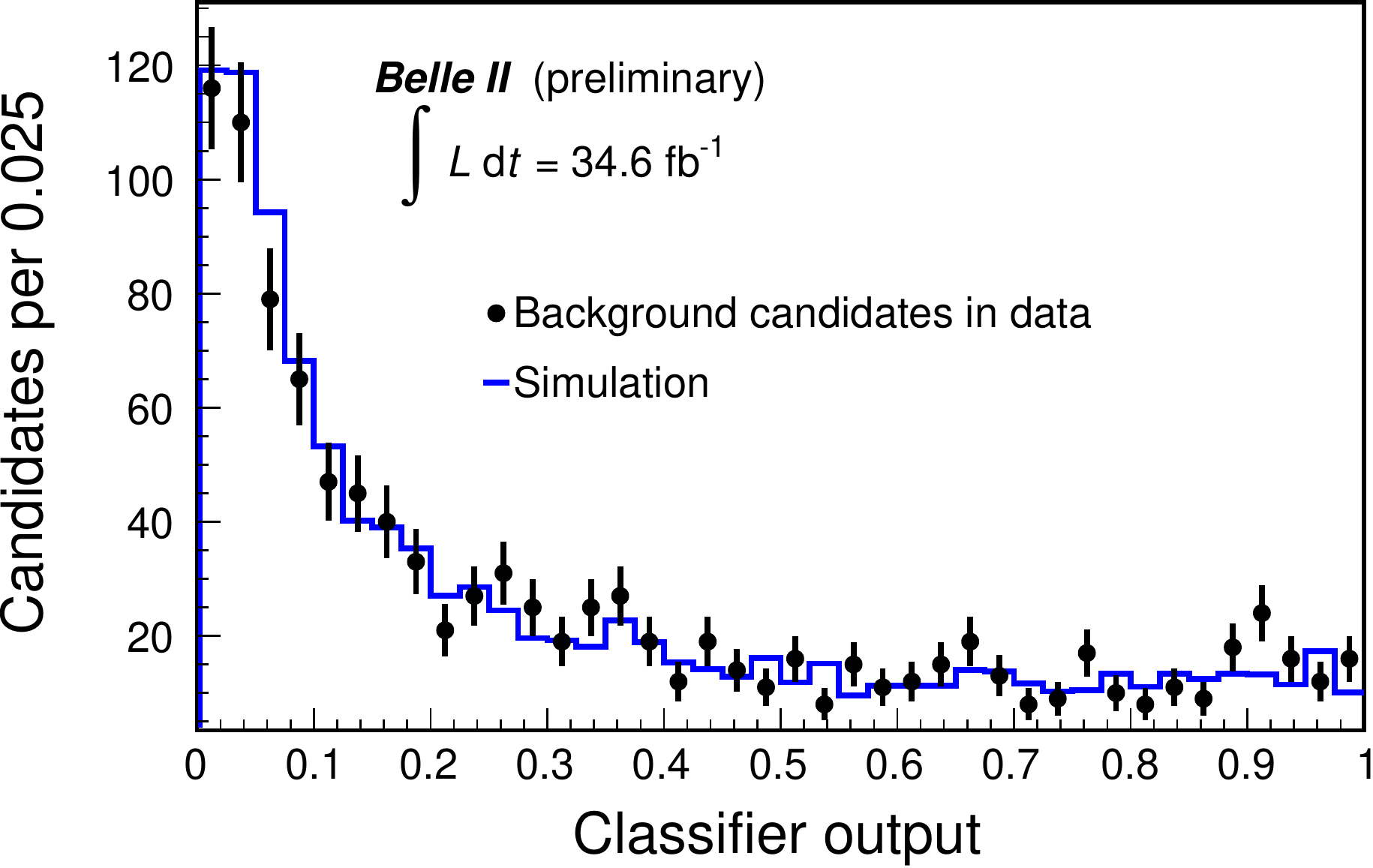}} \hfill
    \subfigure{\includegraphics[width=0.49\textwidth]{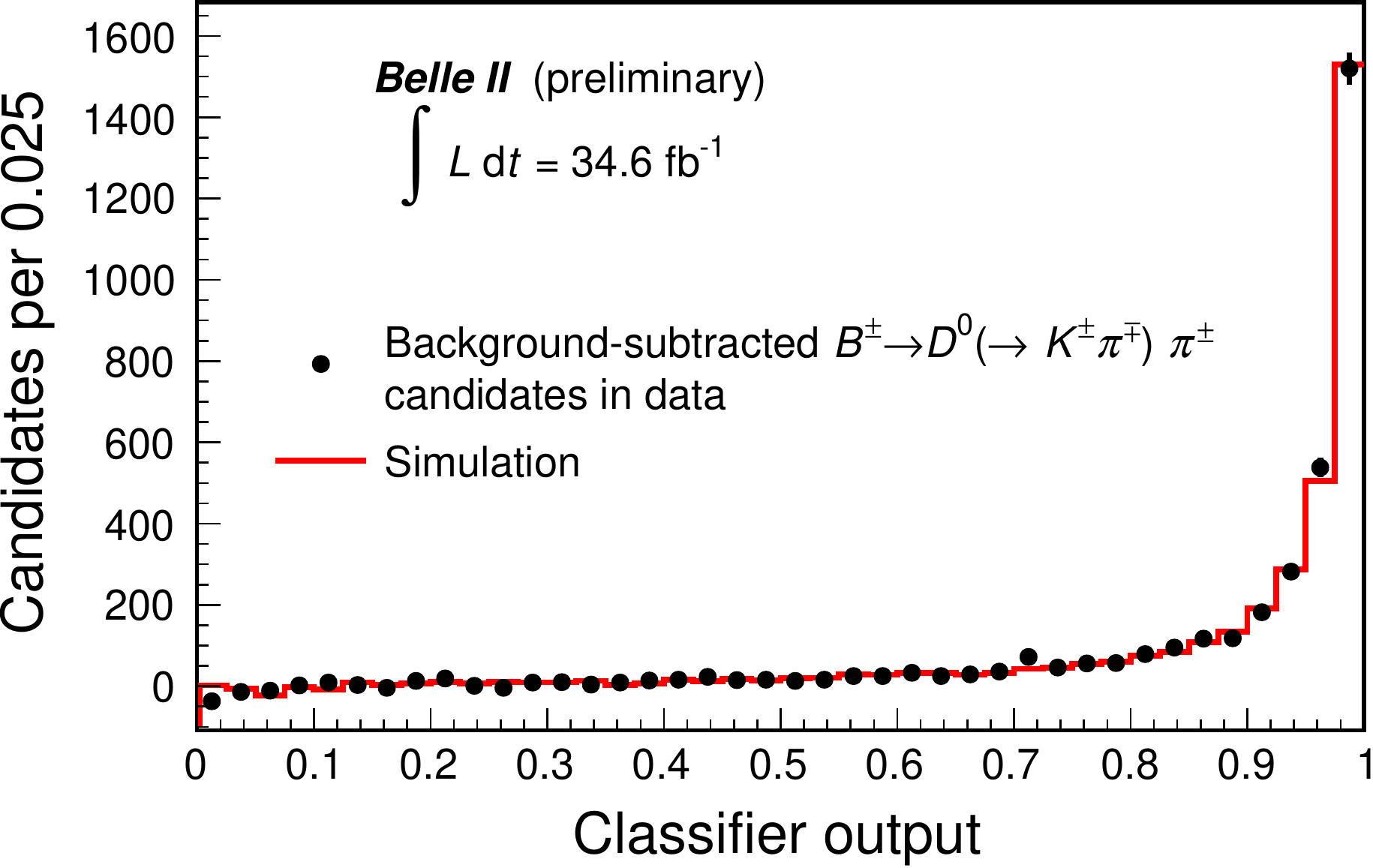}} \\
 \caption{Data-simulation comparison of the output of the boosted decision-tree classifier on (left)~side-band and (right)~side-band-subtracted $\PBplus\to\APD^{0}(\to \PKp\Pgpm)\,\Pgpp$~candidates in the signal region.}
 \label{fig:outputData_Kpi}
\end{figure}

We validate the input and output distributions of the classifier by comparing  data with simulation using control samples.
Figure~\ref{fig:outputData_Kpi} shows the distribution of the output for  \mbox{$\PBplus\to\APD^{0}(\to \PKp\Pgpm)\,\Pgpp$}~candidates reconstructed in data and simulation. No inconsistency is observed. 

\section{Optimization of the signal selection}
\label{sec:optimization}
For each channel, we optimize the selection to isolate abundant, low-background signals using simulated and control-sample data. We vary the selection criteria on continuum-suppression output, charged-particle identification information, and choice of $\pi^0$ (when appropriate) to maximize ${\rm S}/\sqrt{{\rm S}+{\rm B}}$, where ${\rm S}$ and ${\rm B}$ are signal and background yields, respectively, estimated in the same signal-rich region used in the analysis. Continuum-suppression and particle-identification requirements are optimized simultaneously using simulated data. The $\pi^0$ selection is optimized independently by using control \mbox{$B^+ \to \overline{D}^0(\to K^+\pi^-\pi^0)\pi^+$} decays in which S is the \mbox{$B^+ \to \overline{D}^0(\to K^+\pi^-\pi^0)\pi^+$} signal yield, scaled to the expected \mbox{$B^+ \to K^+\pi^0$} yield, and B is the background observed in an $M_{\rm bc}$ sideband of \mbox{$B^+ \to K^+\pi^0$}.


\section{Determination of signal yields}
\label{sec:yields}
More than one candidate per event populates the resulting $\Delta E$ distributions, with average multiplicities ranging from 1.0 to 1.2. We restrict to one candidate per event as follows. For channels with $\pi^0$, we first select the $\pi^0$ candidate with the highest $p$-value of the mass-constrained diphoton fit. If more than one candidate remains, and for all other channels, we select a single $B$ candidate randomly. \par Signal yields are determined with maximum likelihood fits of the unbinned $\Delta E$ distributions of candidates restricted to the signal region $M_{\rm bc} > 5.27$\,GeV/$c^2$ and $-0.15 < \Delta E < 0.15$ GeV. Fit models are determined empirically from simulation, with the only additional flexibility of a global shift of peak positions determined in data when suggested by likelihood-ratio tests. Because of the small sample size, in fits of $B^0 \to \PKzS \pi^0$ candidates the global shift is Gaussian-constrained to the value observed in $B^+ \to K^+ \pi^0$ candidates. Similarly, the $B^+ \to K^+ \pi^0$ and  $B^+ \to \pi^+ \pi^0$ yields are determined through a simultaneous fit of two independent data sets.\par
\begin{figure}[htb]
 \centering
 \includegraphics[width=0.8\textwidth]{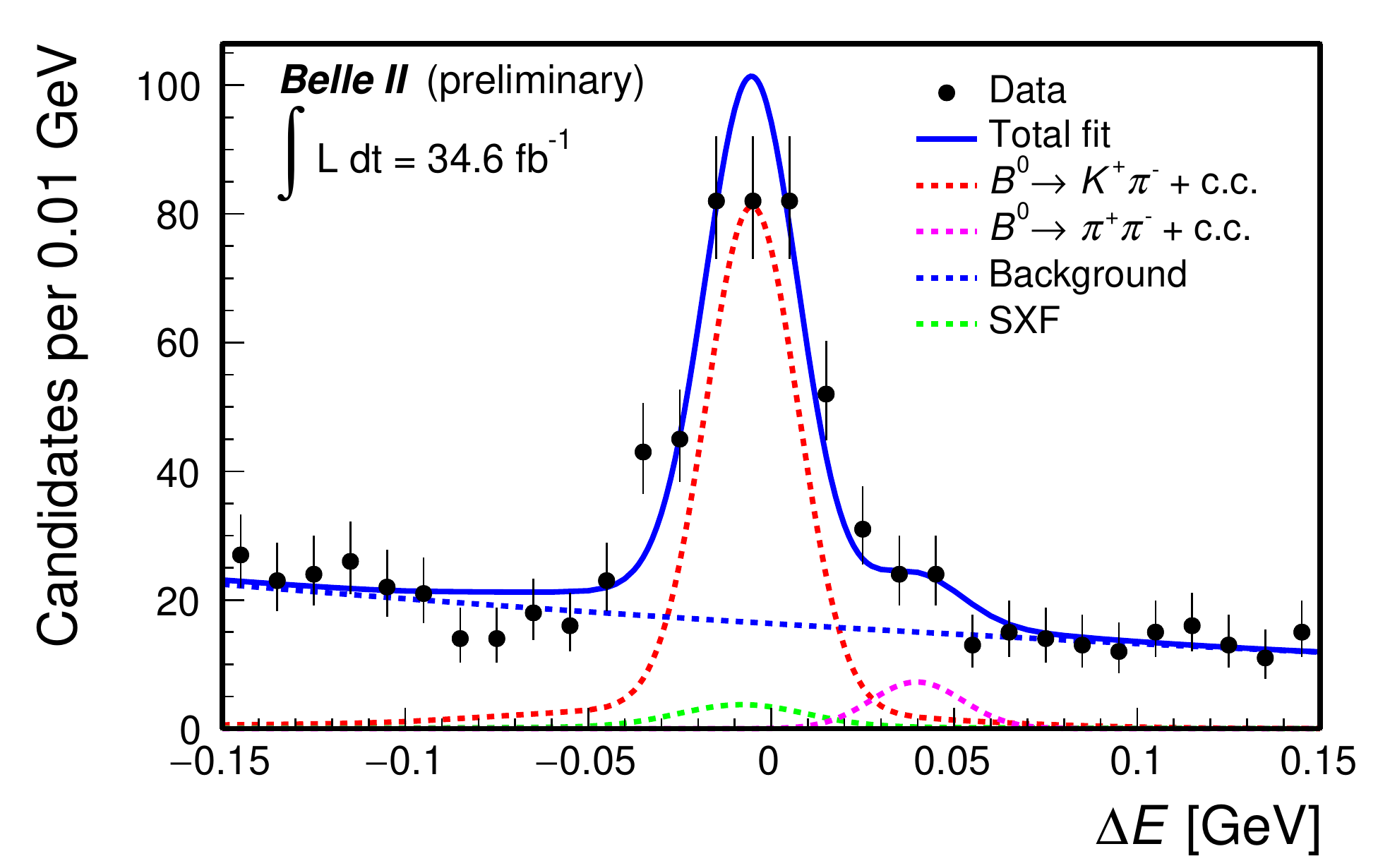}
 \caption{Distribution of $\Delta E$ for $B^0 \to K^+\pi^-$  candidates reconstructed in 
 2019--2020 Belle~II data selected through the baseline criteria with an optimized continuum-suppression and kaon-enriching selection, and further restricted to $M_{\rm bc} > 5.27$\,GeV/$c^2$. A misreconstructed $\pi^+\pi^-$~component modeled with a Gaussian is included with a displacement from the $K^+\pi^-$ peak fixed to the known value. The global position of the two peaks is determined by the fit. The `SxF'(self cross-feed) label indicate candidates formed by misidentified (swapped mass assignments) signal particles. The projection of an unbinned maximum likelihood fit is overlaid.}
 \label{fig:K+pi-}
\end{figure}
\begin{figure}[htb]
 \centering
 \includegraphics[width=0.8\textwidth]{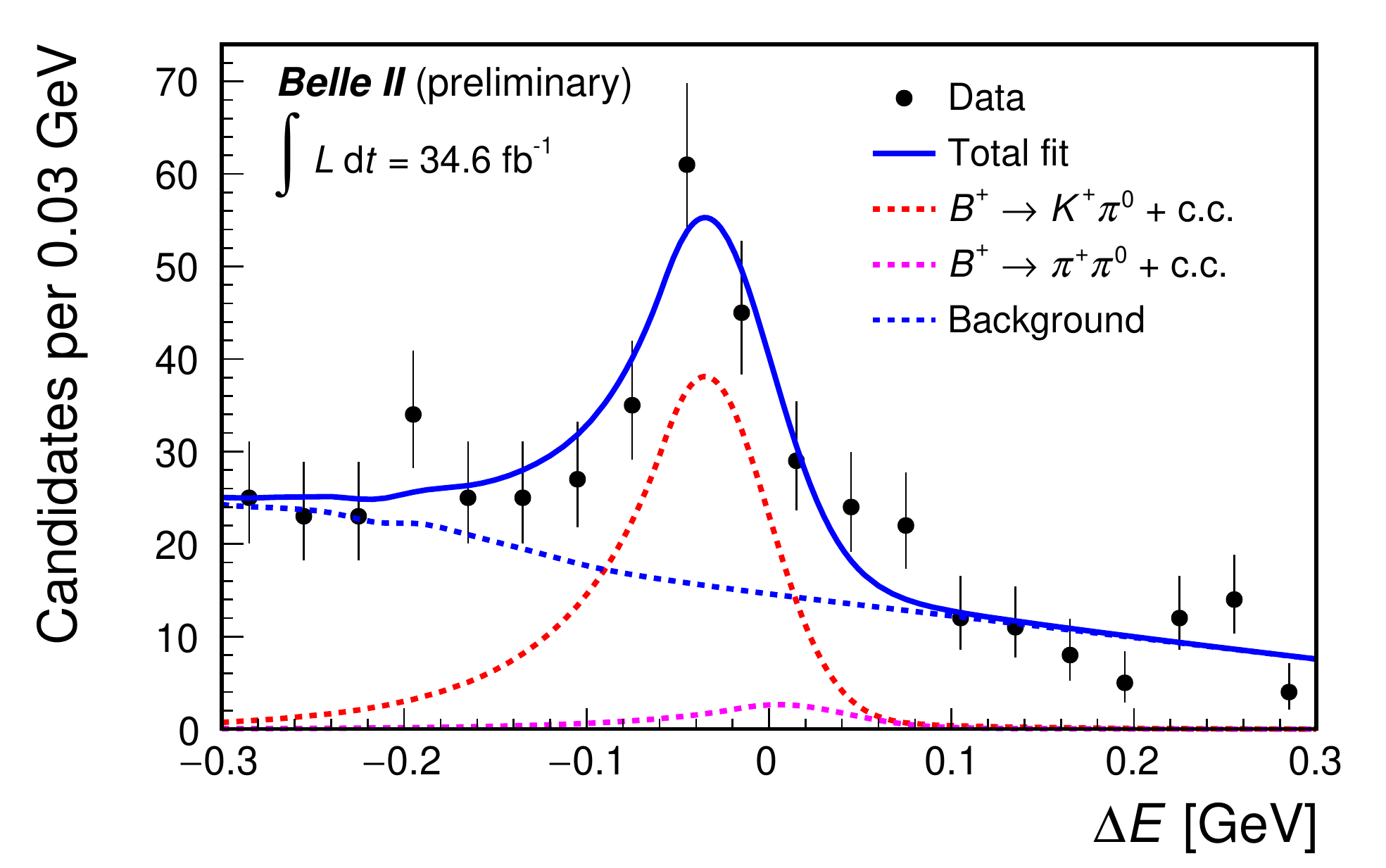}
 \caption{Distribution of $\Delta E$ for $B^+ \to K^+\pi^0$  candidates reconstructed in 
 2019--2020 Belle~II data selected through the baseline criteria with an optimized continuum-suppression and kaon-enriching selection, and further restricted to $M_{\rm bc} > 5.27$\,GeV/$c^2$. The projection of an unbinned maximum likelihood fit is overlaid.}
 \label{fig:K+pi0}
\end{figure}
\begin{figure}[htb]
 \centering
 \includegraphics[width=0.8\textwidth]{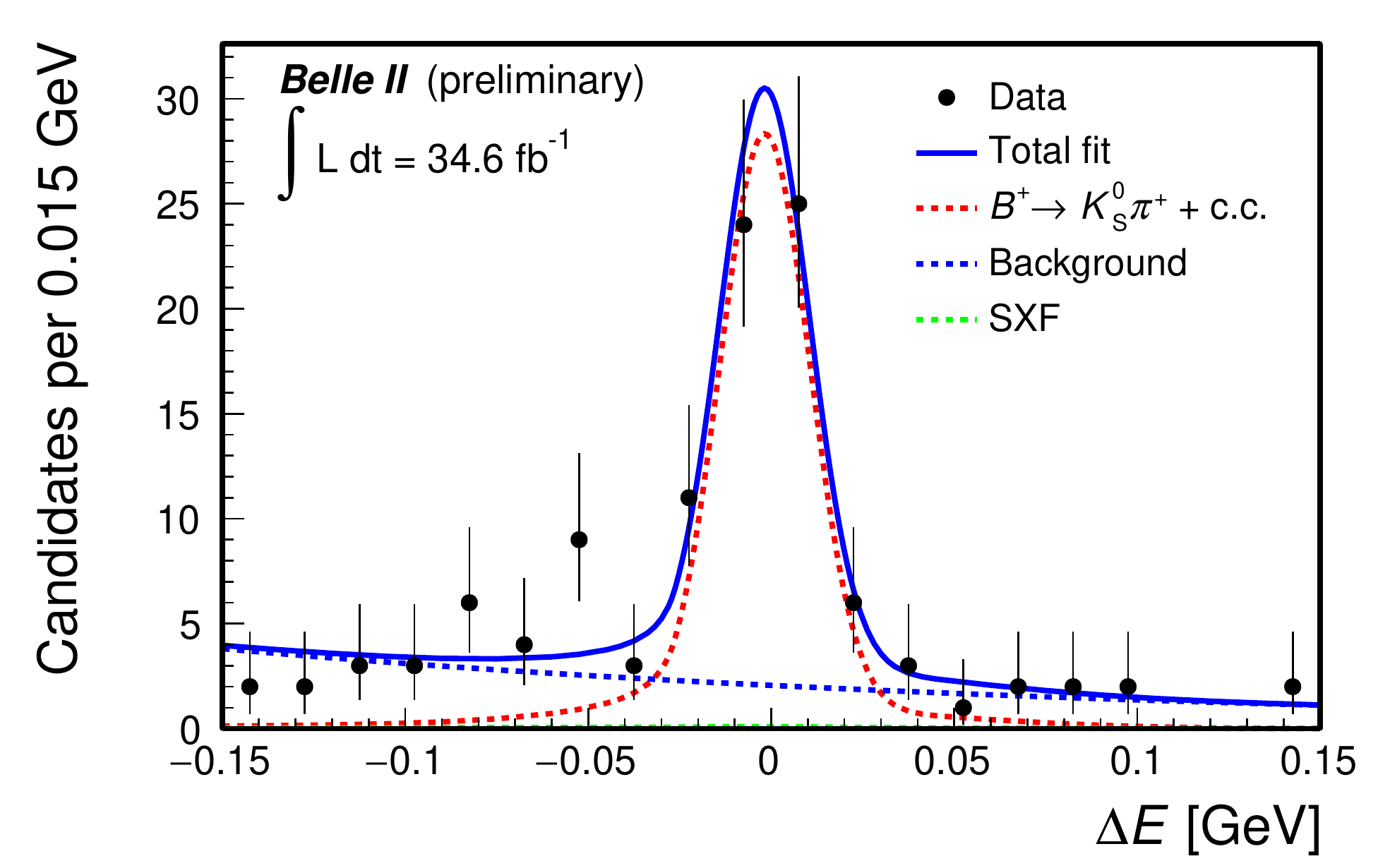}
 \caption{Distribution of $\Delta E$ for $B^+ \to \PKzS\pi^+$  candidates reconstructed in 
 2019--2020 Belle~II data selected through the baseline criteria with an optimized continuum-suppression and kaon-enriching selection, and further restricted to $M_{\rm bc} > 5.27$\,GeV/$c^2$. The projection of an unbinned maximum likelihood fit is overlaid.}
 \label{fig:KSpi+}
\end{figure}
\begin{figure}[htb]
 \centering
 \includegraphics[width=0.8\textwidth]{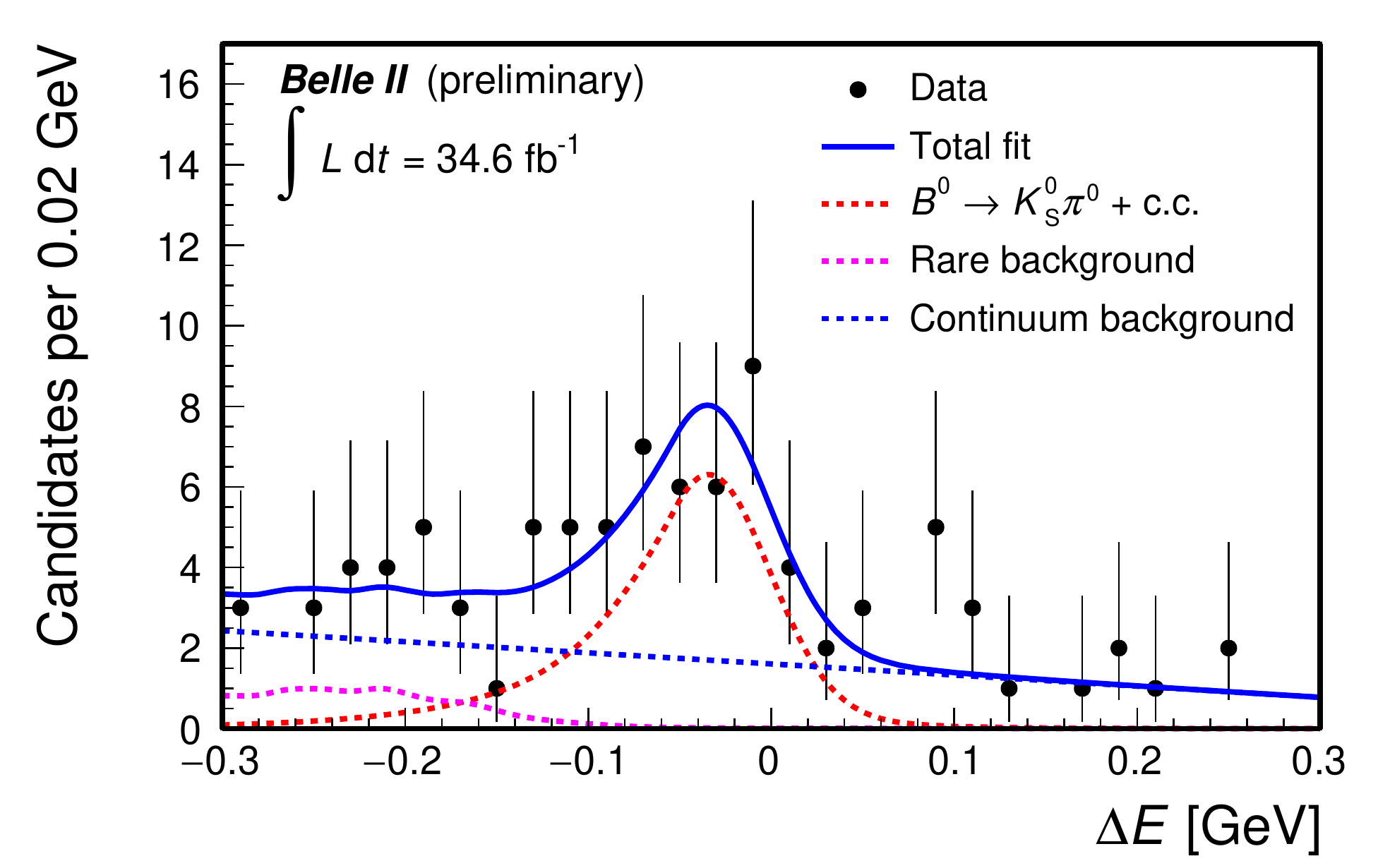}
 \caption{Distribution of $\Delta E$ for $B^0 \to \PKzS\pi^0$  candidates reconstructed in 
 2019--2020 Belle~II data selected through the baseline criteria with an optimized continuum-suppression and kaon-enriching selection, and further restricted to $M_{\rm bc} > 5.27$\,GeV/$c^2$. The `SxF'(self cross-feed) label indicate candidates formed by misidentified (swapped mass assignments) signal particles.  The projection of an unbinned maximum likelihood fit is overlaid.}
 \label{fig:KSpi0}
\end{figure}
\begin{figure}[htb]
 \centering
 \includegraphics[width=0.8\textwidth]{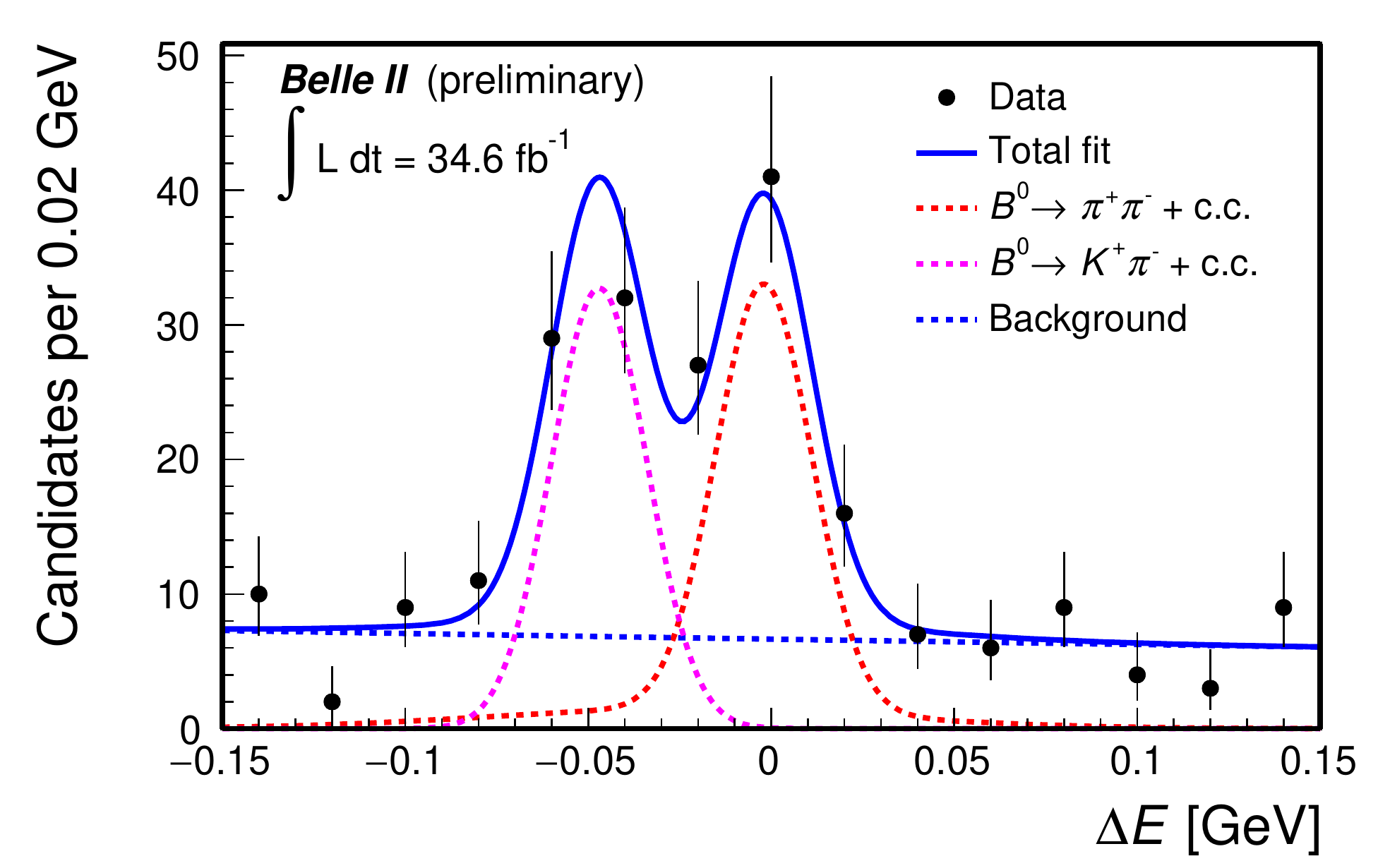}
 \caption{Distribution of $\Delta E$ for $B^0 \to \pi^+\pi^-$  candidates reconstructed in 
 2019--2020 Belle~II data selected through the baseline criteria with an optimized continuum-suppression and kaon-enriching selection, and further restricted to $M_{\rm bc} > 5.27$\,GeV/$c^2$. The projection of an unbinned maximum likelihood fit is overlaid.}
 \label{fig:pi+pi-}
\end{figure}
\begin{figure}[htb]
 \centering
 \includegraphics[width=0.8\textwidth]{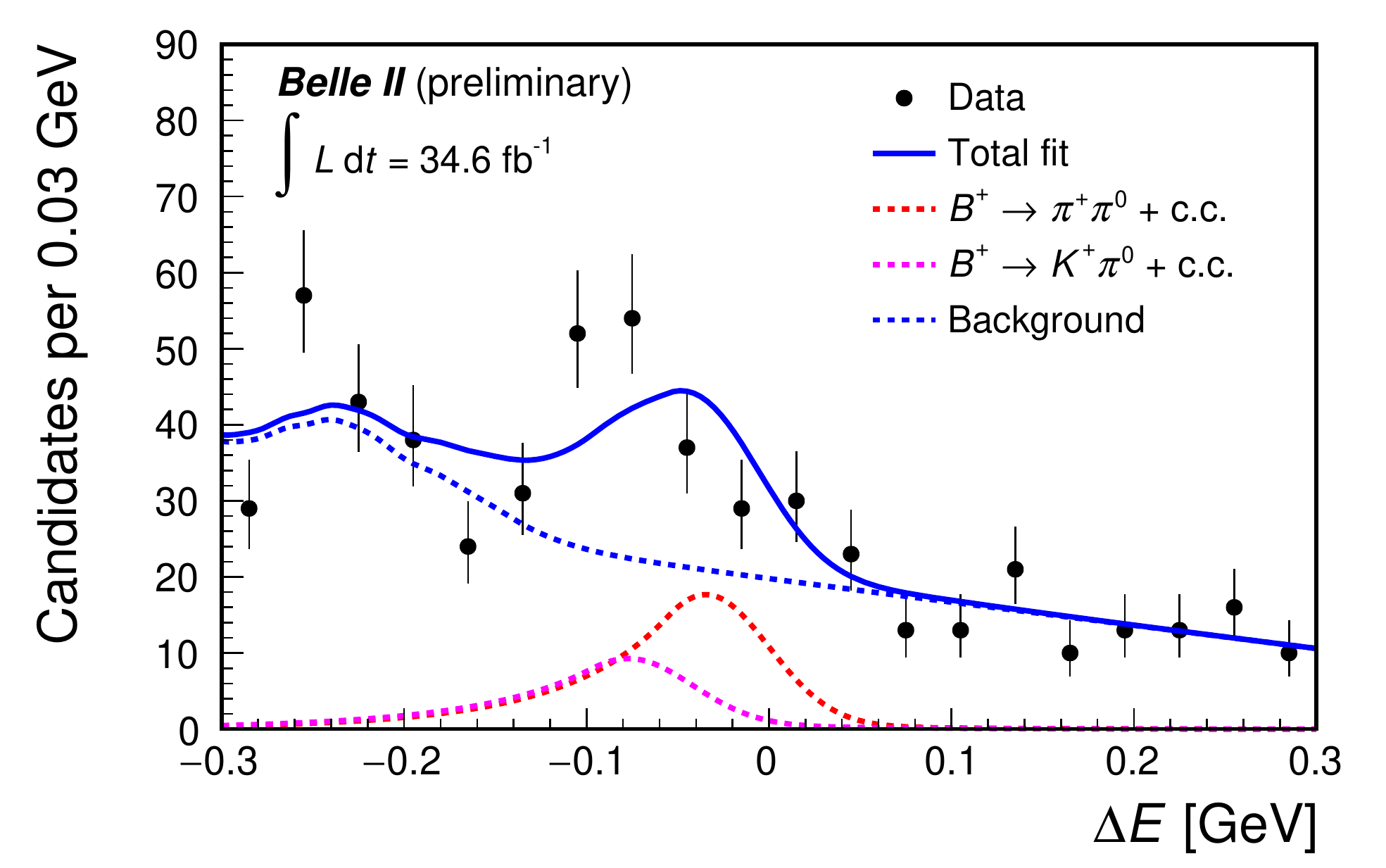}
 \caption{Distribution of $\Delta E$ for $B^+ \to \pi^+\pi^0$  candidates reconstructed in 
 2019--2020 Belle~II data selected through the baseline criteria with an optimized continuum-suppression and kaon-enriching selection, and further restricted to $M_{\rm bc} > 5.27$\,GeV/$c^2$. The projection of an unbinned maximum likelihood fit is overlaid.}
 \label{fig:pi+pi0}
\end{figure}
\begin{figure}[htb]
 \centering
 \includegraphics[width=0.8\textwidth]{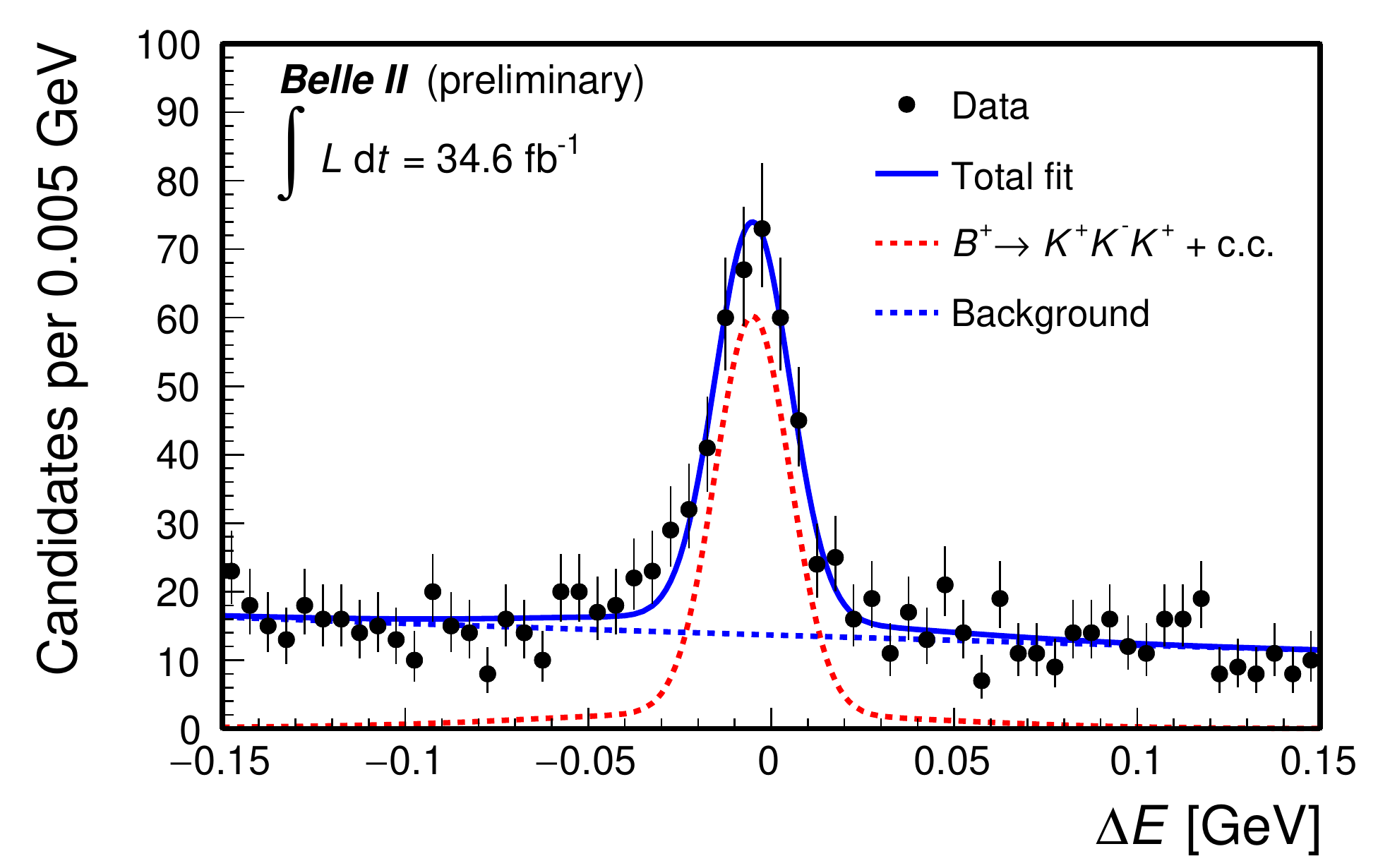}
 \caption{Distribution of $\Delta E$ for $B^+ \to K^+K^-K^+$  candidates reconstructed in (left) simulated data and (right) 2019--2020 Belle~II data, selected through the baseline criteria with an  optimized continuum-suppression and kaon-enriching selection, further restricted to $M_{\rm bc} > 5.27$\,GeV/$c^2$.  The projection of an unbinned maximum likelihood fit is overlaid.}
 \label{fig:KKK}
\end{figure}
\begin{figure}[htb]
 \centering
 \includegraphics[width=0.8\textwidth]{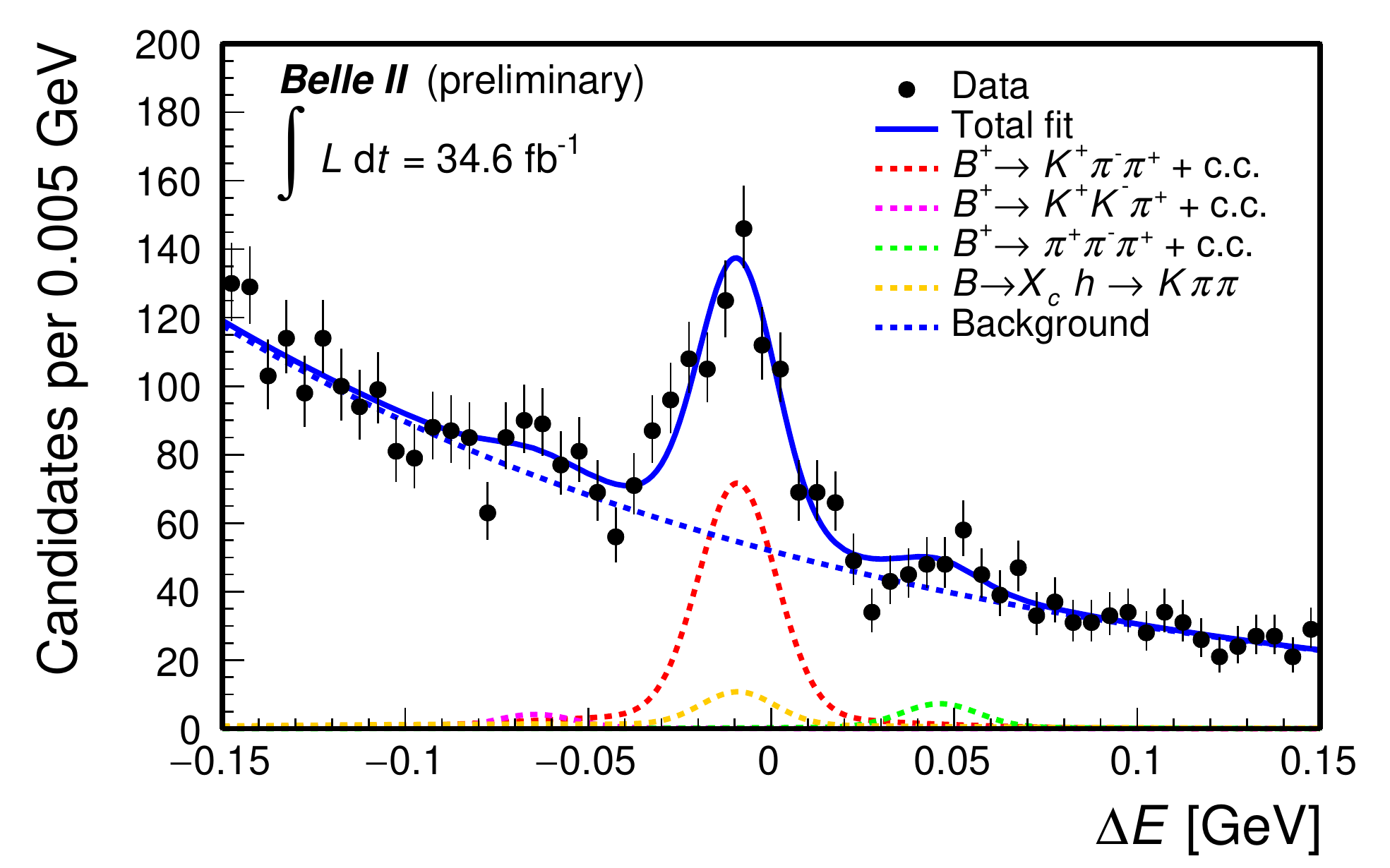}
 \caption{Distribution of $\Delta E$ for $B^+ \to K^+\pi^-\pi^+$  candidates reconstructed in 
 2019--2020 Belle~II data, selected through the baseline criteria with an  optimized continuum-suppression and kaon-enriching selection, further restricted to $M_{\rm bc} > 5.27$\,GeV/$c^2$. Vetoes for peaking backgrounds are applied.
 Misreconstructed  $K^+K^-\pi^+$ and $\pi^+\pi^-\pi^+$ components have the $K^+\pi^-\pi^+$ shape and displacements from the $K^+\pi^-\pi^+$ peak fixed to the known values.  The global position of the three peaks is determined by the fit. The projection of an unbinned maximum likelihood fit is overlaid.}
 \label{fig:Kpipi}
\end{figure}

We use the sum of a single or double Gaussian and a Crystal Ball model~\cite{Skwarnicki:1986xj} for all signals and exponential or straight-line functions, with parameters determined in data, for continuum backgrounds.
We model subleading charmless signals arising from misidentification of final-state particles, or $B \to DX$ signals escaping our vetoes with simplifications of the shapes used for signal (Gaussian, or a Gaussian plus Crystal Ball). The normalizations of such components are determined by the fit for misidentified final states or Gaussian-constrained from simulation otherwise. 
We use sums of Gaussian functions or kernel-density estimated models constrained from simulation for inclusive $B\bar{B}$ backgrounds. 
The $\Delta E$ distributions with fit projections overlaid are shown in Figs.~\ref{fig:K+pi-}--\ref{fig:Kpipi}. 
Prominent narrow signals are visible overlapping smooth backgrounds dominated by continuum. Final states including a $\pi^0$ show a low-$\Delta E$ tail, due to resolution effects in $\pi^0$ reconstruction. Subleading signals from kinematically similar misreconstructed decays are visible in the $B^0 \to K^+\pi^- $, $B^0 \to \pi^+\pi^-$, and $B^+ \to K^+\pi^-\pi^-$ decays. 

\clearpage
\begin{figure}[htb]
 \centering
 \includegraphics[width=0.475\textwidth]{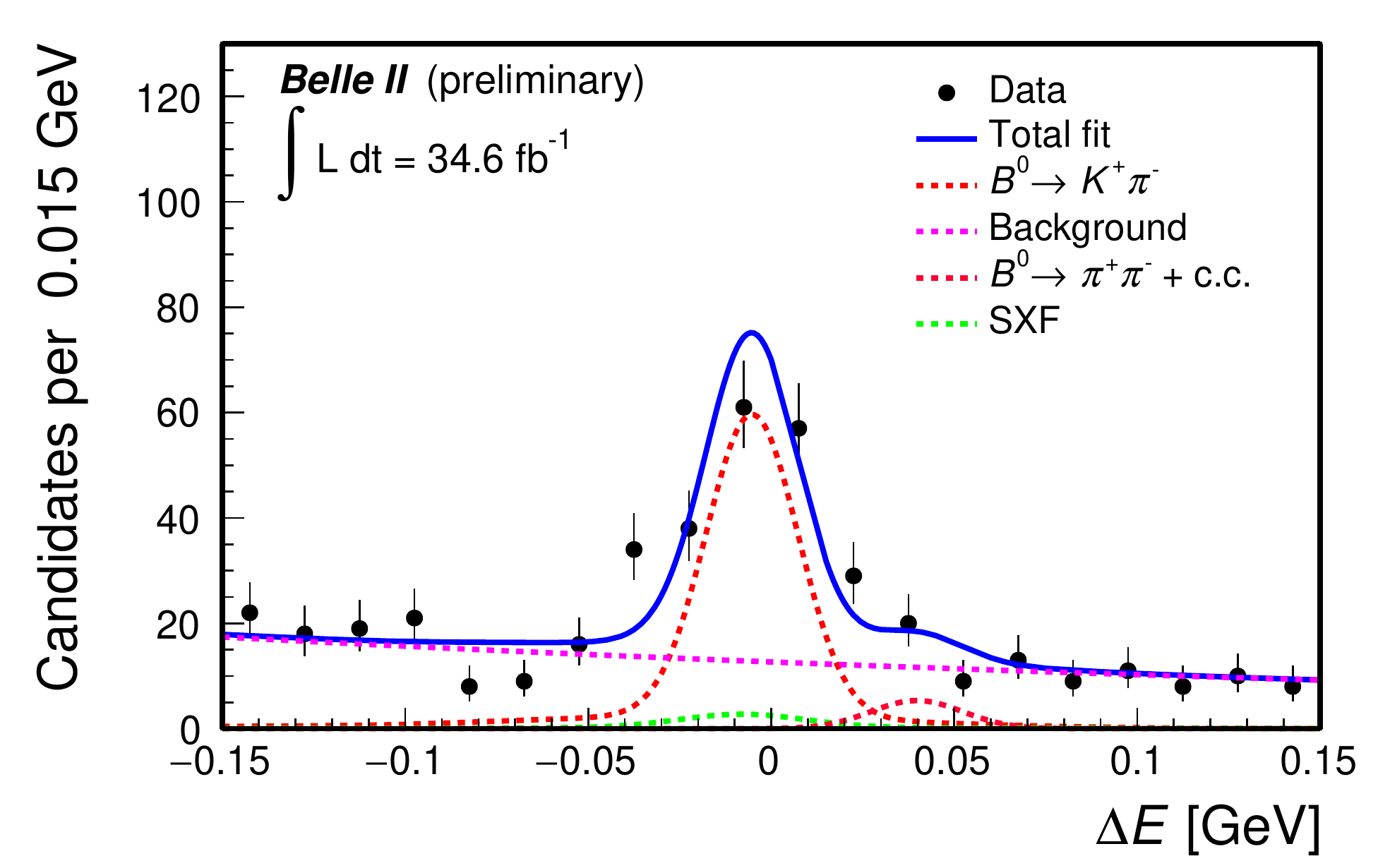}
 \includegraphics[width=0.475\textwidth]{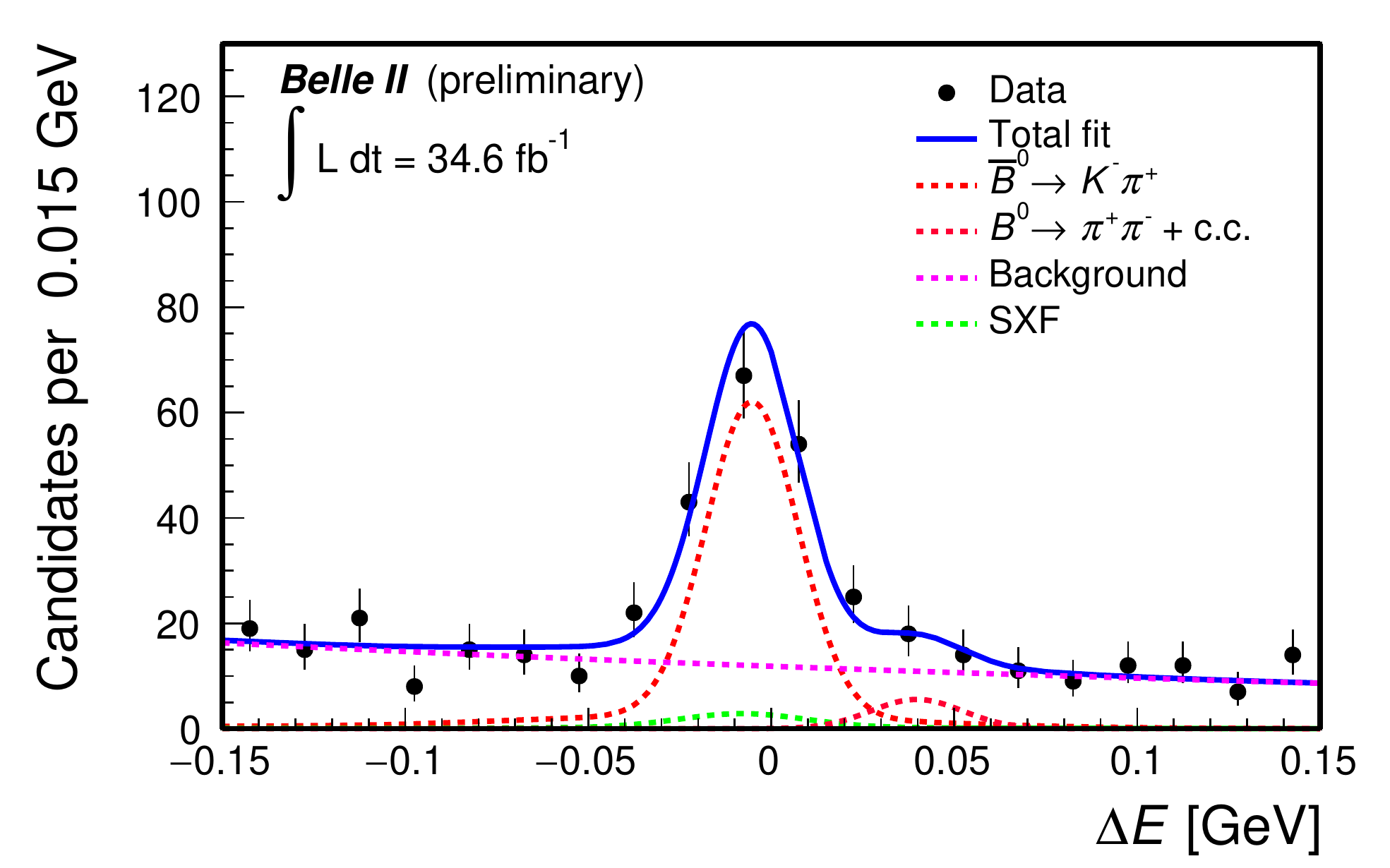}
 \caption{Distributions of $\Delta E$ for (left) $B^0 \to K^+\pi^-$ and (right) $\overline{B}^0 \to K^-\pi^+$ candidates reconstructed in 2019--2020 Belle~II data selected through the baseline criteria with an  optimized continuum-suppression and kaon-enriching selection, and further restricted to $M_{\rm bc} > 5.27$\,GeV/$c^2$. The projection of an unbinned maximum likelihood fit to the charge asymmetry is overlaid.}
 \label{fig:ACP_K+pi-}
\end{figure}

\begin{figure}[htb]
 \centering
 \includegraphics[width=0.475\textwidth]{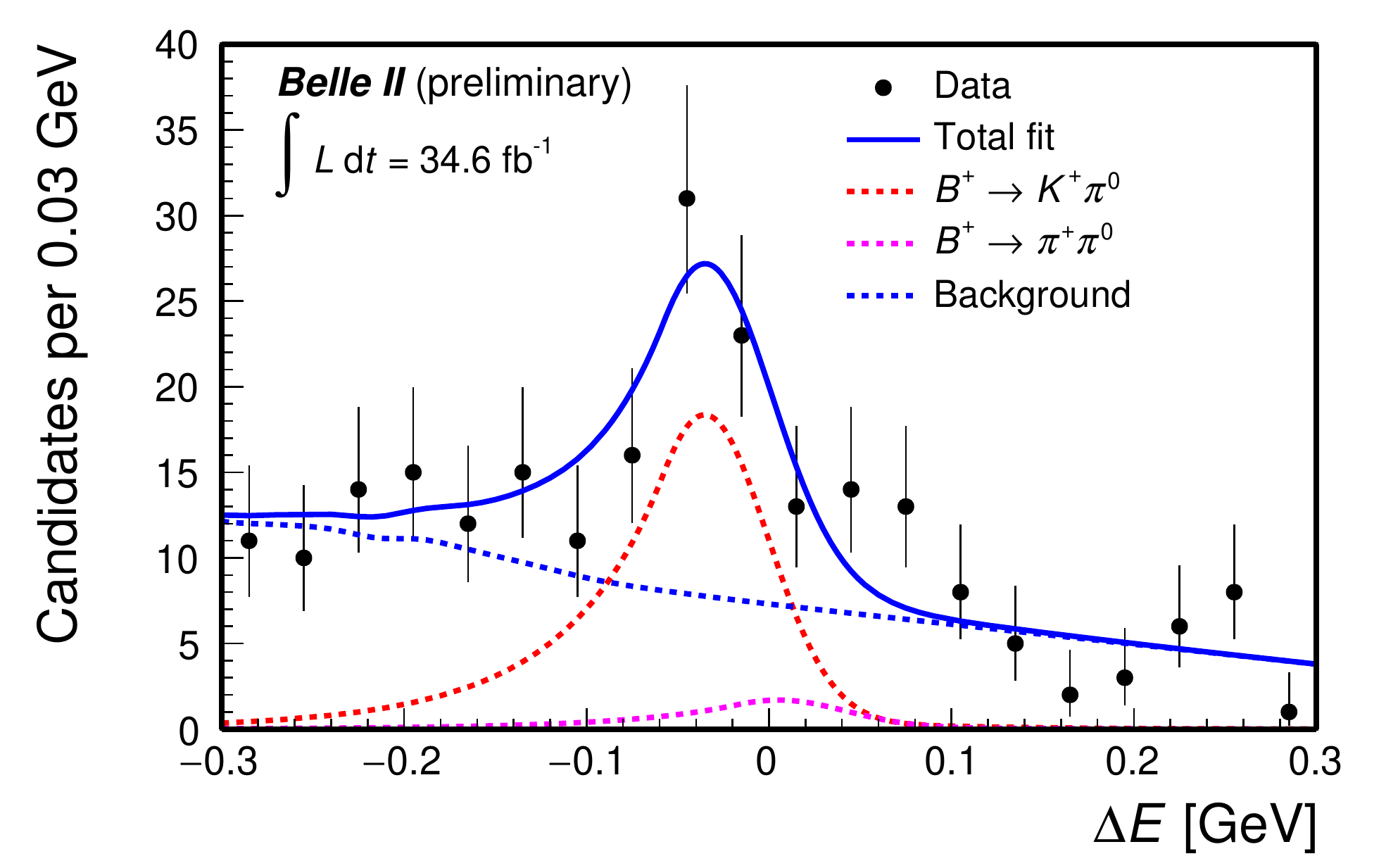}
 \includegraphics[width=0.475\textwidth]{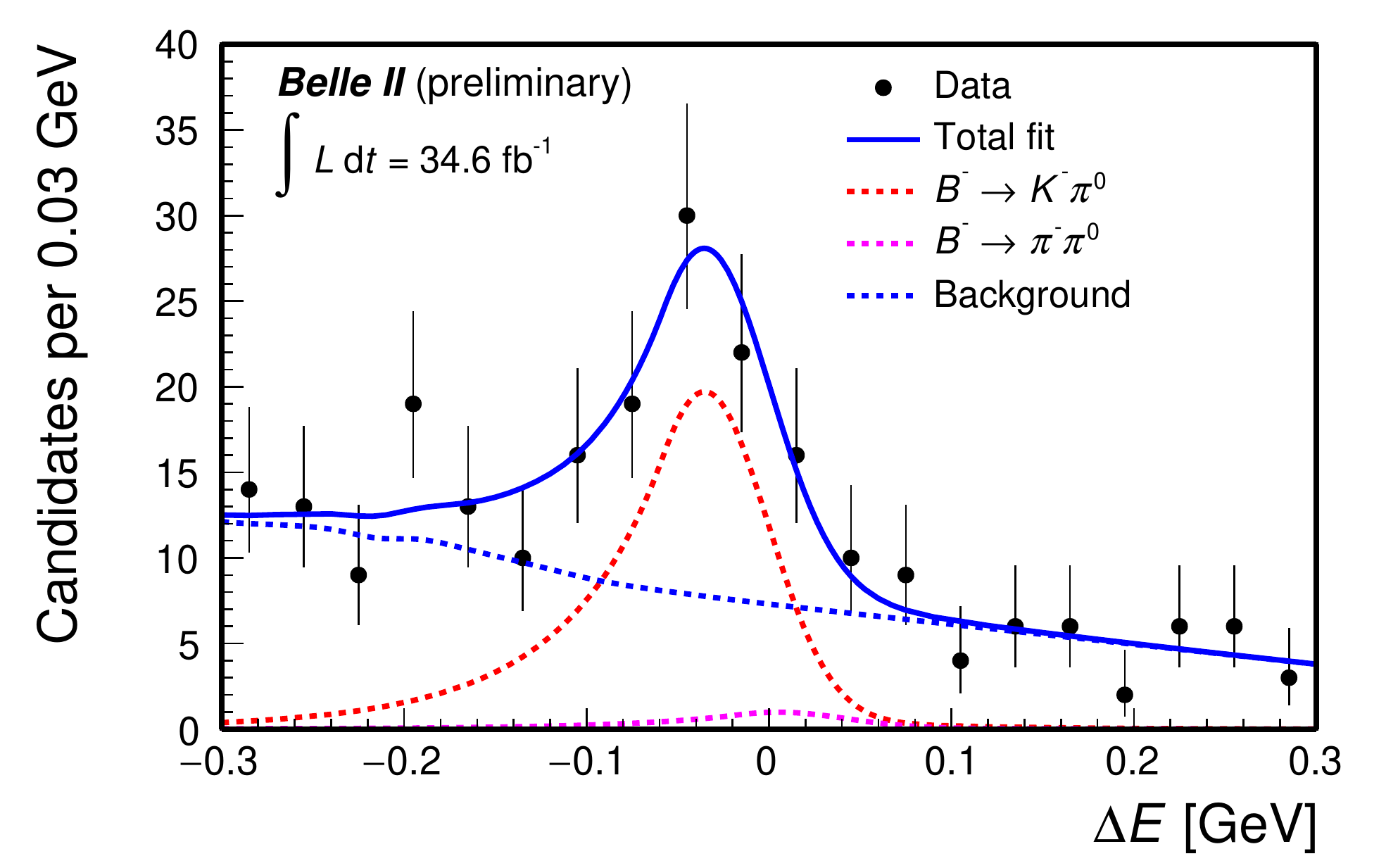}
 \caption{Distributions of $\Delta E$ for (left) $B^+ \to K^+\pi^0$ and (right) $B^- \to K^-\pi^0$ candidates reconstructed in 2019--2020 Belle~II data selected through the baseline criteria with an optimized continuum-suppression and kaon-enriching selection, and further restricted to $M_{\rm bc} > 5.27$\,GeV/$c^2$. The projection of an unbinned maximum likelihood fit to the charge asymmetry is overlaid.}
 \label{fig:ACP_K+pi0}
\end{figure}

\begin{figure}[htb]
 \centering
 \includegraphics[width=0.475\textwidth]{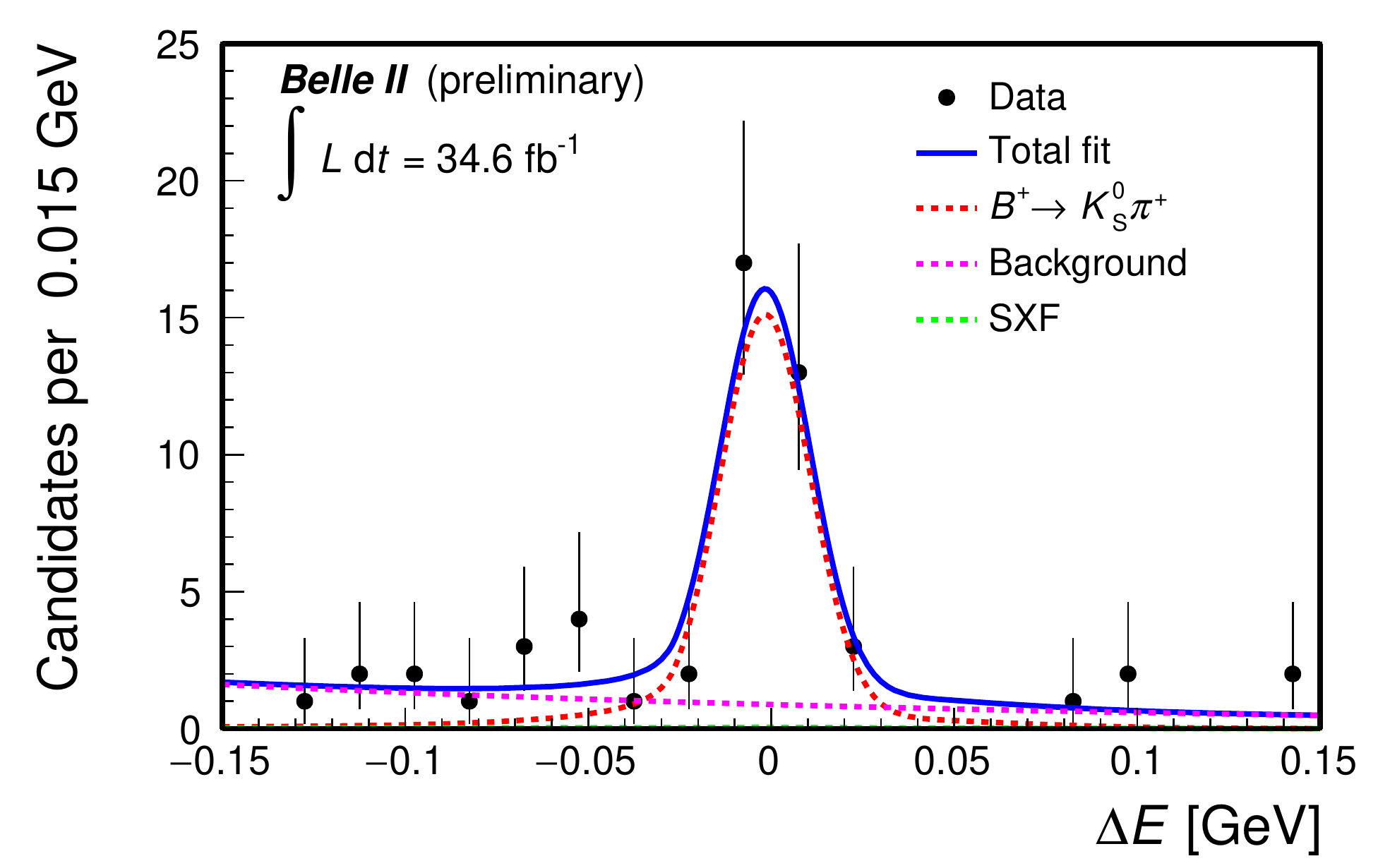}
 \includegraphics[width=0.475\textwidth]{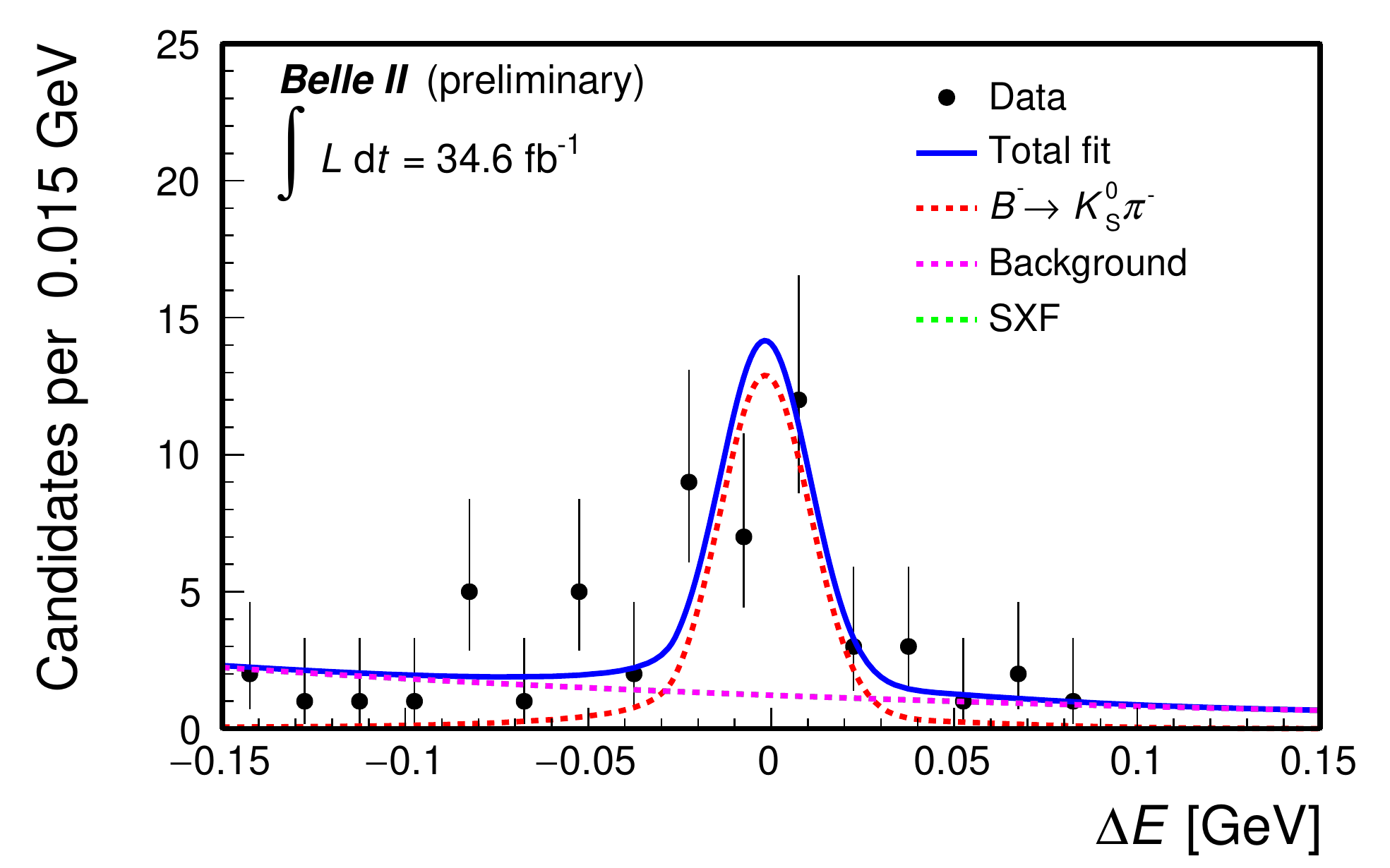}
 \caption{Distributions of $\Delta E$ for (left) $B^+ \to \PKzS\pi^+$ and (right) $B^- \to \PKzS\pi^-$ candidates reconstructed in 2019--2020 Belle~II data selected through the baseline criteria with an optimized continuum-suppression and kaon-enriching selection, and further restricted to $M_{\rm bc} > 5.27$\,GeV/$c^2$. The projection of an unbinned maximum likelihood fit to the charge asymmetry is overlaid.}
 \label{fig:ACP_KSpi-}
\end{figure}

\begin{figure}[htb]
 \centering
 \includegraphics[width=0.475\textwidth]{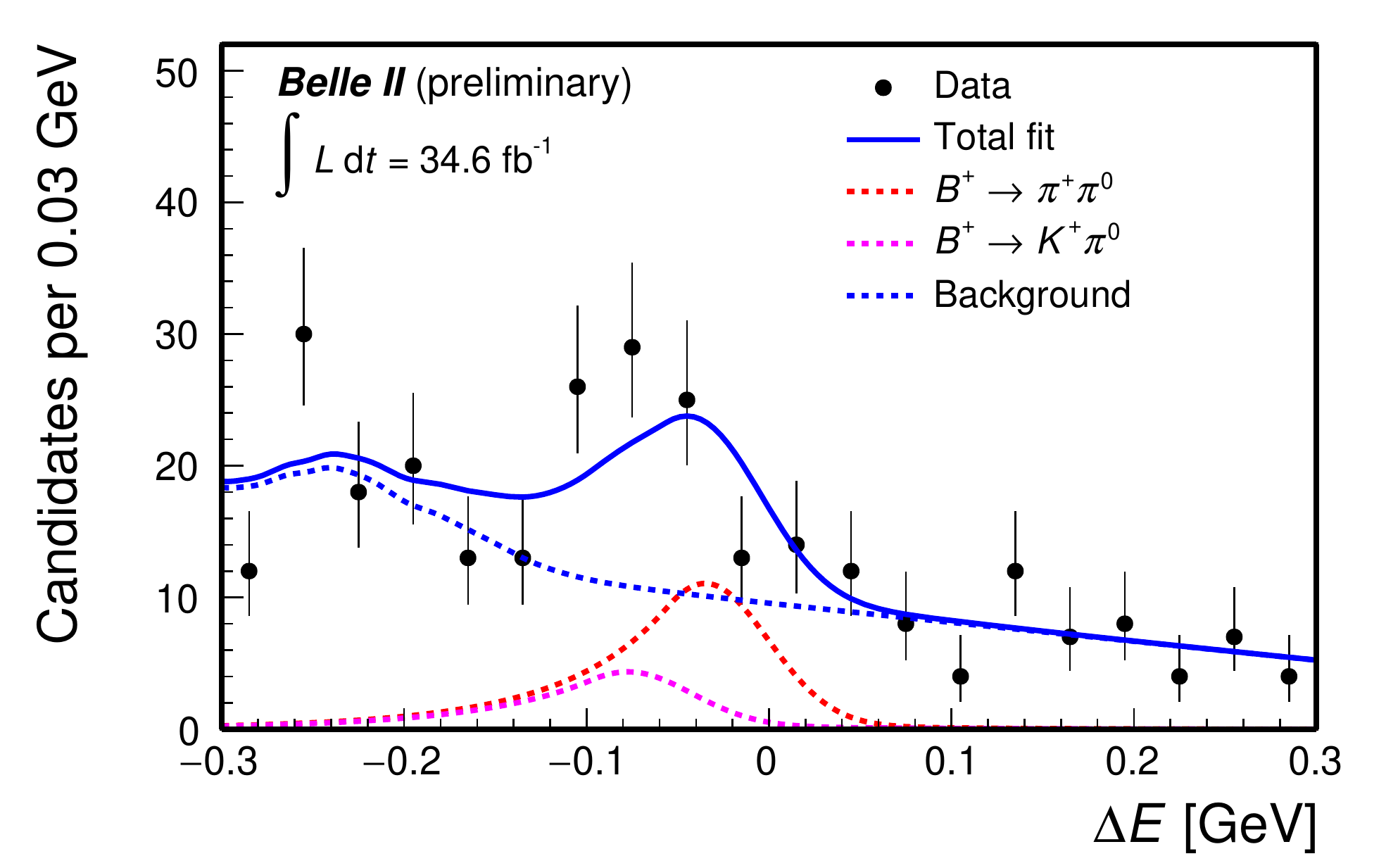}
 \includegraphics[width=0.475\textwidth]{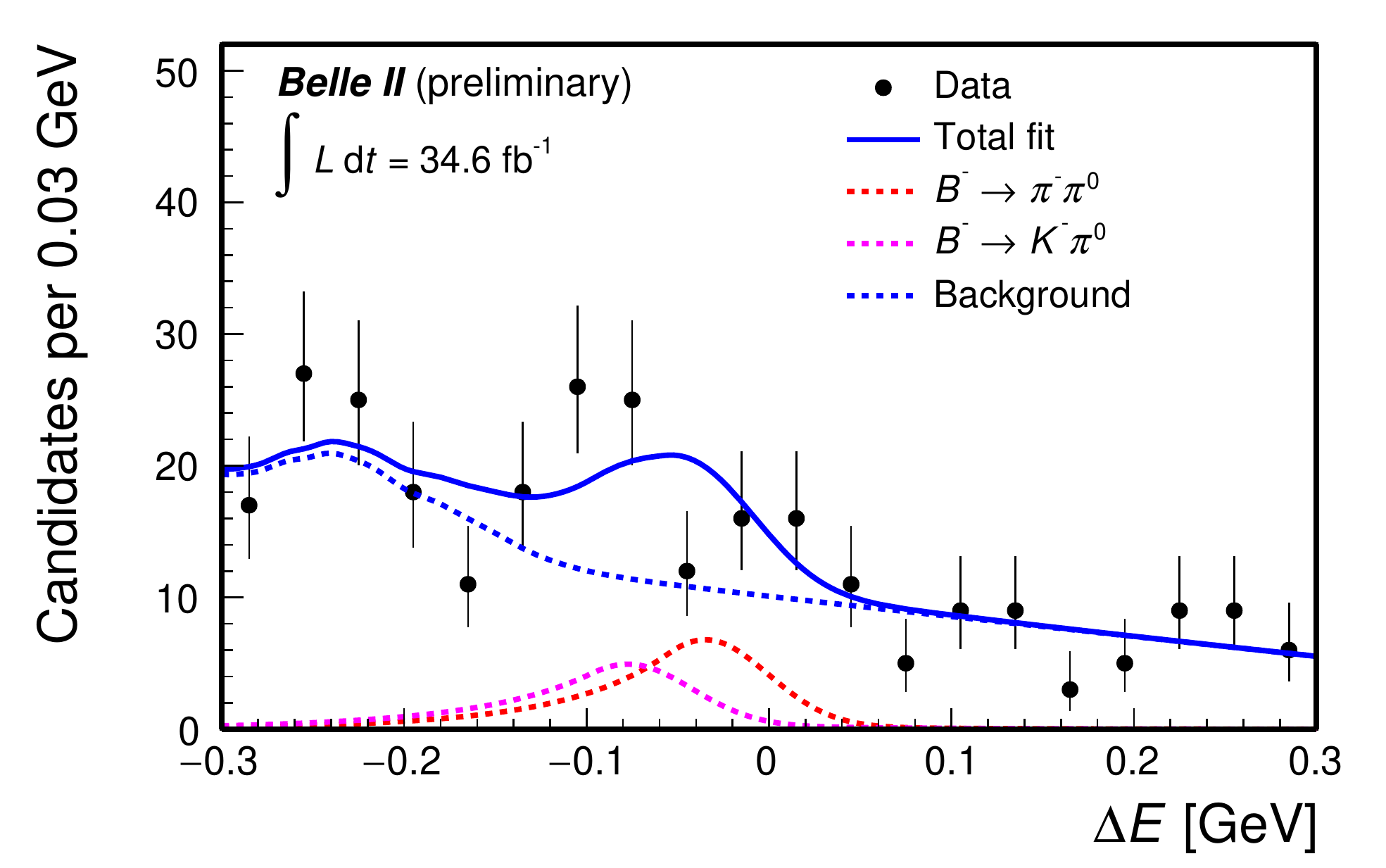}
 \caption{Distributions of $\Delta E$ for (left) $B^+ \to \pi^+\pi^0$ and (right) $B^- \to \pi^-\pi^0$ candidates reconstructed in 2019--2020 Belle~II data selected through the baseline criteria with an  optimized continuum-suppression and kaon-enriching selection, and further restricted to $M_{\rm bc} > 5.27$\,GeV/$c^2$. The projection of an unbinned maximum likelihood fit to the charge asymmetry is overlaid.}
 \label{fig:ACP_pi+pi0}
\end{figure}

\begin{figure}[htb]
 \centering
 \includegraphics[width=0.475\textwidth]{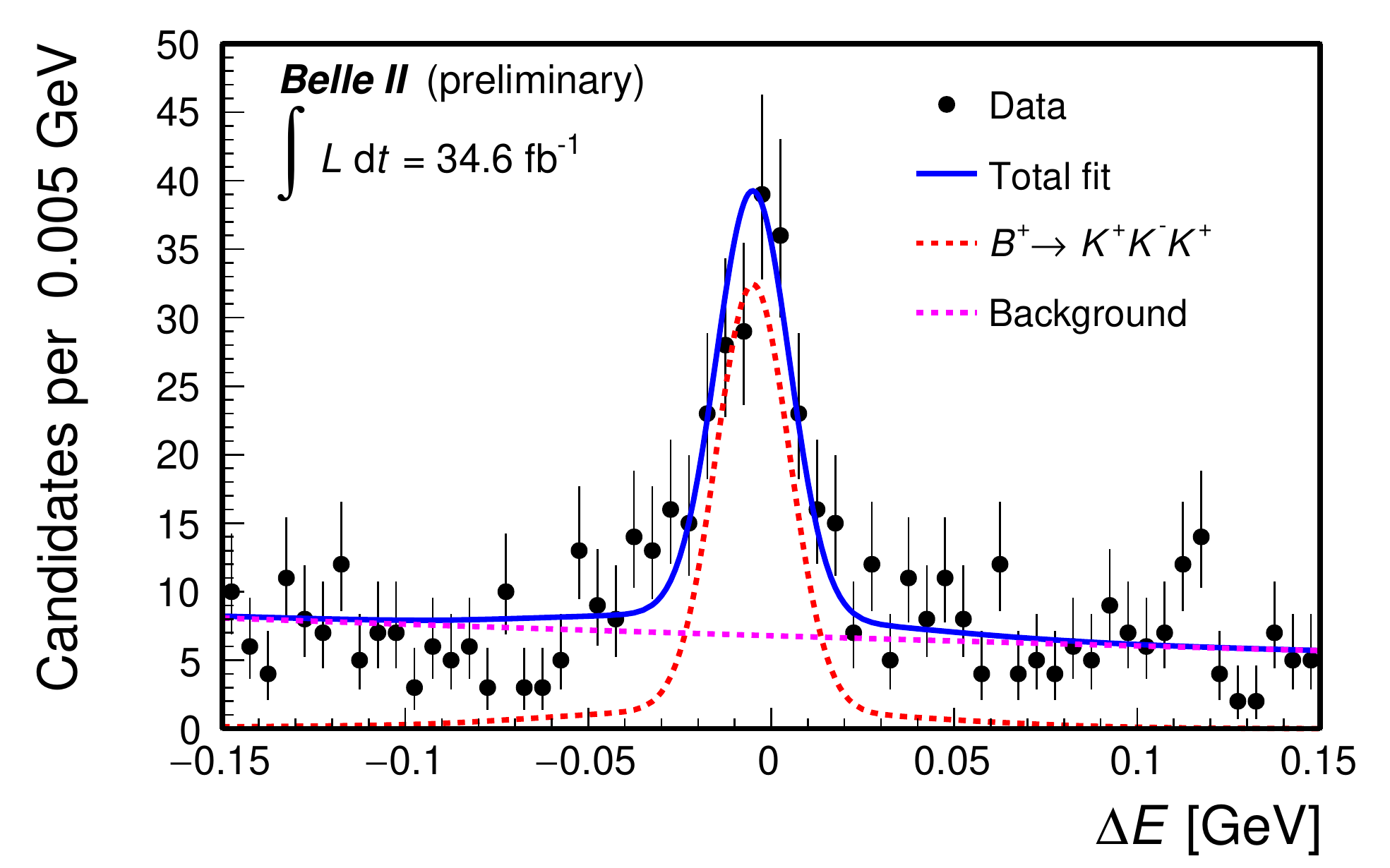}
 \includegraphics[width=0.475\textwidth]{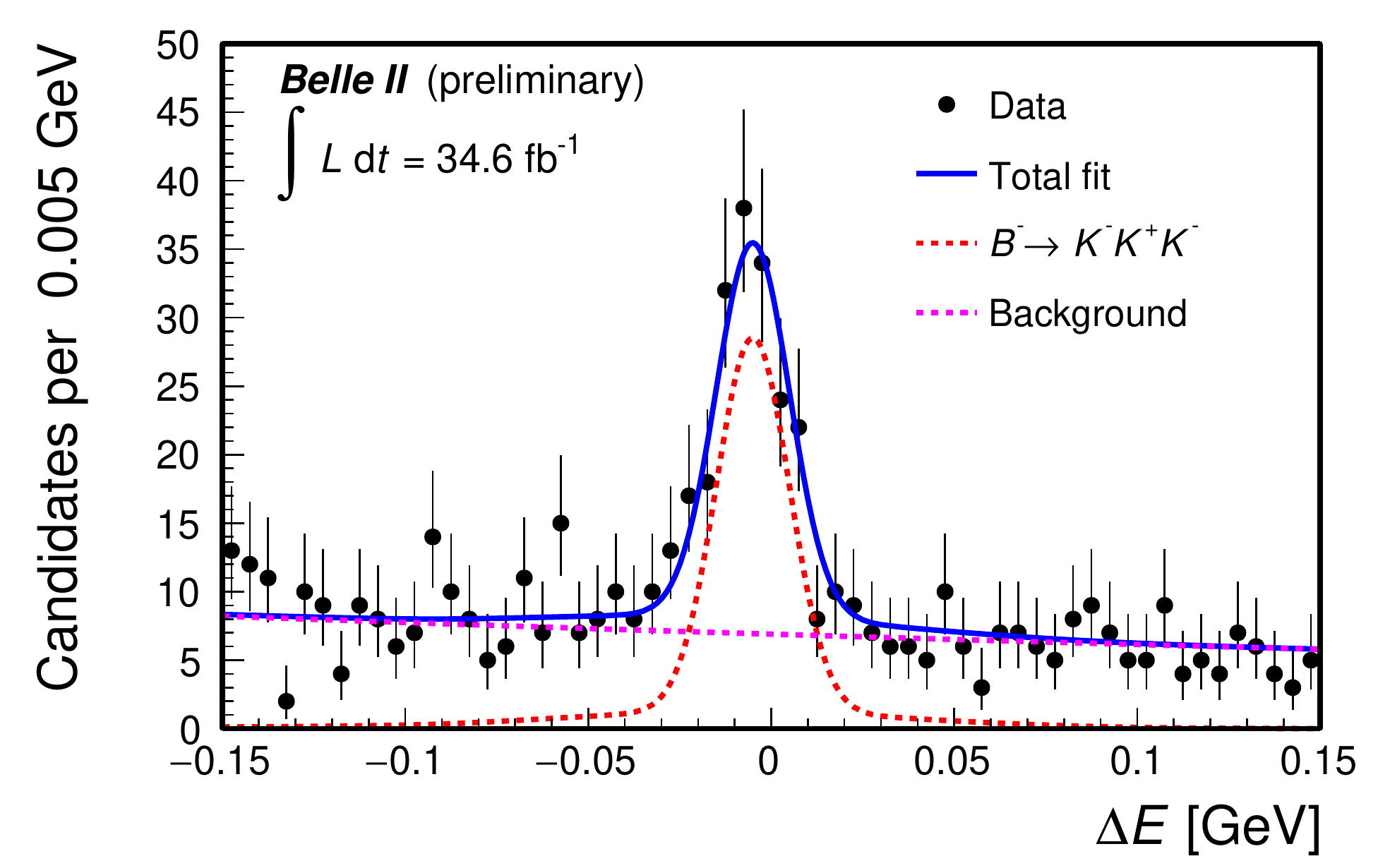}
 \caption{Distributions of $\Delta E$ for (left) $B^+ \to K^+K^-K^+$ and (right) $B^- \to K^- K^+K^-$ candidates reconstructed in 2019--2020 Belle~II data selected through the baseline criteria with an  optimized continuum-suppression and kaon-enriching selection, and further restricted to $M_{\rm bc} > 5.27$\,GeV/$c^2$. The projection of an unbinned maximum likelihood fit to the charge asymmetry is overlaid.}
 \label{fig:ACP_KKK}
\end{figure}

\begin{figure}[htb]
 \centering
 \includegraphics[width=0.475\textwidth]{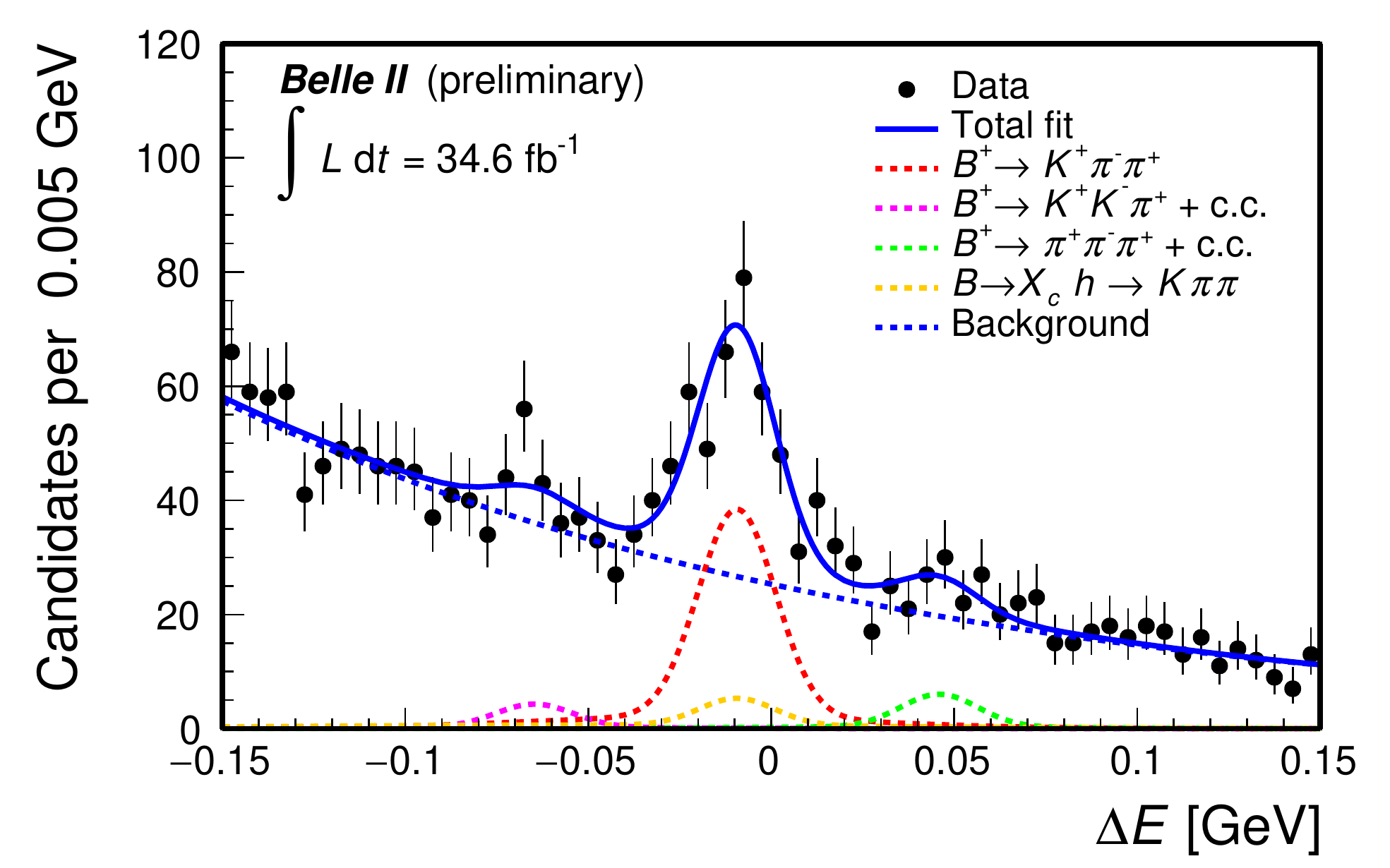}
 \includegraphics[width=0.475\textwidth]{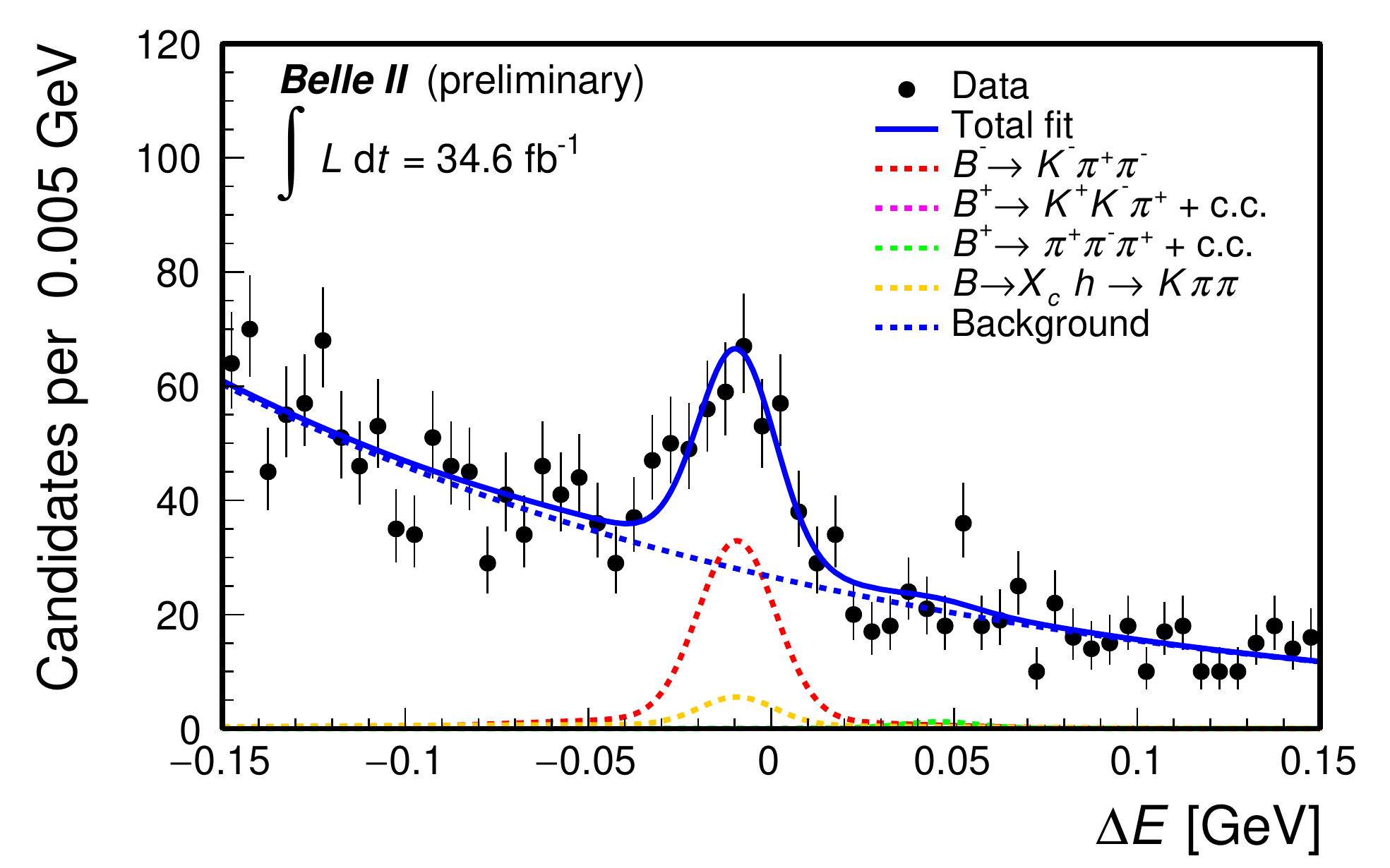}
 \caption{Distributions of $\Delta E$ for (left) $B^+ \to K^+\pi^-\pi^+$ and (right) $B^- \to K^- \pi^+\pi^-$ candidates reconstructed in 2019--2020 Belle~II data selected through the baseline criteria with an optimized continuum-suppression and kaon-enriching selection, and further restricted to $M_{\rm bc} > 5.27$\,GeV/$c^2$. The projection of an unbinned maximum likelihood fit to the charge asymmetry is overlaid.}
 \label{fig:ACP_Kpipi}
\end{figure}

\clearpage
In addition, we use a nonextended likelihood to fit simultaneously the unbinned $\Delta E$ distributions of bottom and antibottom  candidates decaying in flavor-specific final states for measurements of direct CP violation. We use the same signal and background models as used for branching-fraction measurements and use the raw partial-decay-rate asymmetry as a fit parameter,
\begin{equation*}
    \mathcal{A}=\frac{N(b)-N(\bar{b})}{N(b)+N(\bar{b})},
\end{equation*}
where $N$ are signal yields and $b$ ($\bar{b}$) indicates the meson containing a bottom (antibottom) quark. 
Charge-specific $\Delta E$ distributions are shown in Figs.~\ref{fig:ACP_K+pi-}--\ref{fig:ACP_Kpipi} with fit projections overlaid.

 \begin{table}[!ht]
     \centering
 \begin{tabular}{l  r  r  r }
 \hline\hline
 \multicolumn{1}{c}{} & 
 \multicolumn{2}{c}{Yield} &   
 \multicolumn{1}{c}{Raw asymmetry} \\
 Decay & \multicolumn{1}{c}{$B^+$} & \multicolumn{1}{c}{$B^-$} & \multicolumn{1}{c}{}  \\ \hline
   $B^0 \to K^+\pi^-$    &	$142 \pm 13$ &	$147 \pm 13$ & $0.020 \pm 0.064$ \\ 
   $B^+ \to K^+ \pi^0$    &	$69 \pm 14$&	$75 \pm 15 $ &	$0.037 ^{+0.121}_{-0.119}$ \\ 
   $B^+ \to \PKzS \pi^+$    &	$35 \pm  5$ &	$30^{+4}_{-5}$ &	$-0.079^{+0.109}_{-0.114}$ \\
   $B^+ \to \pi^+\pi^0$    &	$43 ^{+19}_{-20}$ &	$24 ^{+13}_{-14}$ &	$-0.275 ^{+0.249}_{-0.322}$ \\ 
    $B^+ \to K^+ K^- K^+$    &	$191 \pm 16$ &	$168\pm 16$ &	$-0.064\pm 0.063$ \\ 
     $B^+ \to K^+ \pi^- \pi^+$    &	$241 \pm 26$ &	$206\pm 26$ &	$-0.078\pm 0.081$  \\ 
   \hline
 \end{tabular}
     \caption{Summary of charge-specific signal yields for the measurement of CP-violating asymmetries in 2019-2020 Belle II data. Only the statistical contributions to the uncertainties are given here.} 
     \label{tab:ACP}
\end{table}{}
\clearpage
\section{Efficiencies and corrections}
\begin{figure}[htb]
 \centering
 \includegraphics[width=0.475\textwidth]{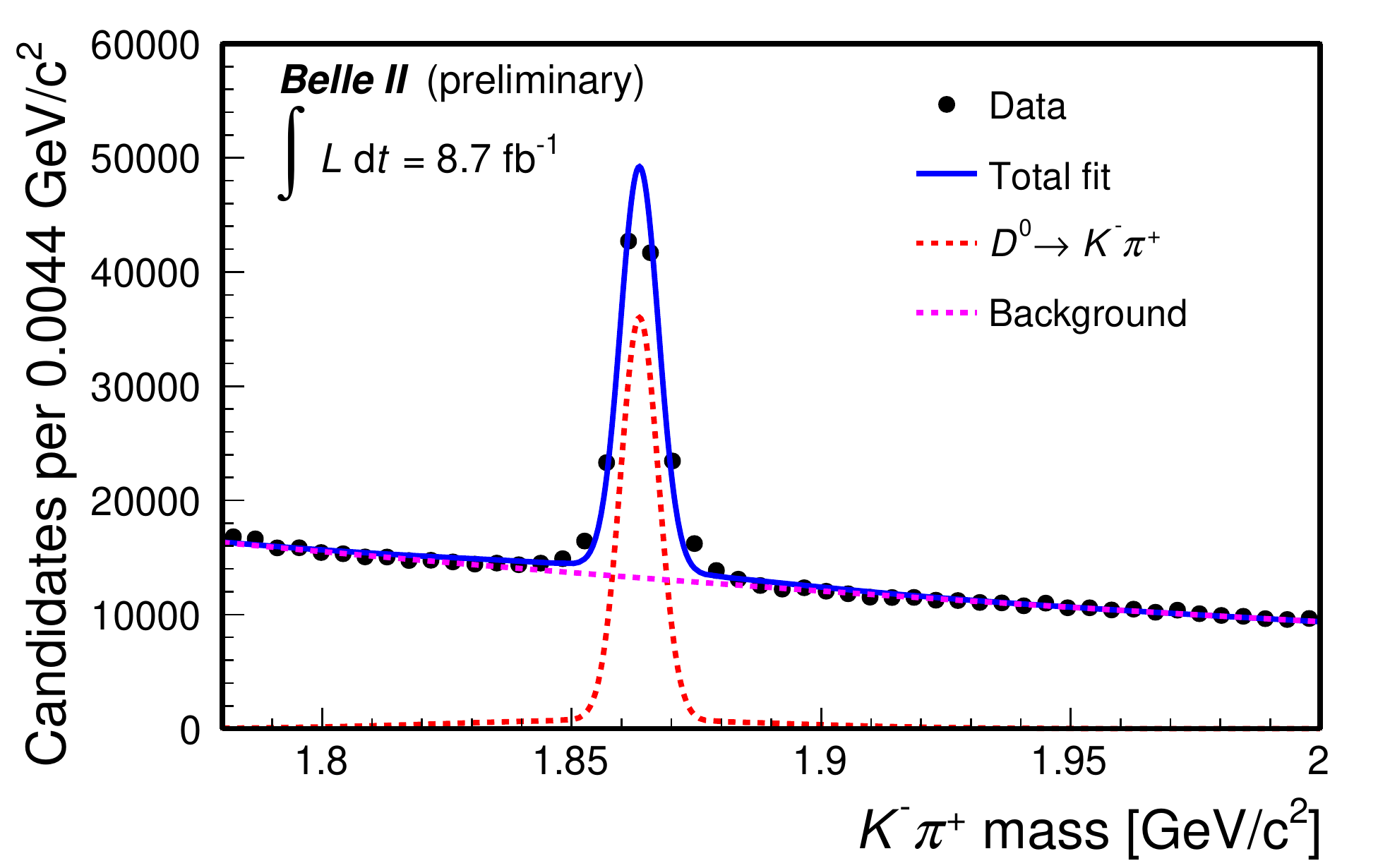}
 \includegraphics[width=0.475\textwidth]{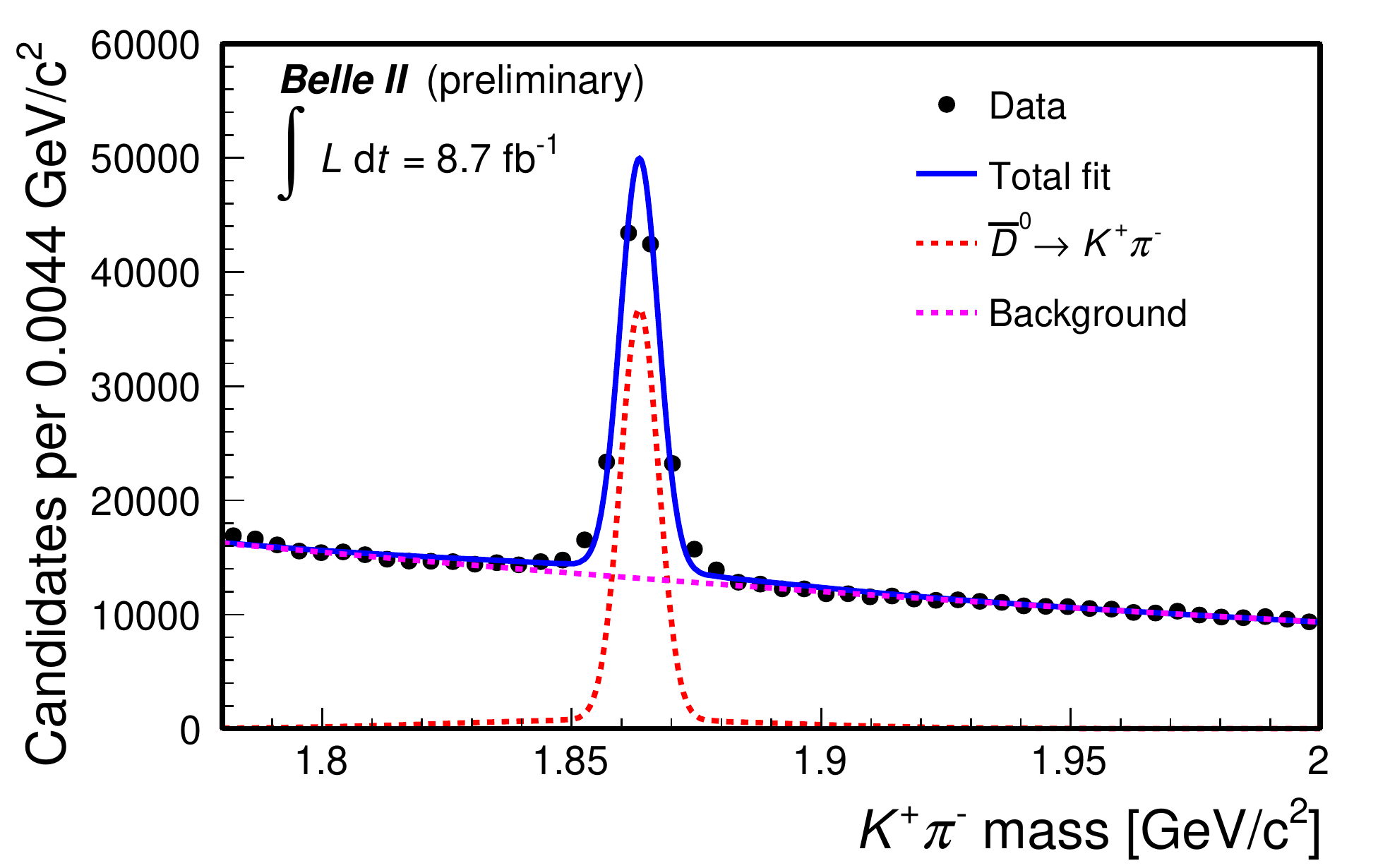}
 \caption{Distributions of $m(K\pi)$ for (left) $D^0 \to K^-\pi^+$ and (right) $\overline{D}^0 \to  K^+\pi^-$ candidates reconstructed in 2019--2020 Belle~II data selected through the baseline criteria with an optimized continuum-suppression and kaon-enriching selection. The projection of an unbinned maximum likelihood fit is overlaid.}
 \label{fig:fit_invM_DtoKpi}
\end{figure}

\begin{figure}[htb]
 \centering
 \includegraphics[width=0.475\textwidth]{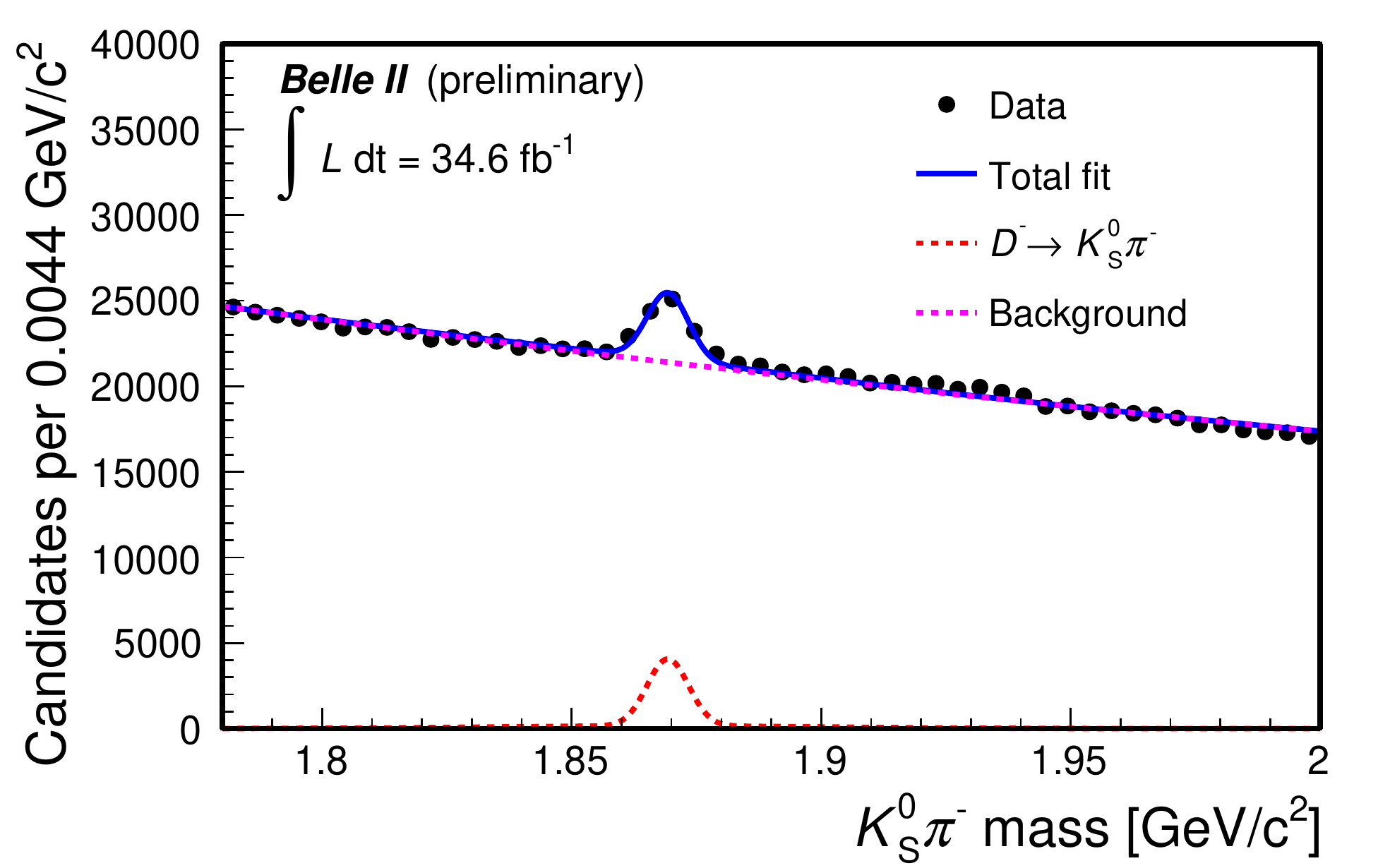}
 \includegraphics[width=0.475\textwidth]{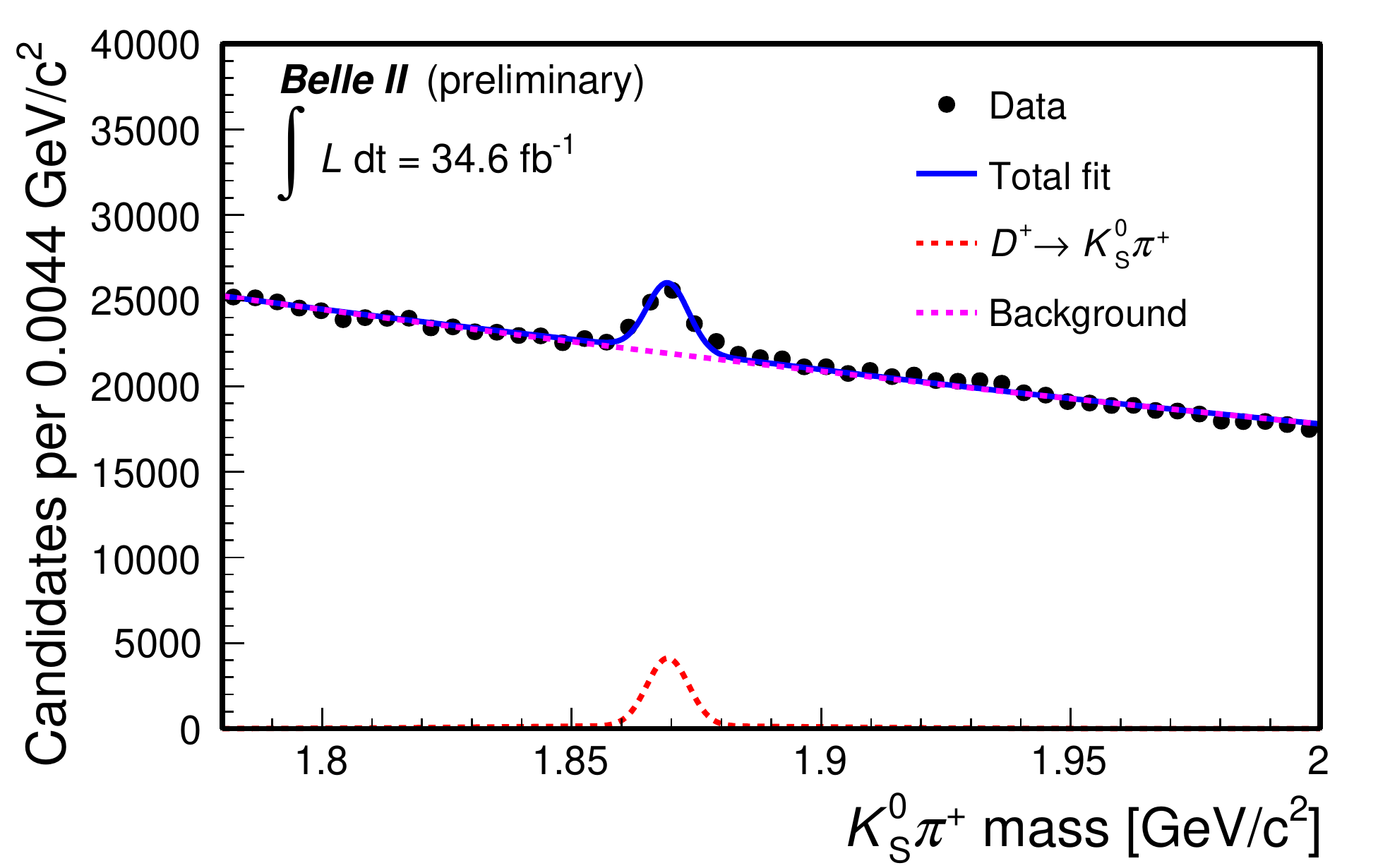}
 \caption{Distributions of $m(\PKzS\pi)$ for (left) $D^- \to \PKzS\pi^-$ and (right) $D^+ \to  \PKzS\pi^+$ candidates reconstructed in 2019--2020 Belle~II data selected through the baseline criteria with an optimized continuum-suppression selection. The projection of an unbinned maximum likelihood fit is overlaid.}
 \label{fig:fit_invM_DtoK0pi}
\end{figure}

The raw event yields observed in data are corrected for selection and reconstruction effects to obtain physics quantities. 
For the measurements of branching fractions, we divide the observed yields by selection and reconstruction efficiencies. The efficiencies are determined from simulation and range between $22\%$ and $41\%$ with typical statistical uncertainties around $0.03\%$.
For those factors of the efficiencies where simulation may not accurately model data, we perform dedicated checks on control samples of data and assess systematic uncertainties (see next section).\\
In measurements of CP-violating asymmetries,  the observed charge-specific raw event yield asymmetries $\mathcal{A}$ are in general due to the combination of genuine CP-violating effects in the decay dynamics and instrumental asymmetries due to differences in interaction or reconstruction probabilities between opposite-charge hadrons. Such combination is additive for small asymmetries,  $\mathcal{A}=\mathcal{A}_{\rm CP}+\mathcal{A}_{\rm det}$, with
\begin{equation*}
    \mathcal{A}_{\rm det}(X)=\frac{X-\bar{X}}{X+\bar{X}},
    \end{equation*}
where $X$ corresponds to a given final  state and $\overline{X}$ to its charge-conjugate.
Hence, observed raw charge-specific decay yields need be corrected for instrumental effects to determine the genuine CP-violating asymmetries.
We estimate the instrumental asymmetry associated with the reconstruction of $K^\pm\pi^\mp$ pairs by measuring the charge-asymmetry in an abundant sample of $D^0 \to K^- \pi^+$ decays. For these decays, direct CP violation is expected to be smaller than 0.1\%, if any~\cite{PDG}. We therefore attribute any nonzero asymmetry to instrumental charge asymmetries. Figure~\ref{fig:fit_invM_DtoKpi} shows the $K^\pm\pi^\mp$-mass  distributions   for $D^0 \to K^-\pi^+$ and $\overline{D}^0 \to  K^+\pi^-$ candidates with fit projections overlaid. The resulting $K^\pm\pi^\mp$ asymmetry is directly applied to the raw measurements of charge-dependent decay rates in $B^0 \to K^+\pi^-$ to extract the physics asymmetry.\\
We correct the observed raw yield asymmetry of $B^+ \to \PKzS \pi^+$ decays using the yield asymmetry observed in an abundant sample of $D^+ \to \PKzS \pi^+$ decays~(Fig.~\ref{fig:fit_invM_DtoK0pi}), in which direct CP violation in $D^+ \to \PKzS \pi^+$ decays is expected to vanish. We correct the observed raw yield asymmetry of $B^+ \to \pi^+ \pi^0$ decays for possible $\pi^+/\pi^-$ reconstruction asymmetries by using the same sample of $D^+ \to \PKzS \pi^+$ decays and subtracting the component $\mathcal{A}(\PKzS)$~deriving from CP violation in neutral kaons, estimated by using the results obtained by the LHCb collaboration~\cite{LHCbInstr:2018}. We finally estimate the instrumental asymmetry related to charged kaon reconstruction alone by combining all inputs in the relationship \mbox{$\mathcal{A}_{\rm det}(K)=\mathcal{A}_{\rm det}(K\pi)-\mathcal{A}_{\rm det}(\PKzS\pi)+\mathcal{A}(\PKzS)$}.  In each case, control channel selections are tuned to reproduce the kinematic conditions of the charmless final states that receive the corrections.  Table~\ref{tab:InstrChargeAsym} shows the resulting corrections.
\begin{table}[htb]
\begin{tabular}{l  c  c}
\hline\hline
Instrumental asymmetry & Value \\
\hline
$\mathcal{A}_{\rm det}(K^+\pi^-)$   & $-0.010 \pm 0.003$ \\
$\mathcal{A}_{\rm det}(\PKzS\pi^+)$   & $-0.007 \pm 0.022$ \\
$\mathcal{A}_{\rm det}(K^+)$   & $-0.015 \pm 0.022$ \\
$\mathcal{A}_{\rm det}(\pi^+)$   & $-0.007 \pm 0.022$ \\
\hline\hline
\end{tabular}
\caption{Instrumental charge-asymmetries associated with $K^\pm\pi^\mp$, $\PKzS\pi^\pm$,  $K^\pm$, and $\pi^\pm$  reconstruction, obtained using samples of $D^0 \to K^- \pi^+$ and $D^+ \to \PKzS \pi^+$ decays.}
\label{tab:InstrChargeAsym}
\end{table}
\clearpage
\section{Determination of branching fractions and CP-violating asymmetries}

We determine each branching fraction as 

\begin{equation*}
    \mathcal{B} = \frac{N}{\varepsilon\times 2\times N_{\PB\APB}},
\end{equation*}
where $N$ is the signal yield obtained from the fits, $\varepsilon$ is the reconstruction and selection efficiency, and $N_{\PB\APB}$ is the number of produced ${\PB\APB}$~pairs, corresponding to $19.7$~million for $\PBplus\PBminus$ and $18.7$~million for $\PBzero\APBzero$ pairs. We obtain the number of ${\PB\APB}$~pairs from the measured integrated luminosity, the \mbox{$\Pep\Pem\to\Upsilon(4{\rm S})$}~cross section~$(1.110 \pm 0.008)\,$nb~\cite{Bevan:2014iga}~(assuming that the $\Upsilon(4{\rm S})$ decays exclusively to ${\PB\APB}$~pairs), and the \mbox{$\Upsilon(4{\rm S})\to\PBzero\APBzero$}~branching fraction \mbox{$f^{00} = 0.487\pm 0.010\pm 0.008$}~\cite{Aubert:2005bq}. For the branching fraction measurement of $\mathcal{B} (\PBzero \to \PKzero \Pgpz)$ and $\mathcal{B} (\PBplus \to \PKzero \pi^+)$, we consider a $0.5$ factor to account for the $\PKzero \to \PKzS$ probability. We use the known value 69.20\% for $\mathcal{B}(\PKzS \to \pi^+\pi^-)$.

The determination of CP-violating asymmetries is more straightforward because all factors that impact symmetrically bottom and antibottom rates cancel, and only flavor-specific yields and flavor-specific efficiency corrections are relevant.
\begin{table}[!ht]
    \centering
\begin{tabular}{l   r  r  r  r }
\hline\hline
Decay & \multicolumn{1}{c}{$\varepsilon\,[\%]$} & \multicolumn{1}{c}{$\mathcal{B}_{\rm s}\,[\%]$} & \multicolumn{1}{c}{Yield} & \multicolumn{1}{c}{$\mathcal{B}\,[10^{-6}]$} \\\hline
  $B^0 \to K^+\pi^-$        & $40.9\quad$ &	& $289^{+22}_{-21}\;\;$  &	$18.9 \pm 1.4\;$ \\ 
  
  $B^+ \to K^+\pi^0$        & $28.9\quad$ &	& $144 ^{+25}_{-24}\;\;$ &    $12.7 ^{+2.2}_{-2.1}\;$	 \\ 
   $B^+ \to K^0 \pi^+$      & $21.9\quad$ &  $34.6\;\;$	& $65^{+10}_{-9}\;\;$    &  $21.8 ^{+3.3}_{-3.0}\;$ \\   
  $B^0 \to K^0\pi^0$        & $24.8\quad$ &  $34.6\;\;$   & $35 \pm 9\;\;\;$         & 		$10.9^{+2.9}_{-2.6}\;$ \\ 
   $B^0 \to \pi^+\pi^-$     & $29.3\quad$ &	& $62^{+11}_{-10}\;\;$   &	$5.6 ^{+1.0}_{-0.9}\;$ \\ 
  
   $B^+ \to \pi^+\pi^0$     & $30.1\quad$ &	& $68 \pm 27$        &	$5.7 \pm 2.3$ \\ 
   
    $B^+ \to K^+ K^- K^+$   & $28.5\quad$ &	& $359\pm 25$        &	$32.0 \pm 2.2$ \\ 
  
  $B^+ \to K^+ \pi^- \pi^+$ & $23.8\quad$ &	& $449\pm 37$        &	$48.0 \pm 3.8$ \\ 
   \hline\hline
\end{tabular}
    \caption{Summary of signal efficiencies $\varepsilon$, fraction of $K^0$ mesons reconstructed in the $\pi^+\pi^-$~final state~ $\mathcal{B}_{\rm s} = f(K^0 \to K^0_S)\times \mathcal{B}(\PKzS \to \pi^+\pi^-)= 0.5\times 0.692$, decay yields in 2019-2020 Belle~II data, and resulting branching fractions. Only the statistical contributions to the uncertainties are given here.} 
    \label{tab:EffYieldBFSummary}
\end{table}{}

\section{Systematic uncertainties}

We consider several sources of systematic uncertainties. We assume the sources to be independent and add in quadrature the corresponding uncertainties. An overview of the effects considered follows. A summary of the fractional size of systematic uncertainties is Tables~\ref{tab:SystematicsBF_overview} and \ref{tab:SystematicsAcp_overview}.

\subsection{Tracking efficiency} 
We assess a systematic uncertainty associated with possible data-simulation discrepancies in the reconstruction of charged particles~\cite{Bertacchi:2020eez}.
The tracking efficiency in data agrees with the value observed in simulation within a $0.91\%$ uncertainty, which we (linearly) add as systematic uncertainty for each final-state charged particle.

\subsection{\PKzS~reconstruction efficiency} 
A small decrease, approximately linear with flight length, in $\PKzS$ reconstruction efficiency was observed in early Belle~II data with respect to simulation.
We assess a systematic uncertainty based on dedicated studies performed for the $B \to \phi K^{(*)}$~analysis~\cite{Ale:2020}. We apply an uncertainty of $1\%$ for each centimeter of average flight length of the $\PKzS$~candidate, resulting in a 12\% total systematic uncertainty, approximately. This source contributes the dominant systematic uncertainty for the measurements of \mbox{$\PBplus\to\PKzS\Pgpp$} and \mbox{$\PBzero\to\PKzS\Pgpz$} branching fractions.

\subsection{\Pgpz~reconstruction efficiency} 
We assess a systematic uncertainty associated with possible data-simulation discrepancies in the $\Pgpz$ reconstruction and selection using the decays \mbox{$B^0 \to D^{*-}(\to \overline{D}^0 (\to K^+ \pi^- \pi^0)\, \pi^-)\, \pi^+$} and \mbox{$B^0 \to D^{*-}(\to \overline{D}^0 (\to K^+ \pi^-)\, \pi^-)\, \pi^+$} where the selection of charged particle is identical and all distributions are weighted so as the $\pi^0$ momentum matches the $\pi^0$ momentum in charmless channels. We compare the yields obtained from fits to the $\Delta E$ distribution of reconstructed $\PB$~candidates~(see~App.~\ref{sec:sysapp}) and obtain an efficiency $\epsilon_{\rm data}^{\Pgpz}$ in data that agree with the value observed in simulation within a $6\%$ uncertainty, which is used as systematic uncertainty. This is the dominant source of systematic uncertainty for the measurements of $\PBplus\to \PKp\Pgpz$ and~$\Pgpp\Pgpz$ branching fractions.

\subsection{Particle-identification and continuum-suppression efficiencies} We evaluate possible data-simulation discrepancies in the particle identification and in the continuum-suppression distributions using the control channel \mbox{$\PBplus\to\APDzero(\to\PKp\Pgpm\Pgpz)\,\Pgpp$} for decay modes including neutral pions and \mbox{$\PBplus\to\APDzero(\to\PKp\Pgpm)\,\Pgpp$} for all others~(see~App.~\ref{sec:sysapp}). We find that the selection efficiencies obtained in data and simulation agree within $2-4\%$ uncertainties (depending on the selection), which are taken as systematic uncertainties.

\subsection{Number of $\PB\APB$~pairs} We  assign  a  $2.7\%$  systematic  uncertainty on the number of $\PB\APB$~pairs,  which  includes  the uncertainty  on  cross-section,  integrated  luminosity~\cite{Abudinen:2019osb}, and  potential  shifts  from  the peak center-of-mass energy during the run periods.

\subsection{Signal modeling} 
Because we used empirical fit models for signal, we assess a systematic uncertainty associated with the model choice. In the branching-fraction measurements, we repeat the measurements using alternative signal models that reproduce data with similar accuracy and quote the difference in fit results as systematic uncertainties. In addition, we assess a systematic uncertainty due to  imperfections in the signal modeling associated with the simulation of hit~multiplicity in the drift~chamber, which impacts $\Delta E$ signal resolutions. We repeat the measurements using models determined after weighting the hit multiplicity in simulation to match data, or with various hit-multiplicity requirements, and quote the largest observed difference with respect to the default results as systematic uncertainty. The contributions due to signal modeling and hit~multiplicity add in quadrature to an uncertainty of typically $2\%$.

For measurements of CP~asymmetries, we evaluate the impact of signal modeling by comparing the results obtained by fitting with charge-symmetric or charge-specific models and taking the difference between results as uncertainty, which has typical size of $0.5\%$.  

\subsection{Continuum background modeling} For branching fraction measurements, we perform fits with alternative background models that reproduce data with similar accuracy and take the difference between fit results as systematic uncertainty, which is typically $3\%$. 

\subsection{Peaking and $\PB\APB$~background model} In measurements of branching fractions of  \mbox{$\PBplus\to\PKp\Pgpz$, and~$\Pgpp\Pgpz$}, we evaluate the effect of the $\PB\APB$~background by varying the fit range from the default $\vert \Delta E\vert < 0.3$~GeV window to $-0.1 < \Delta E < 0.3$ and taking the difference between fit results as uncertainty. For branching fraction measurements of \mbox{$\PBzero\to\PKp\Pgpm$ and~$\PBzero\to\Pgpp\Pgpm$}, we compare results of fits done by floating and by Gaussian-constraining the peaking-background yields according to simulation, and take the difference between fit results as uncertainty. For branching fraction measurements of \mbox{$\PBplus\to\PKp\PKm\PKp$, and~$\PKp\Pgpp\Pgpm$}, we compare results of fits done by fixing and by constraining the peaking-background yields according to simulation, and take the difference between fit results as uncertainty. The uncertainties due to peaking and $\PB\APB$ background bias are typically $0.3\%$. 

For measurements of CP~asymmetries, we perform fits with the charge-conjugate peaking background yields fixed to the expected proportions from simulation, and fixed to exactly half of the total yield, and take the $0.3\%$ difference between results as systematic uncertainty. 

\subsection{Instrumental asymmetries}
We consider the uncertainty on the values of  $\mathcal{A}_{\rm det}$~(Table~\ref{tab:InstrChargeAsym}) as systematic uncertainty due to instrumental asymmetry corrections in measurements of CP~asymmetries. This source is dominant for systematic uncertainties in three-body decays and $\PBplus\to\PKp\Pgpz$.

\begin{table}[h]
\centering
\footnotesize
\caption{Summary of the (fractional) systematic uncertainties of the branching-fraction measurements.}
\begin{tabular}{l c c c c c c c c}
\hline\hline
 Source & $\PKp\Pgpm$ & $K^+ \pi^0$ &$\PKzero\Pgpp$ & $K^0 \pi^0$
 & $\Pgpp\Pgpm$ &  $\pi^+ \pi^0$ & $K^+ K^- K^+$ & $K^+ \pi^- \pi^+$\\
\hline
Tracking             & 1.8\%   & 0.9\%   & 2.7\%  & 1.8\%   & 1.8\%  & 0.9\% & 2.7\% & 2.7\% \\
 $\PKzS$ efficiency      & -        & -      & 12.5\% & 11.6\%  & -       & -  & - &-\\
 $\Pgpz$ efficiency      & -        & 6.5\% &   -     & 6.5\%   & -       &6.5\% & - &- \\
PID and continuum-supp. eff.               & 1.1\%   & 2.6\% & 0.9\%  &  1.4\%  & 1.3\%  & 2.7\% & 2.3\%   &
1.0\% \\
 $N_{B\bar{B}}$      & 2.7 \%  & 2.7\% & 2.7\%  &  2.7\%  & 2.7\%  & 2.7\% & 2.7\%   & 2.7\% \\
  Signal model & 1.1\%   & 2.3\% & $<0.1$\% & $<0.1$\%   & 4.5\%  & 0.5\% & 0.6\%   & 3.5\%\\
 Continuum bkg. model         &  4.2\%   & 3.1\% & 1.5\%  & 4.8\%  & $<0.1$\%  & 3.6\% & 0.3\%   & 4.6\% \\
 $\PB\APB$ bkg. model &  0.4\%   & $<0.1$\% &  -   &  -      & 1.6\% & 0.4\% & -        & 0.2\%\\
\hline
 Total  & 5.5\% &  8.5\% & 13.2\% & 14.6\% & 5.9\% & $8.4\%$ & 4.5\%   & 7.0\% \\
\hline\hline
\end{tabular} 

\label{tab:SystematicsBF_overview}
\end{table}

\begin{table}[h]
\centering
\footnotesize
\caption{Summary of (absolute) systematic uncertainties in the $\mathcal{A_{\rm CP}}$ measurements.}
\begin{tabular}{l c c c c c c}
\hline\hline
 Source & $\PKp\Pgpm$ & $K^+ \pi^0$ &$\PKzero\Pgpp$ & $\pi^+ \pi^0$ & $K^+ K^- K^+$ & $K^+ \pi^- \pi^+$\\
\hline
 Signal model                    & 0.005 & 0.001 &  0.007   & 0.005 & 0.001 & 0.003\\
 Pkg./$\PB\APB$/s$\times$f background model & 0.005 & -     &  0.006   & 0.120 & -     & 0.004  \\
Instrumental asymmetry corrections                    & 0.003 & 0.022 &   0.022      & 0.022     & 0.022 & 0.022\\
\hline
 Total                           & 0.008  & 0.022 &  0.024   & 0.123 & 0.022 & 0.023 \\
\hline\hline
\end{tabular} 

\label{tab:SystematicsAcp_overview}
\end{table}

\section{Results and summary}
\label{sec:summary}
\begin{figure}[htb]
 \centering
 \includegraphics[width=0.7\textwidth]{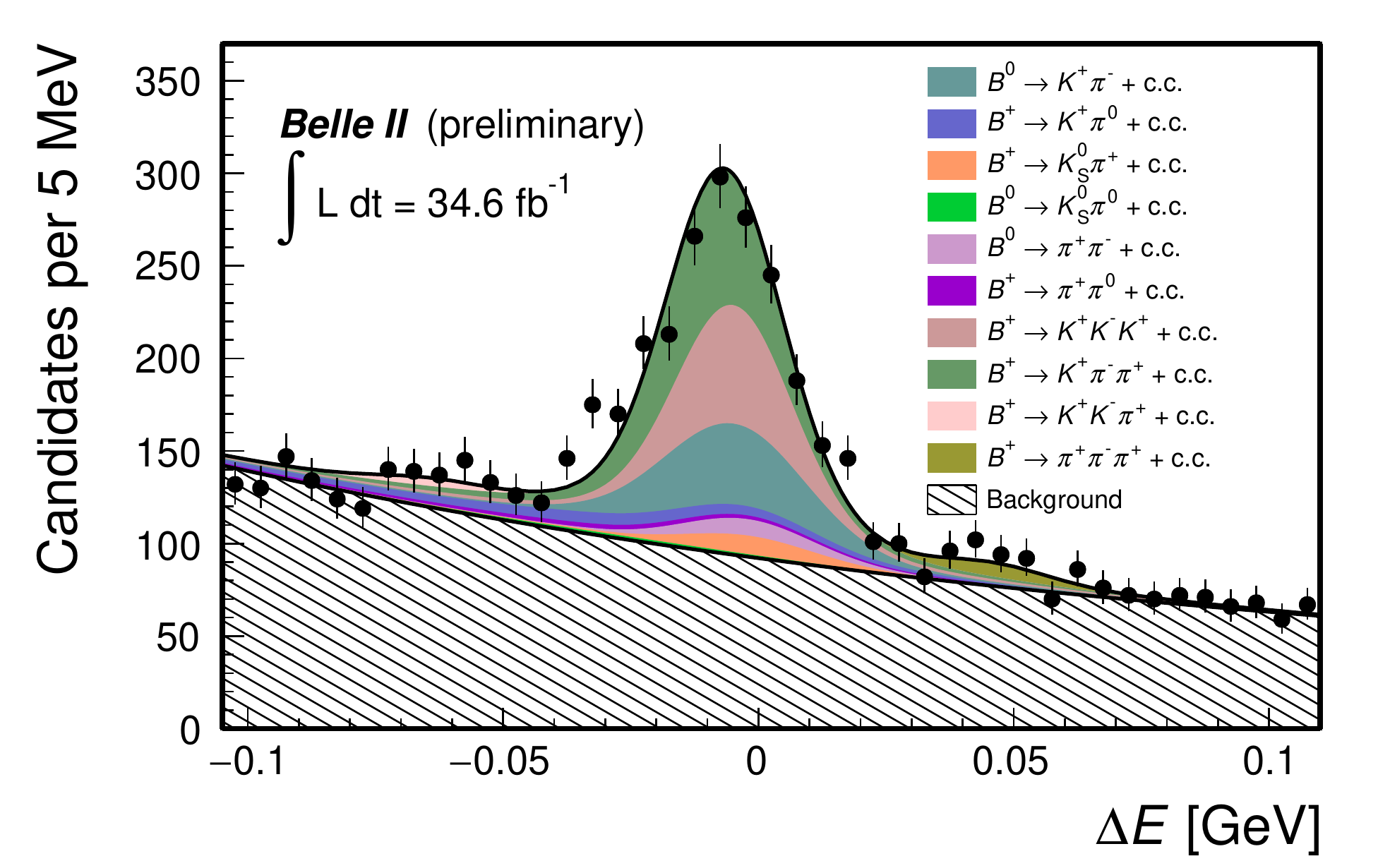}
\caption{Stacked $\Delta E$ distributions of charmless channels reconstructed in the Belle~II data set collected up to mid May 2020 with summed fit projections overlaid.}
 \label{fig:hype}
\end{figure}

We report on first measurements of  branching fractions ($\mathcal{B}$) and CP-violating charge asymmetries ($\mathcal{A}$) in charmless $B$ decays at Belle~II.  We use a sample of 2019 and 2020 data corresponding to $34.6\,\si{fb^{-1}}$ of integrated luminosity. We use simulation to devise optimized event selections. The $\Delta E$ distributions of the resulting samples, restricted in $M_{\rm bc}$, are fit to determine signal yields of approximately 290, 140, 65, 35, 60, 70, 360 and 450 for the channels \mbox{$B^0 \to K^+\pi^-$},
 \mbox{$B^+ \to K^+\pi^0$},
\mbox{$B^+ \to \PKzS\pi^+$},
\mbox{$B^0 \to \PKzS\pi^0$},
\mbox{$B^0 \to \pi^+\pi^-$}, 
\mbox{$B^+ \to \pi^+\pi^0$}, 
\mbox{$B^+ \to K^+K^-K^+$},  and \mbox{$B^+ \to K^+\pi^-\pi^+$}, totaling nearly 1500~charmless $B$~decays~(Fig.~\ref{fig:hype}). Signal yields are corrected for efficiencies determined from simulation and control data samples to obtain the following results,
\begin{center}
$\mathcal{B}(B^0 \to K^+\pi^-) = [18.9 \pm 1.4(\rm stat) \pm 1.0(\rm syst)]\times 10^{-6}$,
\end{center}
\begin{center}
$\mathcal{B}(B^+ \to K^+\pi^0) = [12.7 ^{+2.2}_{-2.1} (\rm stat)\pm 1.1(\rm syst)]\times 10^{-6}$,
\end{center}
\begin{center}
$\mathcal{B}(B^+ \to K^0\pi^+) = [21.8 ^{+3.3}_{-3.0}(\rm stat) \pm 2.9(\rm syst)]\times 10^{-6}$,
\end{center}
\begin{center}
$\mathcal{B}(B^0 \to K^0\pi^0) = [10.9^{+2.9}_{-2.6} (\rm stat)\pm 1.6(\rm syst)]\times 10^{-6}$, 
\end{center}
\begin{center}
$\mathcal{B}(B^0 \to \pi^+\pi^-) = [5.6 ^{+1.0}_{-0.9}(\rm stat) \pm 0.3(\rm syst)]\times 10^{-6}$,
\end{center}
\begin{center}
$\mathcal{B}(B^+ \to \pi^+\pi^0) = [5.7 \pm 2.3 (\rm stat)\pm 0.5(\rm syst)]\times 10^{-6}$,
\end{center}
\begin{center}
$\mathcal{B}(B^+ \to K^+K^-K^+) = [32.0 \pm 2.2(\rm stat.) \pm 1.4 (\rm syst)]\times 10^{-6}$, 
\end{center}
\begin{center}
$\mathcal{B}(B^+ \to K^+\pi^-\pi^+) = [48.0 \pm 3.8 (\rm stat)\pm 3.3 (\rm syst)]\times 10^{-6}$,
\end{center}
\begin{center}
$\mathcal{A}_{\rm CP}(B^0 \to K^+\pi^-) = 0.030 \pm 0.064 (\rm stat) \pm 0.008(\rm syst)$,
\end{center}
\begin{center}
$\mathcal{A}_{\rm CP}(B^+ \to K^+\pi^0) = 0.052 ^{+0.121}_{-0.119} (\rm stat)\pm 0.022(\rm syst)$,
\end{center}
\begin{center}
$\mathcal{A}_{\rm CP}(B^+ \to \PKzero\pi^+) = -0.072 ^{+0.109}_{-0.114}(\rm stat) \pm 0.024(\rm syst)$,
\end{center}
\begin{center}
$\mathcal{A}_{\rm CP}(B^+ \to \pi^+\pi^0) = -0.268 ^{+0.249}_{-0.322} (\rm stat)\pm 0.123(\rm syst)$,
\end{center}
\begin{center}
$\mathcal{A}_{\rm CP}(B^+ \to K^+K^-K^+) = -0.049 \pm 0.063(\rm stat) \pm 0.022 (\rm syst)$, and 
\end{center}
\begin{center}
$\mathcal{A}_{\rm CP}(B^+ \to K^+\pi^-\pi^+) = -0.063 \pm 0.081 (\rm stat)\pm 0.023(\rm syst)$.
\end{center}
These are the first measurements in charmless decays reported by Belle~II. 
 Results are compatible with known determinations and show detector performance comparable with the best Belle results offering a reliable basis to assess projections for future reach. All the inputs to verify the $K\pi$ isospin sum rule are now available except for $\mathcal{A}_{\rm CP}(B^0 \to K^0_{\rm S} \pi^0)$. Similarly, only the reconstruction of the $B^0 \to \pi^0\pi^0$ mode is missing for the $\alpha/\phi_2$ determination through $B \to \pi \pi$ decays. 

\clearpage

\appendix
\section{Improvements in baseline selection and continuum suppression}

\begin{figure}[htb]
 \centering
 \includegraphics[width=0.7\textwidth]{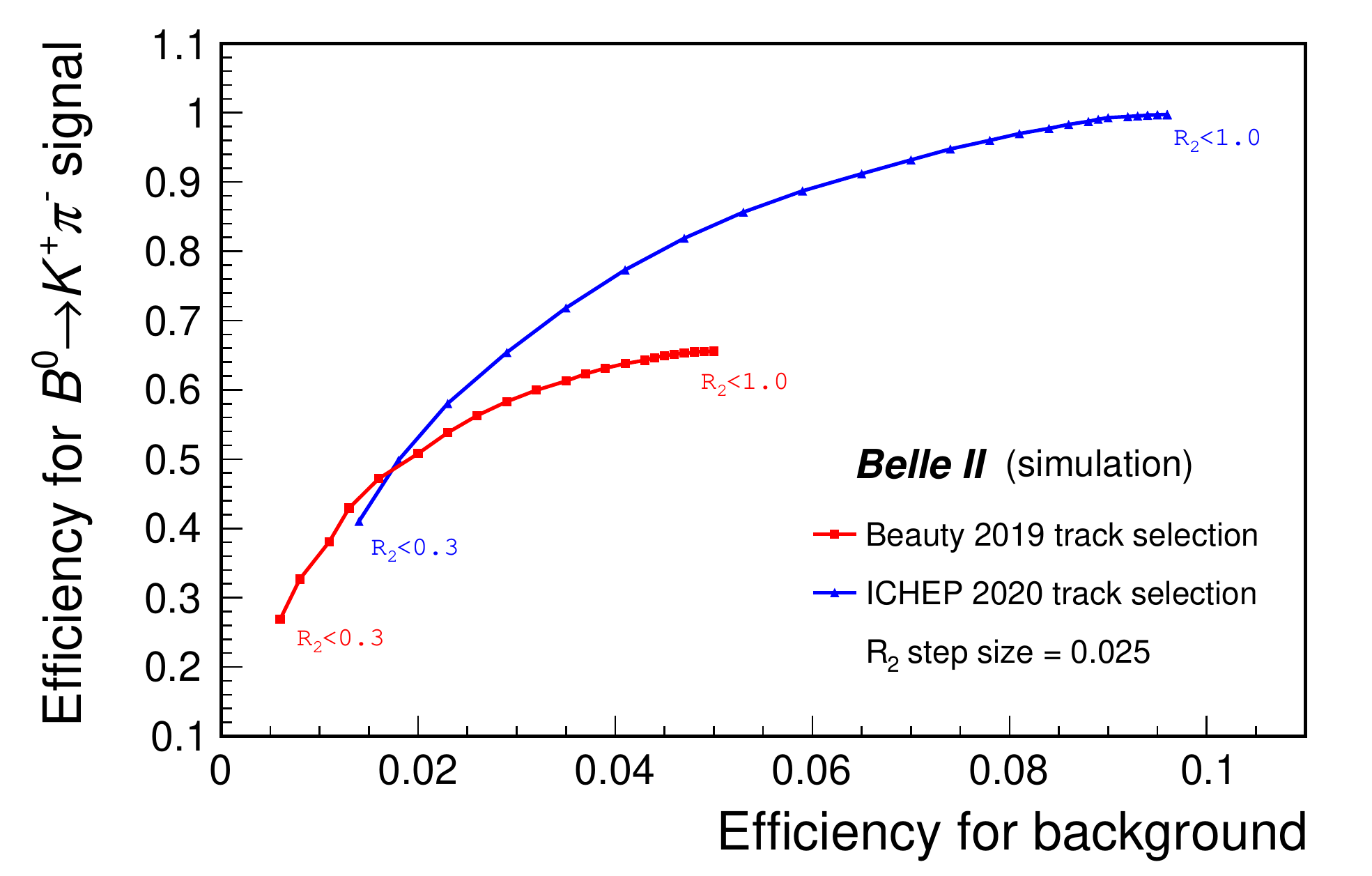}
\caption{Signal efficiency as a function of background efficiency for the benchmark decay mode $\PB\to\PKp\Pgpm$. The baseline selection criteria used for the first reconstruction shown at the (red)~Beauty~2019 conference is compared with the criteria used for the current results (blue). Every point corresponds to a different selection on the topological $R_2$~event variable. The improved performance of the blue curve is due to refined criteria for selection of lower-level physics primitives as charged-particle candidates.}
 \label{fig:Skim}
\end{figure}

Since the first reconstruction of $\PBzero\to\PKp\Pgpm$ shown at the Beauty~2019 conference~\cite{Benedikt:2019}, we refined the baseline selection criteria for charged particles and other physics primitives ($\pi^0$ candidates, $K^0_S$ candidates). Figure~\ref{fig:Skim} shows an example of the resulting performance improvement in terms of signal efficiency  as a function of background efficiency for the benchmark decay mode~\mbox{$\PBzero\to\PKp\Pgpm$}. 

\begin{figure}[htb]
\centering
  \includegraphics[width=0.7\textwidth]{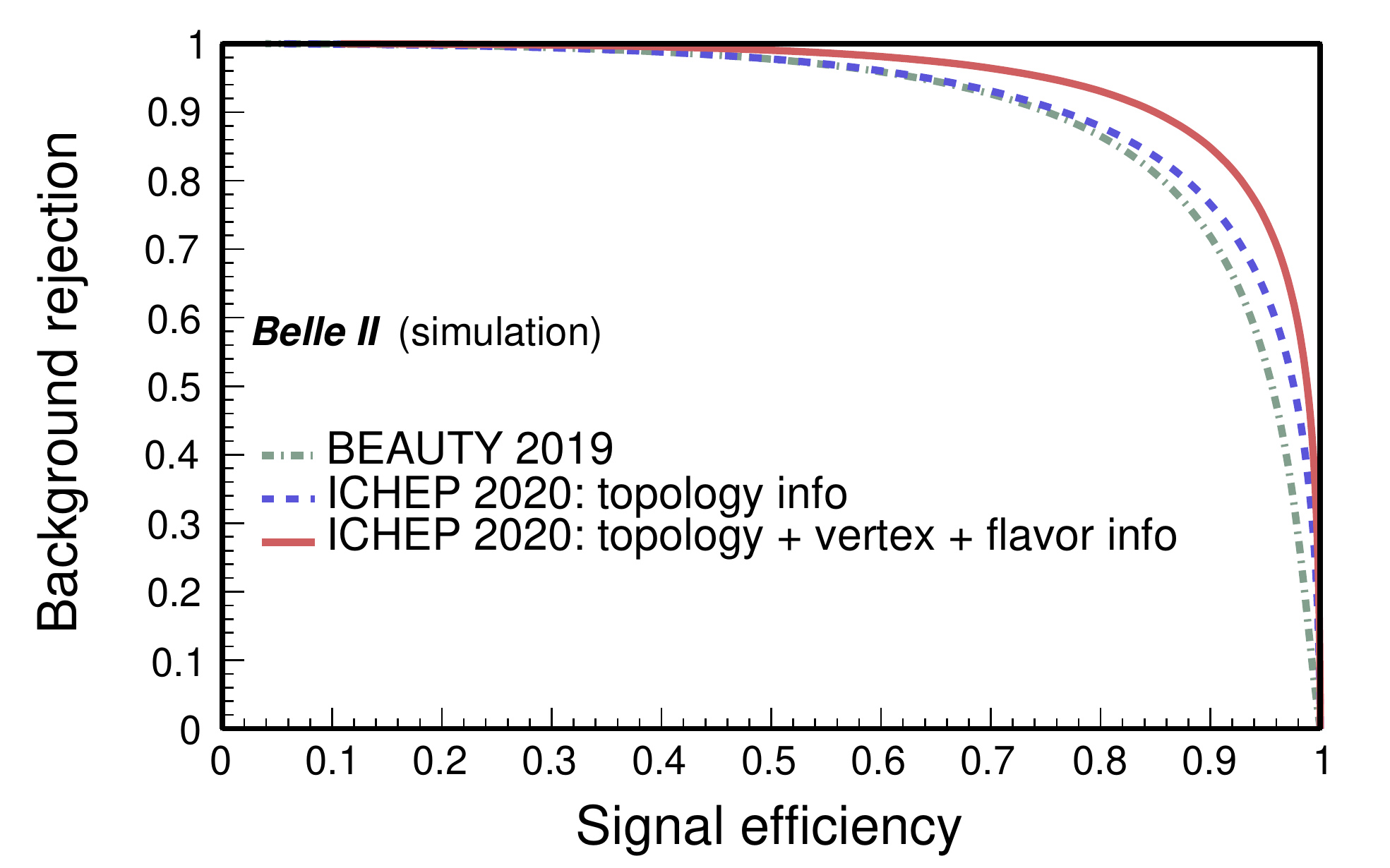}
 \caption{Receiver operating characteristic of the offline continuum-suppression classifier used for selecting $B^0 \to K^+\pi^-$ decays in the current results compared with the classifier used for the Beauty~2019 conference.}
 \label{fig:roc_Kpi}
\end{figure}

In addition, we achieved a 10\% improvement in continuum-background suppression by using additional input information on event topology together with flavor and vertex separation and vertex quality information. Figure~\ref{fig:roc_Kpi} compares the performance of the continuum suppression classifier used for the 2019 reconstruction of the first Belle II $B^0 \to K^+\pi^-$ signal with the performance of the classifier used for the current results. The performance of the current classifier is shown for two configurations, one using only event topology information, and one using event topology together with flavor, vertex separation and vertex quality information. 

\clearpage

\section{Examples of systematic-uncertainty validation}
\label{sec:sysapp}
All systematic uncertainties of the analysis are validated on data. Two examples of such validations follow. 

\begin{figure}[htb]
 \centering
 \includegraphics[width=0.475\textwidth]{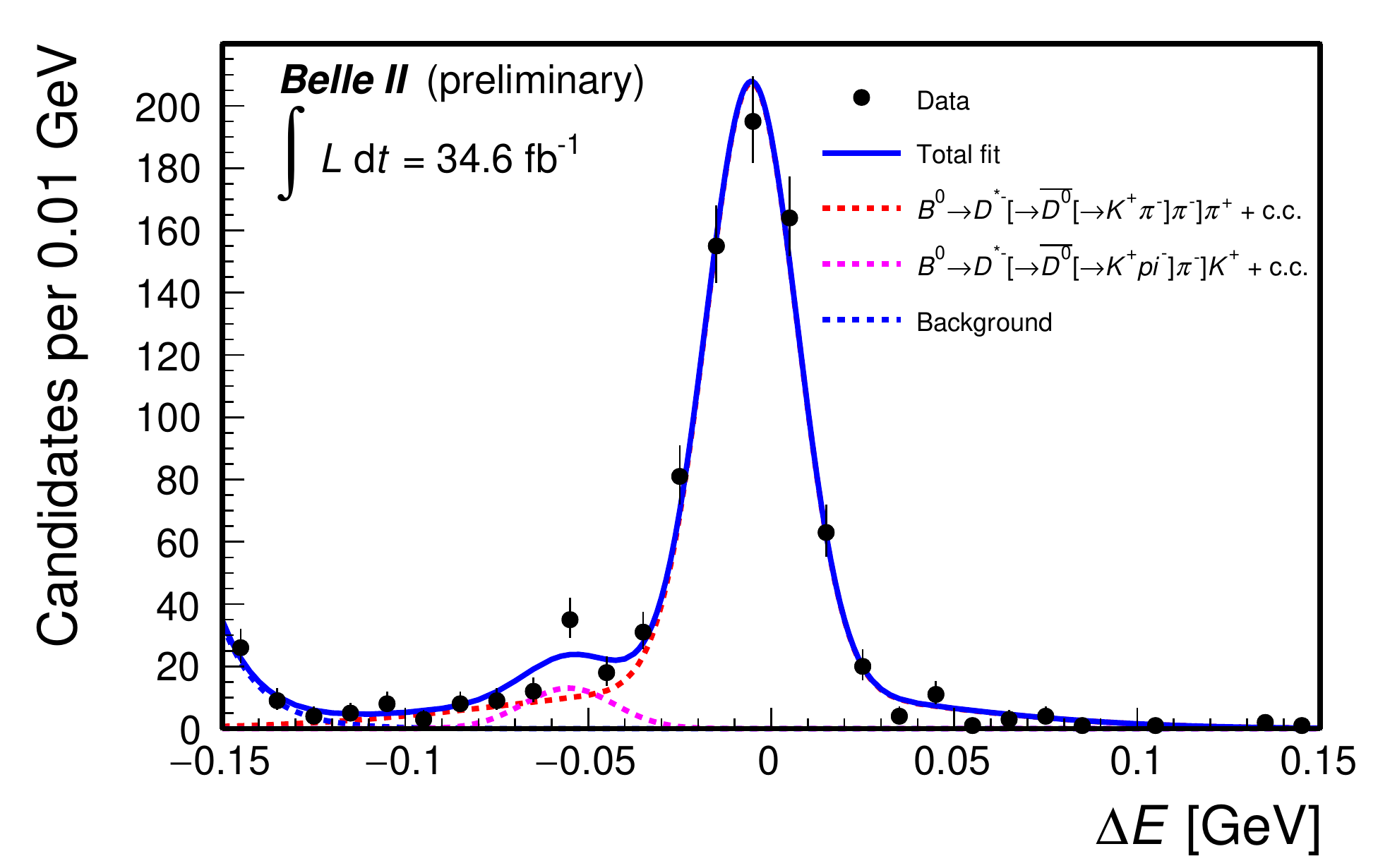}
 \includegraphics[width=0.475\textwidth]{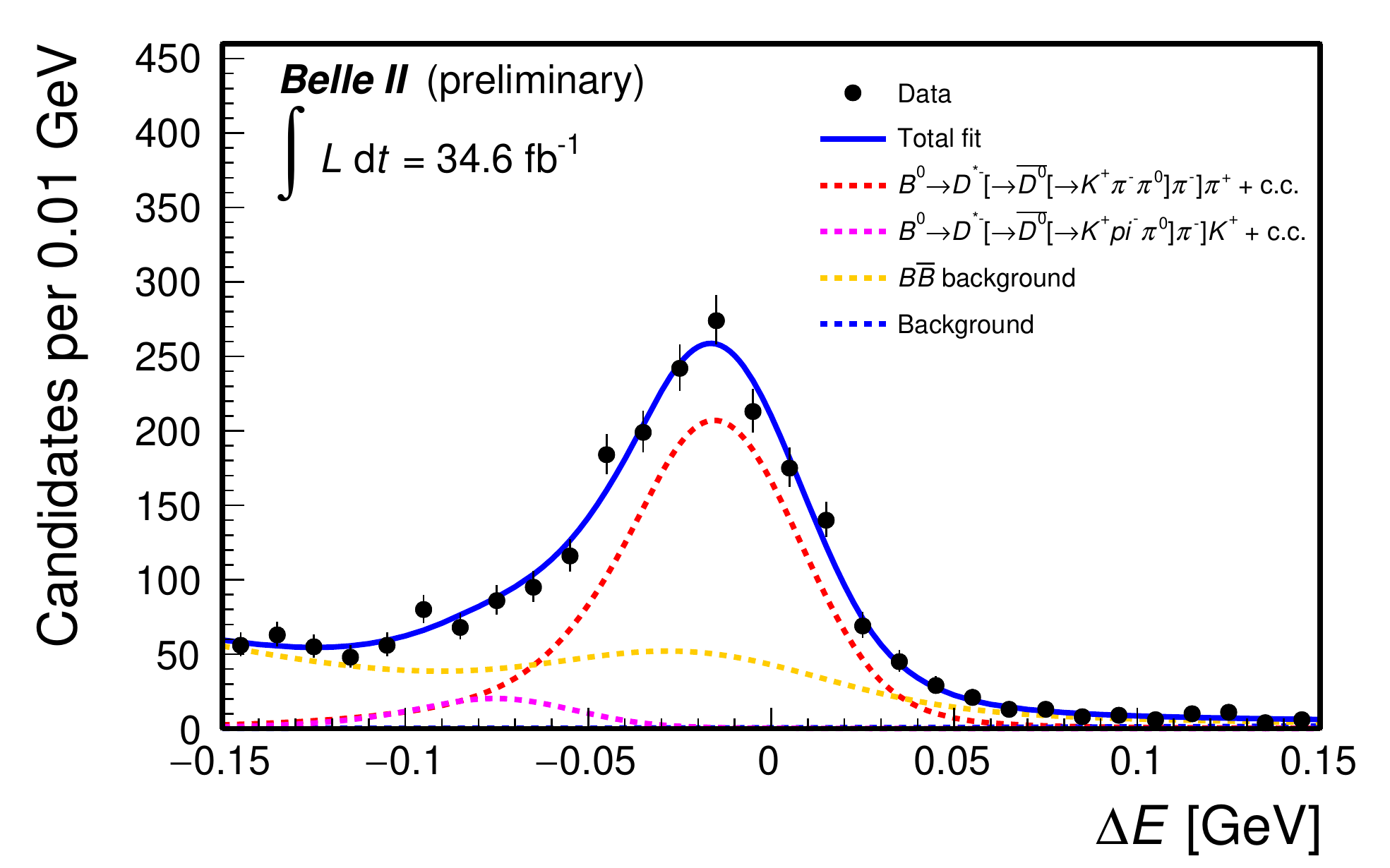}
 \caption{Distributions of $\Delta E$ for (left)~$\overline{B}^0 \to D^{*+}(\to D^0(\to K^-\pi^+)\pi^+)\pi^-$ and 
 (right)~$\overline{B}^0 \to D^{*+}(\to D^0(\to K^-\pi^+\pi^0)\pi^+)\pi^-$ candidates reconstructed in 2019--2020 Belle~II data selected through the baseline criteria with an optimized continuum-suppression and kaon-enriching selection, and further restricted to $M_{\rm bc} > 5.27$\,GeV/$c^2$. The projection of an unbinned maximum likelihood fit is overlaid.}
 \label{fig:fit_dE_Kpipi0}
\end{figure}

We use the ratio of reconstructed $\overline{B}^0 \to D^{*+}(\to D^0(\to K^-\pi^+\pi^0)\pi^+)\pi^-$ and \linebreak
$\overline{B}^0 \to D^{*+}(\to D^0(\to K^-\pi^+)\pi^+)\pi^-$ yields to obtain the \Pgpz-reconstruction efficiency in data.  Figure~\ref{fig:fit_dE_Kpipi0} shows the $\Delta E$ distributions  with fit projections overlaid used for yield determinations. 

\begin{figure}[htb]
 \centering
 \includegraphics[width=0.475\textwidth]{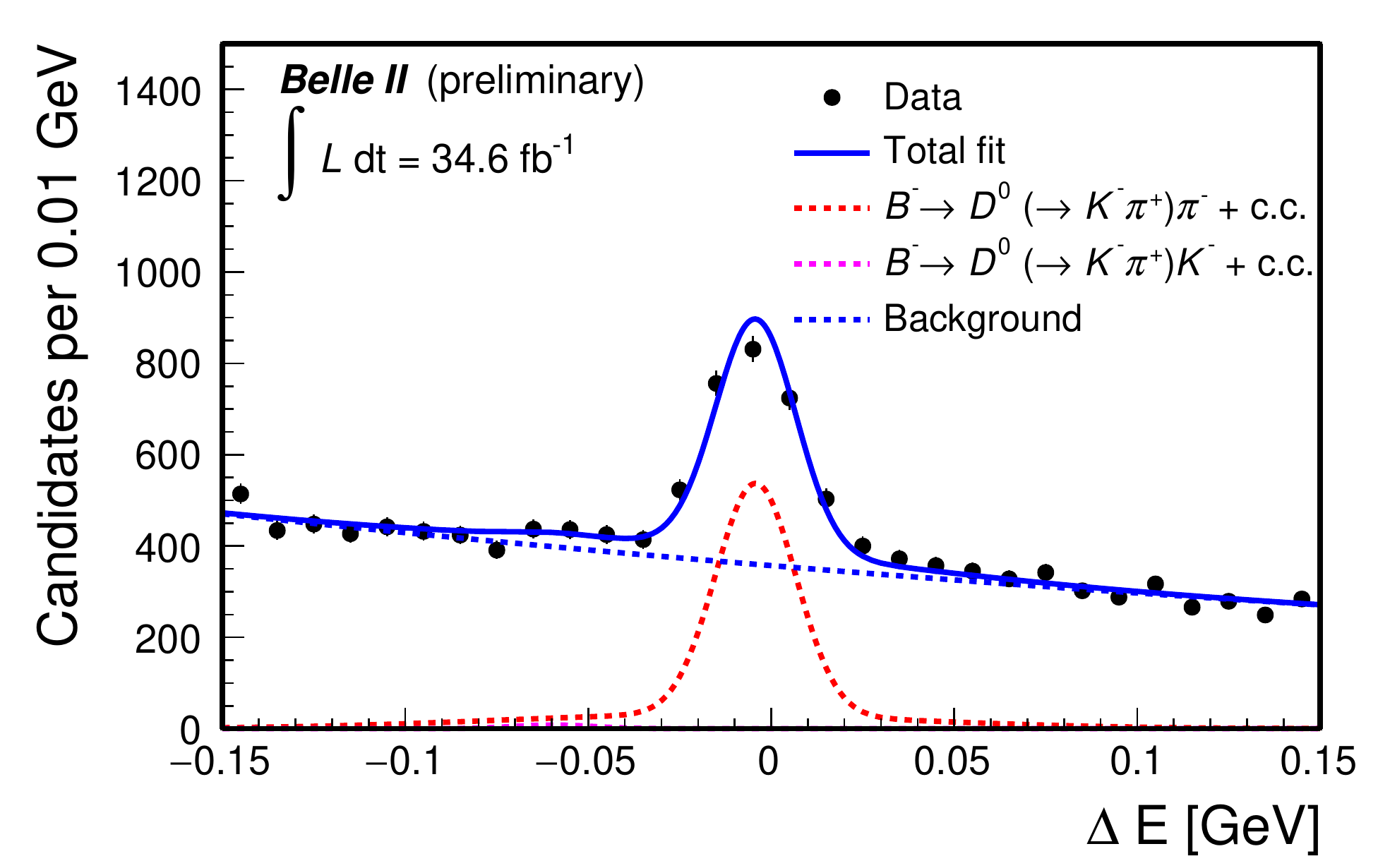}
 \includegraphics[width=0.475\textwidth]{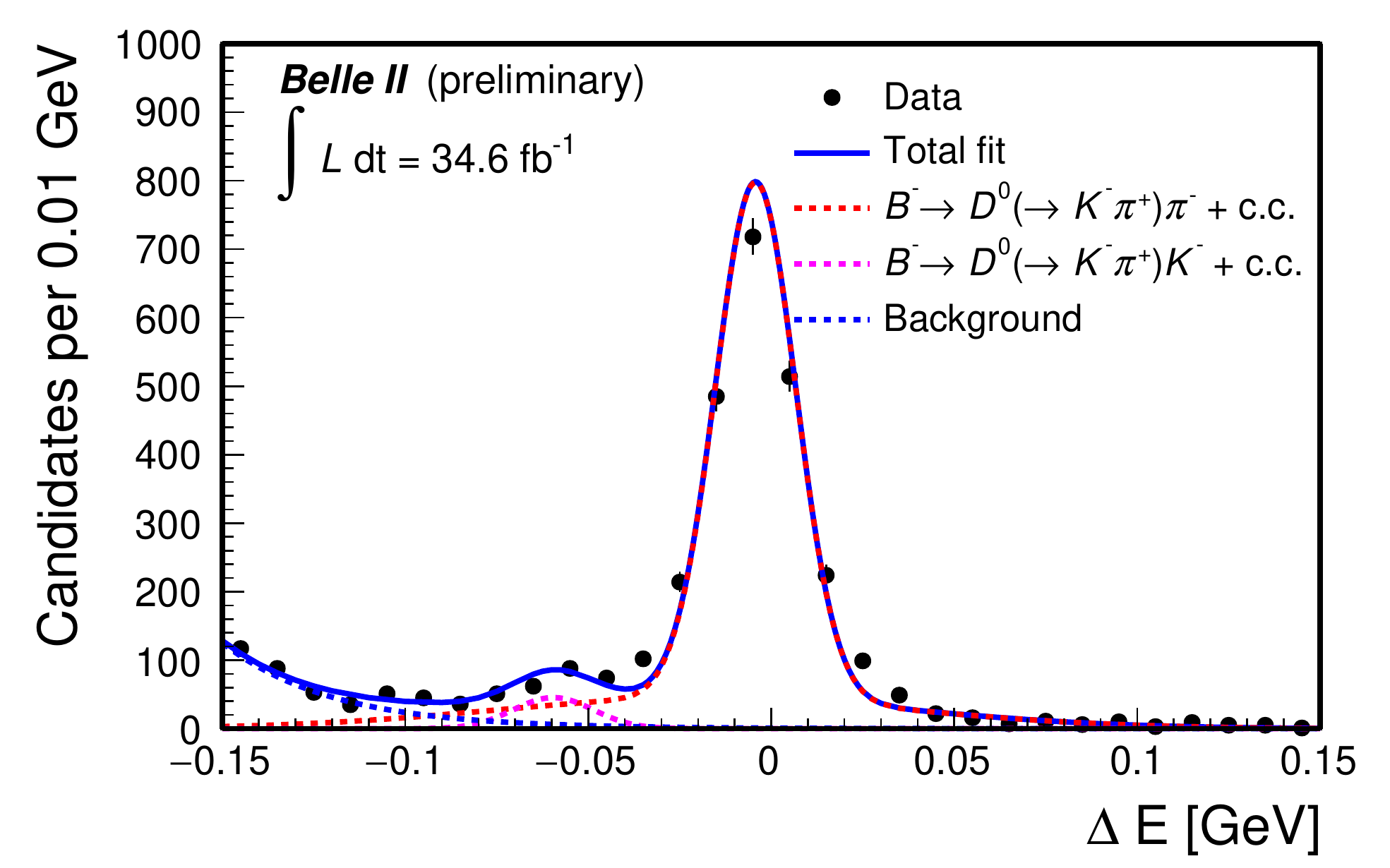}
 \caption{Distributions of $\Delta E$ for $B^- \to D^0(\to K^-\pi^+)\pi^-$ candidates reconstructed in 2019--2020 Belle~II data that (left) fail and (right) pass the optimized continuum-suppression and kaon-enriching selection. The projection of an unbinned maximum likelihood fit is overlaid.}
 \label{fig:fit_PIDCS}
\end{figure}

We use the fraction of reconstructed $B^- \to D^0(\to K^-\pi^+)\pi^-$ candidates that pass the kaon-enriching selection and the continuum-background selection to validate the corresponding efficiencies in data.  Figure~\ref{fig:fit_PIDCS} shows the corresponding $\Delta E$ distributions with fit projections overlaid for candidates that (left) failed and (right) met the continuum-suppression and kaon-enriching selection optimized for $\PBp\to\PKp\Pgpm\Pgpp$. We obtain the efficiency of the selection from a simultaneous fit to these two disjoint samples. 

\clearpage

\section*{Acknowledgments}

We thank the SuperKEKB group for the excellent operation of the
accelerator; the KEK cryogenics group for the efficient
operation of the solenoid; and the KEK computer group for
on-site computing support.
This work was supported by the following funding sources:
Science Committee of the Republic of Armenia Grant No. 18T-1C180;
Australian Research Council and research grant Nos.
DP180102629, 
DP170102389, 
DP170102204, 
DP150103061, 
FT130100303, 
and
FT130100018; 
Austrian Federal Ministry of Education, Science and Research, and
Austrian Science Fund No. P 31361-N36; 
Natural Sciences and Engineering Research Council of Canada, Compute Canada and CANARIE;
Chinese Academy of Sciences and research grant No. QYZDJ-SSW-SLH011,
National Natural Science Foundation of China and research grant Nos.
11521505,
11575017,
11675166,
11761141009,
11705209,
and
11975076,
LiaoNing Revitalization Talents Program under contract No. XLYC1807135,
Shanghai Municipal Science and Technology Committee under contract No. 19ZR1403000,
Shanghai Pujiang Program under Grant No. 18PJ1401000,
and the CAS Center for Excellence in Particle Physics (CCEPP);
the Ministry of Education, Youth and Sports of the Czech Republic under Contract No.~LTT17020 and 
Charles University grants SVV 260448 and GAUK 404316;
European Research Council, 7th Framework PIEF-GA-2013-622527, 
Horizon 2020 Marie Sklodowska-Curie grant agreement No. 700525 `NIOBE,' 
and
Horizon 2020 Marie Sklodowska-Curie RISE project JENNIFER2 grant agreement No. 822070 (European grants);
L'Institut National de Physique Nucl\'{e}aire et de Physique des Particules (IN2P3) du CNRS (France);
BMBF, DFG, HGF, MPG, AvH Foundation, and Deutsche Forschungsgemeinschaft (DFG) under Germany's Excellence Strategy -- EXC2121 ``Quantum Universe''' -- 390833306 (Germany);
Department of Atomic Energy and Department of Science and Technology (India);
Israel Science Foundation grant No. 2476/17
and
United States-Israel Binational Science Foundation grant No. 2016113;
Istituto Nazionale di Fisica Nucleare and the research grants BELLE2;
Japan Society for the Promotion of Science,  Grant-in-Aid for Scientific Research grant Nos.
16H03968, 
16H03993, 
16H06492,
16K05323, 
17H01133, 
17H05405, 
18K03621, 
18H03710, 
18H05226,
19H00682, 
26220706,
and
26400255,
the National Institute of Informatics, and Science Information NETwork 5 (SINET5), 
and
the Ministry of Education, Culture, Sports, Science, and Technology (MEXT) of Japan;  
National Research Foundation (NRF) of Korea Grant Nos.
2016R1\-D1A1B\-01010135,
2016R1\-D1A1B\-02012900,
2018R1\-A2B\-3003643,
2018R1\-A6A1A\-06024970,
2018R1\-D1A1B\-07047294,
2019K1\-A3A7A\-09033840,
and
2019R1\-I1A3A\-01058933,
Radiation Science Research Institute,
Foreign Large-size Research Facility Application Supporting project,
the Global Science Experimental Data Hub Center of the Korea Institute of Science and Technology Information
and
KREONET/GLORIAD;
Universiti Malaya RU grant, Akademi Sains Malaysia and Ministry of Education Malaysia;
Frontiers of Science Program contracts
FOINS-296,
CB-221329,
CB-236394,
CB-254409,
and
CB-180023, and SEP-CINVESTAV research grant 237 (Mexico);
the Polish Ministry of Science and Higher Education and the National Science Center;
the Ministry of Science and Higher Education of the Russian Federation,
Agreement 14.W03.31.0026;
University of Tabuk research grants
S-1440-0321, S-0256-1438, and S-0280-1439 (Saudi Arabia);
Slovenian Research Agency and research grant Nos.
J1-9124
and
P1-0135; 
Agencia Estatal de Investigacion, Spain grant Nos.
FPA2014-55613-P
and
FPA2017-84445-P,
and
CIDEGENT/2018/020 of Generalitat Valenciana;
Ministry of Science and Technology and research grant Nos.
MOST106-2112-M-002-005-MY3
and
MOST107-2119-M-002-035-MY3, 
and the Ministry of Education (Taiwan);
Thailand Center of Excellence in Physics;
TUBITAK ULAKBIM (Turkey);
Ministry of Education and Science of Ukraine;
the US National Science Foundation and research grant Nos.
PHY-1807007 
and
PHY-1913789, 
and the US Department of Energy and research grant Nos.
DE-AC06-76RLO1830, 
DE{}-SC0007983, 
DE{}-SC0009824, 
DE{}-SC0009973, 
DE{}-SC0010073, 
DE{}-SC0010118, 
DE{}-SC0010504, 
DE{}-SC0011784, 
DE{}-SC0012704; 
and
the National Foundation for Science and Technology Development (NAFOSTED) 
of Vietnam under contract No 103.99-2018.45.

\bibliography{belle2}
\bibliographystyle{belle2-note}

\end{document}